\pdfoutput=1
\newif\ifcomment
\newif\ifdraft
\newif\ifarxiv
\arxivtrue
\drafttrue
\ifdraft
\documentclass[11pt]{article}
\usepackage{color}
\usepackage{authblk}
\usepackage{lineno}
\else
\documentclass{ar-1col}
\fi
\usepackage[comma,square,numbers,sort&compress]{natbib}
\ifdraft
\usepackage[allcolors=red]{hyperref}
\else
\usepackage[hidelinks]{hyperref}
\fi
\usepackage{graphicx}
\usepackage{url}
\usepackage{amsmath}
\usepackage{amsfonts}
\usepackage{amssymb}
\usepackage[T1]{fontenc}
\usepackage[utf8]{inputenc}
\usepackage{mathrsfs}
\usepackage{calrsfs}
\usepackage{txfonts}
\usepackage{pslatex}
\usepackage{footnote}
\makesavenoteenv{tabular}
\makesavenoteenv{table}
\usepackage{upgreek}

\usepackage[symbol]{footmisc}

\newcommand{\pp}           {pp}
\newcommand{\np}           {np}

\newcommand{\nn}           {nn}
\renewcommand{\AA}         {AA}
\newcommand{\AB}           {AB}
\newcommand{\NN}           {NN}
\newcommand{\pPb}          {pPb}
\newcommand{\pA}           {pA}
\newcommand{\pAu}          {pAu}
\newcommand{\dAu}          {dAu}
\newcommand{\PbPb}         {PbPb}
\newcommand{\CuCu}         {CuCu}
\newcommand{\AuAu}         {AuAu}

\newcommand{\ppbar}        {$\rm p\bar p$}

\newcommand{\pT}           {\ensuremath{p_{\rm T}}}
\newcommand{\pt}           {\ensuremath{p_{\rm T}}}
\newcommand{\pTd}          {{\rm d}\pt}
\newcommand{\TAA}          {\ensuremath{T_{_{\rm AA}}}}
\newcommand{\TAB}          {\ensuremath{T_{_{\rm AB}}}} %{\mbox{\scriptsize{\it{AB}}}}
\newcommand{\TNN}          {\ensuremath{T_{_{\rm NN}}}}
\newcommand{\Tpp}          {\ensuremath{T_{_{\rm pp}}}}
\newcommand{\TA}           {\ensuremath{T_{_{\rm A}}}}
\newcommand{\TB}           {\ensuremath{T_{_{\rm B}}}}
\newcommand{\RAA}          {\ensuremath{R_{\rm AA}}}
\newcommand{\RpA}          {\ensuremath{R_{\rm pA}}}
\newcommand{\RAB}          {\ensuremath{R_{\rm AB}}}

\newcommand{\Lambdaqcd}    {\Lambda_{\ensuremath{\rm QCD}}}
\newcommand{\mean}[1]      {\ensuremath{\left<#1\right>}}

\newcommand{\eg}           {e.g.}
\newcommand{\ie}           {i.e.}
\newcommand{\cm}           {c.m.}
\newcommand{\dd}           {\ensuremath{{\rm d}}}

\newcommand{\snn}          {\ensuremath{\!\sqrt{s_{_{\rm NN}}}}}
\newcommand{\sqrts}        {\ensuremath{\sqrt{s}}}
\newcommand{\sigmapp}      {\ensuremath{\sigma_{\rm pp}}}
\newcommand{\sigmaNN}      {\ensuremath{\sigma_{\rm NN}}}
\newcommand{\sigmapPb}     {\ensuremath{\sigma_{\rm pPb}}}
\newcommand{\sigmaPbPb}    {\ensuremath{\sigma_{\rm PbPb}}}
\newcommand{\sigmapA}      {\ensuremath{\sigma_{\rm pA}}}
\newcommand{\sigmaAB}      {\ensuremath{\sigma_{\rm AB}}}
\newcommand{\sigmaAA}      {\ensuremath{\sigma_{\rm AA}}}
\newcommand{\sigmahard}    {\ensuremath{\sigma_{\rm hard}}}

\newcommand{\Ncoll}        {\ensuremath{N_{\rm coll}}}
\newcommand{\Npart}        {\ensuremath{N_{\rm part}}}

\newcommand{\Nhard}        {\ensuremath{N_{\rm hard}}}
\newcommand{\Nhardnn}      {\ensuremath{N^{\rm hard}_{\rm NN}}}

\newcommand{\bnn}          {\ensuremath{b_{_{\rm NN}}}}
\newcommand{\tnn}          {\TNN} %\ensuremath{T_{\rm NN}}}

\newcommand{\hrefurl}[1]   {\href{#1}{\url{#1}}}
\newcommand{\Refe}[1]      {Ref.~\cite{#1}}
\newcommand{\Refs}[1]      {Refs.~\cite{#1}}
\newcommand{\Tab}[1]       {Table~\ref{#1}}
\newcommand{\Fig}[1]       {Fig.~\ref{#1}}

\newcommand{\Eq}[1]        {Eq.~(\ref{#1})}
\newcommand{\Sec}[1]       {Sec.~\ref{#1}}
\newcommand{\App}[1]       {App.~\ref{#1}}
\newcommand{\Figure}[1]    {Figure~\ref{#1}}

\newcommand{\Section}[1]   {Section~\ref{#1}}
\newcommand{\hide}[1]      {\color{white}#1\color{black}}

\newcommand{\MCG}          {MCG}

\newcommand{\Hpythia}      {{\sc hg-pythia}}
\newcommand{\Hijing}       {{\sc hijing}}
\newcommand{\Pythia}       {{\sc pythia}}
\newcommand{\Herwig}       {{\sc herwig++}}
\newcommand{\epos}         {{\sc epos}}
\newcommand{\qgsjet}       {{\sc qgsjet}}
\newcommand{\dpmjet}       {{\sc dpmjet}}
\newcommand{\ampt}         {{\sc ampt}}
\newcommand{\angantayr}    {{\sc angantayr}}
\newcommand{\dipsy}        {{\sc dipsy}}

\newcommand{\starlight}    {{\sc starlight}}
\newcommand{\superchic}    {{\sc superchic}}
\newcommand{\trento}{T\raisebox{-0.5ex}{R}ENTo}
\newcommand{\MPI}          {MPI}
\newcommand{\com}[1]       {}

\ifdraft
\else
\jname{Annu.Rev.Nucl.Part.Sci.}
\jvol{7i}
\jyear{2020}
\doi{10.1146/annurev-nucl-0iiiii-iiiii}
\fi
\begin{document}

% Page header
\markboth{David d'Enterria and Constantin Loizides}{Progress in the Glauber model at collider energies}

% Title
\title{Progress in the Glauber model at collider energies}
%\title{The Glauber model at collider energies}
%\title{The Glauber model at the LHC}

%Authors, affiliations address.
\ifdraft
\author[1]{David d'Enterria\footnote{email: dde@cern.ch}}
\author[2]{Constantin Loizides\footnote{email: loizides@cern.ch}}
\affil[1]{\small CERN, EP Department, 1211 Geneva, Switzerland}
\affil[2]{\small ORNL, Physics Division, Oak Ridge, TN, USA}
\ifarxiv
 \date{\vspace{-0.5cm}\small\today}
 \enlargethispage{1cm}
\else
 \date{\vspace{-0.2cm}\small\today}
\fi
\maketitle
\vspace{-1.0cm}
\else
\author{David d'Enterria,$^1$ and Constantin Loizides$^2$
\affil{$^1$CERN, EP Department, 1211 Geneva, Switzerland; email: dde@cern.ch}
\affil{$^2$ORNL, Physics Division, Oak Ridge, TN, USA, email: loizides@cern.ch}
}
\fi

\begin{abstract}
We review the theoretical and experimental progress in the Glauber model of multiple nucleon and/or parton scatterings, after the last 10--15 years of operation with proton and nuclear beams at the CERN Large Hadron Collider~(LHC) and with various light and heavy colliding ions at the BNL Relativistic Heavy Ion Collider~(RHIC). The main developments and the state-of-the-art of the field are summarized. These encompass measurements of the inclusive inelastic proton and nuclear cross sections, advances in the description of the proton and nuclear density profiles and their fluctuations, inclusion of subnucleonic degrees of freedom, experimental procedures and issues related to the determination of the collision centrality, validation of the binary scaling prescription for hard scattering cross sections, and constraints on transport properties of quark-gluon matter from varying initial-state conditions in relativistic hydrodynamics calculations.
These advances confirm the validity and usefulness of the Glauber formalism for quantitative studies of QCD matter produced in high-energy collisions of systems, from protons to uranium nuclei, of vastly different size. 
\end{abstract}

\ifdraft
\else
%Keywords, etc.
\begin{keywords}
Glauber model, hadronic collisions, centrality determination, Monte Carlo event generators, inclusive hadronic cross sections
\end{keywords}
\maketitle
\fi

%Table of Contents
\tableofcontents

\ifcomment
The Editorial Committee encourages you to present a critical discussion of the current status of the field, rather than an encyclopedic coverage of papers. In short, we want a compelling and up-to-date presentation that communicates the opportunities and excitement of the subject. We also welcome your personal perspective, especially with respect to what you think is most important and where the field is going—yet the presentation must be balanced.

We ask that you limit the review to 10,200 words of main text, up to 8 figures/tables, and a maximum of 150 references—counts that will produce our desired 25 typeset pages. Please note that each small figure or table requires space approximately equal to 300 words of text and each large one 600 words; thus, for every figure/table, subtract the appropriate number of words from your total allotment. Be sure to include a brief abstract (150 words at maximum) as well as up to 6 keywords associated with your topic; do not include these elements in your word count.

Due Date: October 2, 2020
Length: 10,200 words (including up to 8 figures/tables) plus 150 references (25 pages in the final volume)
Author Handbook and Graphics Guide: http://www.annualreviews.org/page/authors/general-information (select the name of your journal from the drop-down list under Article Preparation and Submission) \fi

%%%%%%%%%%%%%%%%%%%%%%%%%%%%%%%%%%%%%%%%%%%%%%%%%%%%%%%%%%%%%%%%%%%%%%%%%%%%%%%%%%%
\section{INTRODUCTION}
\label{sec:intro}
%%%%%%%%%%%%%%%%%%%%%%%%%%%%%%%%%%%%%%%%%%%%%%%%%%%%%%%%%%%%%%%%%%%%%%%%%%%%%%%%%%%
The central goal of high-energy nucleus-nucleus~(\AA) collisions is to study the collective thermo\-dynamic and transport properties of quarks and gluons, the elementary degrees of freedom of the theory of strong interaction~(Quantum Chromodynamics, QCD)~\cite{Busza:2018rrf}.
Colliding heavy-ions at center-of-mass~(\cm) energies above tens of GeV is the only experimental way known to produce a large multibody system of deconfined partons, the quark-gluon plasma~(QGP), predicted by lattice QCD calculations for energy densities exceeding a critical value of $\upepsilon_{\rm c}\approx$\,0.5\,GeV/fm$^3$~\cite{Borsanyi:2013bia,Bazavov:2014pvz}.
\com{It is worth noting that}Such collisions provide the only available means to study the thermodynamics of a non-Abelian quantum gauge field theory in the laboratory.

%On scales compared to the proton radius, nuclei are rather extended objects.
Larger transverse spatial overlaps of the two incoming ions lead to a larger volume of the created system thereby resulting in ``mesoscopic'' conditions conductive to partial~(local) thermalization and subsequent collective behavior of the produced partons. 
Hence, the interpretation of data from high-energy heavy-ion collisions relies on a detailed knowledge of the initial QGP matter distribution, resulting from the overlap of the two nuclei colliding at a given impact parameter~($b$), complemented with a theoretical modeling of its subsequent spacetime evolution. 
Deriving from the data the transverse size of the created QGP, as well as any event-by-event irregularities\com{(``lumpiness'' or ``hot spots'') present in the matter distribution} arising from 
density and/or color fluctuations, is crucial to interpret the experimental observations, because the total and local depositions of energy in the collision chiefly influence the initial conditions and the evolution of the produced strongly-interacting system.

The generic method to study the properties of QCD matter relies on measuring the distributions of a variety of observables in \AA\ collisions, and compare them to the same measurements in proton-proton~(\pp) and/or proton-nucleus (\pA) collisions, where a QGP is not expected to be produced. Common observables include particle production yields~(light hadrons, heavy-quarks, quarkonia, jets, photons, etc.) as functions of their transverse momentum (\pT), pseudorapidity ($\eta$), azimuthal angle ($\phi$)~\cite{dEnterria:2006mtd}. 
Carrying out such quantitative comparisons among colliding systems of different sizes requires appropriate normalization of the measured distributions by using \eg\ the number of participating nucleons~($\Npart$), the number of independent binary nucleon-nucleon~(\NN) collisions~($\Ncoll$), the medium transverse area ($A_\perp$), or the system eccentricity~$\epsilon_n$ (given by the $n^{\rm th}$ moments of its azimuthal spatial distribution), as described below. 

%Since nuclei are extended objects, with their radii being up to almost an order of magnitude larger than that of nucleons, a natural way to characterize the nuclear overlap region is to compute properties of the interacting or participating nucleons, such as their number~($\Npart$) and eccentricity~($\ep$) as well as higher moments of their spatial distribution.
%Another important quantity obtained from the interacting nucleons is the total number of independent, binary nucleon--nucleon collisions~($\Ncoll$), which represents the increased probability for a rare process to occur in \AA\ relative to proton--proton~(\pp) collisions.

Computing the aforementioned quantities typically relies on describing multiple scatterings of the constituents~(nucleons, themselves characterized by parton densities) inside the colliding nuclei, using the so-called Glauber formalism, named after Roy Glauber's pioneering work on the use of quantum mechanical scattering theory\com{ for composite systems} to calculate cross sections in \pA\ and \AA\ collisions~\cite{Martin:1958zz,Glauber:1970jm}.
The Glauber formalism is based on the geometric or {\it eikonal} approximation, which assumes projectile nucleons to travel along straight lines~(\ie\ with negligible momentum exchanges compared to their longitudinal momenta) and to undergo multiple independent subcollisions with nucleons in the target.
In the original form of the Glauber approach~\cite{Czyz:1969jg,Bialas:1976ed,Bialas:1977pd}, the total hadronic cross section of a collision of two nuclei (with $A$ and $B$ number of nucleons, respectively) is given by a $(2A+2B+1)$-dimensional integral %in the transverse plane
\begin{eqnarray}
\label{eq:sigmaAB}
\sigmaAB = \int {\rm d}^2 b \int {\rm d}^2 {\rm s}^{\rm A}_1 \cdots {\rm d}^2 {\rm s}^{\rm A}_A {\rm d}^2 {\rm s}^{\rm B}_1 \cdots {\rm d}^2 {\rm s}^{\rm B}_B \times \nonumber \\
\TA({\bf s}^{\rm A}_1)\cdots \TA({\bf s}^{\rm A}_A) \TB({\bf s}^{\rm B}_1)\cdots \TB({\bf s}^{\rm B}_B) \times \\
\left\{ 1 - \prod_{j=1}^{B} \prod_{i=1}^{A} \left[ 1 - {\sigma} ({\bf b}-{\bf s}^{\rm A}_i + {\bf s}^{\rm B}_j \right] \right\}\;, \nonumber
\end{eqnarray}
where ${\bf b}$ is the collision impact parameter and ${\bf s}$ denotes a position in the transverse plane. The interaction probability ${\sigma}({\bf s})$ is normalized to give the nucleon--nucleon inelastic cross section $\sigmaNN = \int {\rm d}^2{\rm s}\,{\sigma}({\bf s})$. 
The {\it nuclear thickness function}  ${T}_{\rm A}({\bf b})=\int {\rm d}z\,\rho_A({\bf b},z)$ describes the transverse nucleon density by integrating the nuclear density $\rho$ along the longitudinal direction~($z$).

In the so-called {\it optical} limit of the Glauber model derived in~\cite{Bialas:1977pd}, local density fluctuations and correlations are ignored so that each nucleon in the projectile interacts with the incoming target as a flux tube described with a smooth density. 
The total cross section\com{ in the optical limit} then reduces to
\begin{equation}
\sigmaAB = \int {\rm d}^2 b \left\{ 1-\left[1-\sigmaNN \TAB({\bf b}) \right]^{AB}  \right\}\;,
\label{eq:sigmaAB2}
\end{equation}
where
\begin{equation}
\TAB\left(\bf{b}  \right) = \int { {\rm d}^2 {\rm s}\; \TA \left(\bf{s} \right)\TB \left(\bf{s}  - \bf{b}  \right)}
\label{eq:tab}
\end{equation}
is known as the {\it nuclear overlap} function, normalized as $\int \dd^2b\; \TAB(b)= A\,B$ by integrating over all impact parameters.
%Note that $\hat{T}\left(\bf{b}\right)$ has the unit of inverse area, which can be interpreted as the effective overlap area for which a specific nucleon in A can interact with a given nucleon in B.
Expressions~(\ref{eq:sigmaAB}) and (\ref{eq:sigmaAB2}) give identical results for large enough nuclei and/or for sufficiently small values of $\sigmaNN$.

In contrast to optical calculations, Monte Carlo Glauber~(\MCG) calculations~\cite{Wang:1991hta,Broniowski:2007nz,Alver:2008aq,Alvioli:2009ab,Rybczynski:2013yba,Loizides:2014vua,Loizides:2016djv,Mitchell:2016jio,Loizides:2017ack,Bozek:2019wyr} evaluate the phase space of \Eq{eq:sigmaAB} stochastically by distributing the $A, B$ nucleons of nucleus A and B, respectively, in coordinate space according to their corresponding nuclear densities, separated by an impact parameter $b$ sampled from\footnote{Hereafter, for simplicity, the impact-parameter variable $b$ ($\bnn$) is used to denote the distance between nuclei~(or nucleons).} $\mathrm{d}\sigma/\mathrm{d}b \propto b$.
The nuclear transverse profiles are usually approximated with parametrizations of charge density distributions extracted from low-energy electron-nucleus scattering experiments~\cite{DeJager:1974liz,DeJager:1987qc} that, for large spherical nuclei, are usually described by 2-parameter Fermi~(2pF) (also called Woods-Saxon) distributions, $\rho(r)=\rho_0/(1+\exp(\frac{r-R}{a}))$, with half-density radius~$R$ and diffusivity~$a$. 
Following the eikonal approximation, the collision is treated as a sequence of independent binary nucleon-nucleon collisions, \ie\ the nucleons travel on straight-line trajectories and their interaction probability does not depend on the number of collisions they have suffered before. 
In its simplest form, an interaction takes place between two nucleons if the distance $d$ between their centers satisfies $d < \sqrt{\sigmaNN/\pi}$. 
Alternatives to this so-called ``black disk'' approximation, using for example a Gaussian-like distribution for the nucleon overlap or more complex forms, are also used as discussed below.

\begin{figure}[th!]
\centering
\includegraphics[width=0.49\textwidth]{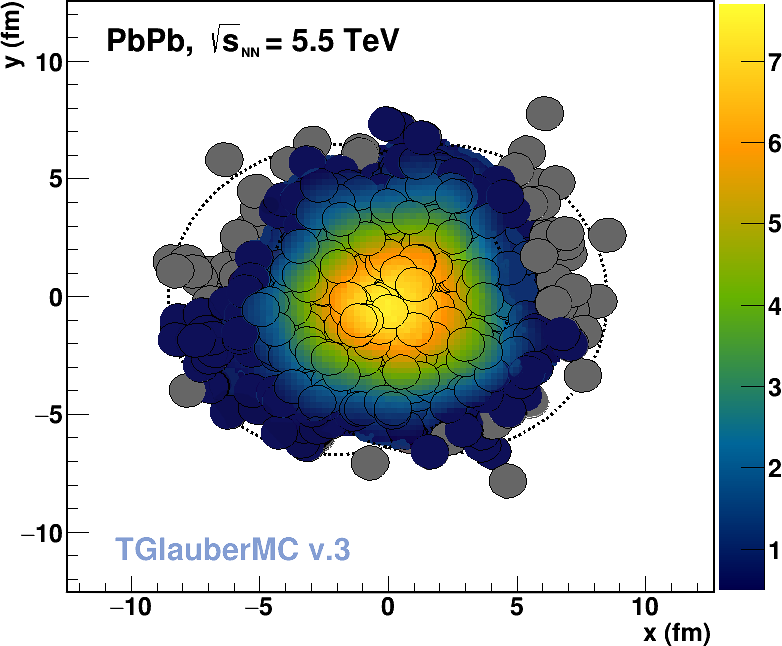}
\hspace{0.1cm}
\includegraphics[width=0.49\textwidth]{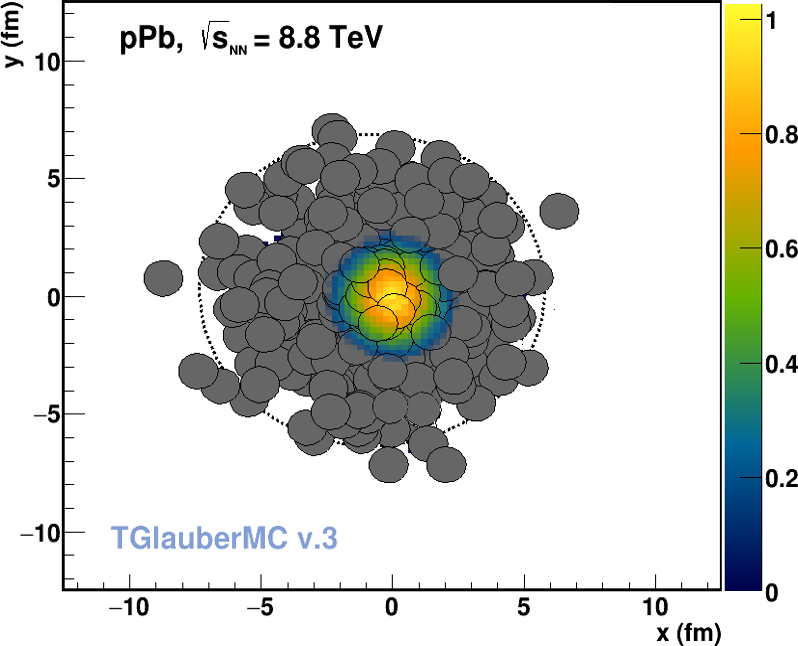}
\caption{Events displays of initial transverse density profiles in the $(x,y)$ plane of \PbPb\ collisions at $\snn = 5.5$\,TeV\com{ impact parameter $b=3$--4~fm)} with a triangular-like shape (5--10\% centrality, left), and \pPb\ at $\snn = 8.8$\,TeV (0--1\% centrality, right) generated with TGlauberMC~\cite{Loizides:2017ack}. 
The colored bidimensional surfaces indicate the local density (fm$^{-2}$), following the weighting given by \Eq{eq:s0_MCG} with $\alpha = 0.2$. 
The grey circles indicate spectator nucleons that do not participate in the collision process.
\label{fig:evt_displays}}
\end{figure}

One of the most typical phenomenological applications of the Glauber model is to provide the initial conditions for the number density~(or entropy or energy densities, related to the former via the QCD equation of state, EoS) of the medium formed in nuclear collisions as input for hydrodynamic calculations of its subsequent space-time evolution. 
The initial entropy profile in the transverse plane at midrapidity ($\eta=0$) is typically assumed to be proportional to a linear combination of the number density of particles produced in soft (with yields assumed to scale as the \Npart\ pairs) and hard (scaling as \Ncoll) scatterings~\cite{Wang:2000bf} %~\cite{Hirano:2012kj}
\begin{equation}
s_0(\vec{r}_\perp) \equiv \frac{dS}{\tau_0\,\dd^2r_\perp \dd\eta}\bigg|_{\eta=0} = \frac{C}{\tau_0}\left( \frac{1-\alpha}{2}\,\rho_{\rm part}(\vec{r}_\perp) + \alpha\,\rho_{\rm coll}(\vec{r}_\perp)  \right)\,,
\label{eq:s0_MCG}
\end{equation}
with the relative weight\footnote{The historical wounded nucleon Model (WNM)~\cite{Bialas:1976ed} for low-energy heavy-ion collisions assumes an entropy deposit only for each nucleon that engages in one or more inelastic collisions, and would correspond to $\alpha = 0$.} often taken as $\alpha = 0.2$ at the LHC. %The \Npart\ and  \Ncoll\ components account mostly for particle production from soft and hard \NN\ scatterings, respectively. 
Two typical snapshots in the $(x,y)$ plane of the medium formed in \PbPb\ and \pPb\ collisions at the LHC are shown in \Fig{fig:evt_displays}. 
Right after the collision, the spectator nucleons (in grey in the figure) continue undisturbed with the original longitudinal momenta inside the beam line, whereas the hot and dense QGP formed at midrapidity at the LHC expands longitudinally (transversely) at about (0.6 times) the speed of light.
%Different features of both snapshots are discussed below.

Among the quantities derived from Glauber models, the initial azimuthal anisotropy of the QGP system is an important one as it directly propagates to various flow components in the final state through collective hydrodynamic pressure gradients, and is thereby particularly sensitive to the transport and thermodynamic properties of QCD matter~\cite{Ollitrault:1992bk}. The harmonics eccentricities $\epsilon_n$ of the produced QGP (with $n=2$ being its ellipticity, $n=3$ triangularity, etc.) are theoretically defined as
\begin{equation}
\epsilon_n \equiv \frac{\left|\int r^n e^{in\phi} \,w(r,\phi) rdrd\phi\right|}{\int r^n \, w(r,\phi) rdrd\phi}\,,
\label{eq:ecc}
\end{equation}
where the weights $w(r,\phi)$ are often taken as the initial density of the medium derived from the Glauber model itself via a linear combination of the underlying \Npart\ and \Ncoll\ distributions, as in \Eq{eq:s0_MCG}. 
Assuming linear hydrodynamics response, the final harmonic flows determined from the azimuthal distributions of the produced hadrons via $\frac{dN}{d\phi}=\frac{N}{2\pi}\left(1+\sum_n(2v_n\cos(n(\phi-\psi_n))\right)$ where $\psi_n$ is the flow plane angle for harmonic flow $v_n$, are proportional to the initial eccentricity: $v_n = \kappa\, \epsilon_n$, with $\kappa\approx 0.2$ depending on the EoS, and deviations being sensitive to the ratio of the medium shear viscosity over entropy density $\upeta/s$~\cite{Romatschke:2009im,Teaney:2009qa,Luzum:2009sb,Schenke:2010rr,Heinz:2013th,Weller:2017tsr}. 
Whereas flow studies historically focused on the elliptic component~($v_2$), analyses of the higher Fourier harmonics have blossomed in the last years because of the large flow signals observed in the RHIC and LHC data~\cite{Alver:2010gr,ALICE:2011ab,Adare:2011tg,ATLAS:2012at,Chatrchyan:2013nka}.
%that are found to be strongly sensitive to the underlying QGP shear viscosity~\cite{Romatschke:2009im,Teaney:2009qa,Luzum:2009sb,Schenke:2010rr,Weller:2017tsr}. 
A typical \PbPb\ event at the LHC, leading to a medium with a triangular-like profile, is displayed in \Fig{fig:evt_displays}~(left).

Another phenomenologically relevant quantity often derived from \MCG\ models is the average QGP path length $L(b)$ traversed by a given perturbative probe, such as an energetic parton, produced in the collision~\cite{Dainese:2004te,Lokhtin:2005px,Djordjevic:2018ita}. 
The $L(b)$ dependence of the energy loss suffered by a parton going through a QGP provides valuable information on the jet quenching mechanism and on the medium properties~\cite{dEnterria:2009xfs}.

All Glauber quantities mentioned above~(\TAA, \Ncoll, \Npart, $A_\perp$, $\epsilon_n$, $L$) depend on the impact parameter $b$, which is not directly measurable by the experiments, but is monotonically correlated with the overall multiplicity of produced particles in the collision:
A smaller impact parameter, \ie\ a more {\it central} collision, will on average lead to a higher particle multiplicity. The reaction {\it centrality} is usually expressed in percentiles of the total inelastic hadronic cross section, with 0\% meaning ``most central'', \ie\ fully head-on collisions at $b = 0$~fm, and 100\%  meaning ``most peripheral'', \ie\ grazing collisions beyond which there is no QCD interaction\footnote{Significant electromagnetic (e.m.) interactions of the ions can also happen in ``ultraperipheral interactions''~\cite{Baltz:2007kq} at impact parameters larger than the sum of their nuclear radii.}~(corresponding to $b \gtrsim R_A + R_B$, where $R_{A}$ and $R_{B}$ are the nuclear radii).
Experimental measurements performed in intervals of multiplicity can be mapped to centrality ranges using Glauber-based simulations, and from there extended to all other relevant quantities.

This review aims at summarizing the main theoretical and experimental progress in the Glauber formalism over the $10$--$15$ years passed since a previous review summarized the status of the field at the BNL RHIC energies~\cite{Miller:2007ri}. 
Key differences at the LHC (with TeV colliding energies per nucleon, up to 50 times larger than those studied at the previous RHIC energy frontier) compared to previous accelerators, are driven by the fact that the nucleon substructure becomes more important, and that even the particle multiplicities produced in small collision systems~(\pp, \pPb) can reach values as large as those measured in \CuCu\ collisions at 200~GeV~\cite{Alver:2010ck}. 
An example of the high-density medium formed in \pPb\ collisions at the LHC is schematically shown in \Fig{fig:evt_displays}~(right).
Many of the Glauber model developments have gone in parallel to experimental and theoretical studies of \pp, \pA, and \AA\ collisions at the CERN LHC, as well as to prospective studies for future facilities such as the Future Circular Collider (FCC)~\cite{Benedikt:2018csr}. 
The document is organized as follows.
\Section{sec:sigma_inel} covers measurements and calculations of the proton and nuclear inelastic cross sections.
\Section{sec:IS} discusses improvements in the description of the proton and nucleus transverse profiles. 
\Section{sec:pheno} reviews phenomenological applications of the Glauber model. 
\Section{sec:experim} examines the experimental methods used to determine the collision centrality and discusses their inherent biases. 
The review ends with a summary of the main conclusions in \Section{sec:summ}.
\ifarxiv
Appendix~A provides some basic technical details of MCG simulations.
\fi
%%%%%%%%%%%%%%%%%%%%%%%%%%%%%%%%%%%%%%%%%%%%%%%%%%%%%%%%%%%%%%%%%%%%%%%%%%%%%%%%%%%%%%%%%%%%%%%%%%%%%
%\section{INCLUSIVE INELASTIC PROTON AND NUCLEAR CROSS SECTIONS}
\section{INELASTIC PROTON AND NUCLEAR CROSS SECTIONS}
\label{sec:sigma_inel}
%%%%%%%%%%%%%%%%%%%%%%%%%%%%%%%%%%%%%%%%%%%%%%%%%%%%%%%%%%%%%%%%%%%%%%%%%%%%%%%%%%%%%%%%%%%%%%%%%%%%%
A key ingredient of all Glauber calculations, via Eqs.~(\ref{eq:sigmaAB}) and (\ref{eq:sigmaAB2}), is the inclusive inelastic nucleon-nucleon cross section, $\sigmaNN$, evaluated at the same \cm\ energy $\snn$ as the \pA\ or \AA\ collision under consideration. 
The value of $\sigmaNN$ receives contributions from both (semi)hard parton-parton scatterings~(aka.\ ``minijets''), computable above a given $\pT\approx$~2\,GeV cutoff by perturbative QCD (pQCD) approaches, as well as from softer ``peripheral'' scatterings of diffractive nature, with a scale not very far from $\Lambdaqcd\approx0.2$\,GeV. 
Due to the latter nonperturbative contributions, that cannot be computed from first-principle QCD calculations today, one resorts to phenomenological fits of the experimental data to predict the evolution of $\sigmaNN$ as a function of $\snn$. 
At high \cm\ energies, above a few tens of GeV, any potential difference due to the valence-quark structure is increasingly irrelevant, as the bulk of the pQCD cross section proceeds through gluon-gluon scatterings, and all experimental measurements of \pp\ and \ppbar\ (as well as \nn\ and \np) collisions can be combined to extract $\sigmaNN$. 

\begin{figure}[ht!]
\centering
\includegraphics[width=0.9\textwidth]{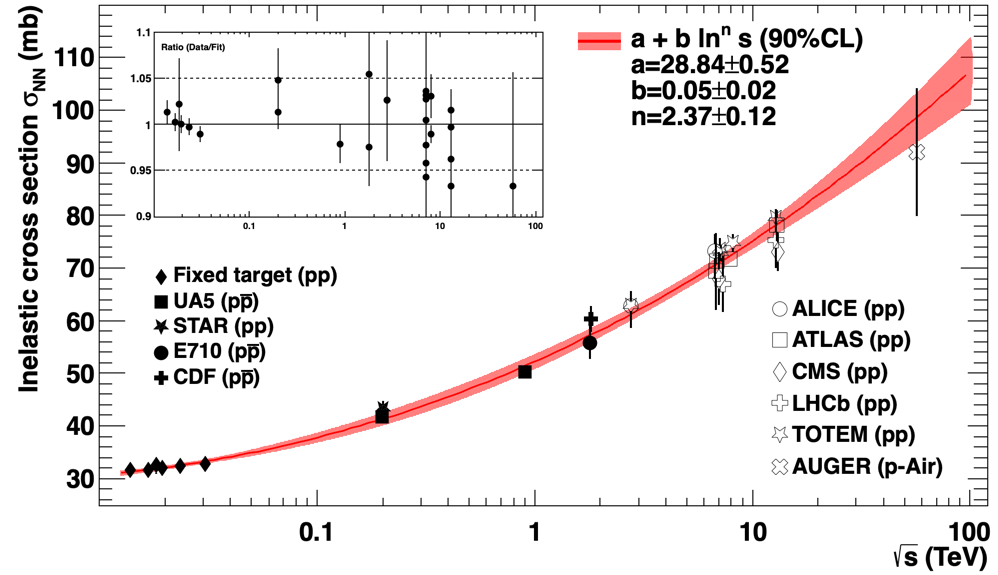} 
\caption{Inelastic \pp\ (\ppbar) cross sections as a function of \cm\ energy measured over $\sqrts\approx10$\,GeV\,--\,$100$\,TeV fitted to Eq.~(\ref{eq:signnfunc}). The inset shows the ratio of the data divided by the value of the fit.}
\label{fig:sigmappvs}
\end{figure}

The collision-energy dependence of the inelastic cross section $\sigmaNN$ is shown in \Fig{fig:sigmappvs} from all available measurements today, starting off with fixed-target studies in the range $\snn \approx 10$--30\,GeV performed in the 1970s--90s~\cite{Tanabashi:2018oca}, combined with data at \ppbar\ (UA5~\cite{Alner:1986iy} at $\sqrts=200$ and $900$\,GeV, E710~\cite{Amos:1990jh,Amos:1991bp} and CDF~\cite{Abe:1993xy,Abe:1993xx} at $\sqrts=1.8$\,TeV) and \pp\ (STAR~\cite{Adam:2020ozo} at $\sqrts=200$\,GeV, ALICE at 7 TeV~\cite{Abelev:2012sea}, ATLAS at 7, 8, and 13\,TeV~\cite{Aad:2011eu,Aad:2014dca,Aaboud:2016ijx,Aaboud:2016mmw}, CMS at 7 and 13\,TeV~\cite{Chatrchyan:2012nj,Sirunyan:2018nqx}, LHCb at 7 and 13\,TeV~\cite{Aaij:2014vfa,Aaij:2018okq}, and TOTEM at 2.76, 7, 8, and 13\,TeV~\cite{Antchev:2011vs,Antchev:2013iaa,Antchev:2013paa,Antchev:2017dia,Cafagna:2020izf}) colliders, as well as the AUGER result at $\sqrts=57$\,TeV derived from cosmic-ray (proton-air) data~\cite{Abreu:2012wt}.
The experimental $\sigmaNN$ values are obtained either (i) via the subtraction $\sigmapp = \sigma_{\rm tot} - \sigma_{\rm el}$, where $\sigma_{\rm tot}$ and $\sigma_{\rm el}$ have been accurately measured in dedicated forward Roman pot detectors (TOTEM~\cite{Antchev:2011vs,Antchev:2013iaa,Antchev:2013paa,Antchev:2017dia,Cafagna:2020izf} and ALFA~\cite{Aad:2014dca,Aaboud:2016ijx,Aaboud:2016mmw}), or (ii) from measurements of inelastic particle production data in the central detectors collected with MB triggers.
The latter measurements are less precise than the former, as they require an extrapolation, dominated by diffractive contributions, to forward regions of phase space not covered by the detectors. Many Glauber MC codes, such as {\sc glissando}~\cite{Broniowski:2007nz,Rybczynski:2013yba,Bozek:2019wyr}, use the COMPETE collision-energy parametrization  of the \NN\ cross section~\cite{Cudell:2002xe}, as provided in the PDG review~\cite{Tanabashi:2018oca}. However, such a relatively complex multiparameter expression is only required when aiming at a reproduction of the \NN\ cross section data over the full collision energy range experimentally measured, down to $\sqrts \approx 1$\,GeV.
For the regime of energies relevant at colliders, the $\sqrts$ dependence of the experimental $\sigmaNN$ results can be fit to a simpler parametrization, as done in~\cite{Loizides:2017ack},
\begin{equation}
\sigmaNN(s) = a + b \,\ln^{n}(s)\,,
\label{eq:signnfunc}
\end{equation}
%with the exponent fixed either to $n=1$ or $n=2$, or otherwise left free in the fit\footnote{The choice $n=2$ represents the asymptotic $\sqrts$-dependence that saturates the Froissart bound~\cite{Froissart:1961ux}.}. 
%The parameter values and goodness-of-fit $\chi^2/N_{\rm dof}$ for the three fits are listed in \Tab{tab:snnfitvals}. 
%The $n=2$ case is used as central value for the interpolation (and extrapolation) of $\sigmaNN$ as a function of $\sqrts$, and the difference~(normalized by 2.4 to account for the full width at half maximum) of the so-derived $\sigmaNN$ values to those obtained with the $n=1,2.6$ fits is assigned as systematic uncertainty~(red band in \Fig{fig:sigmappvs}).
resulting in $a=28.84\pm0.52$, $b=0.0458\pm0.0167$ and $n=2.374\pm0.123$ with goodness-of-fit over number of degrees of freedom $\chi^2/N_{\rm dof}=0.7$~(and $s$ given in GeV$^2$ units). The first row of \Tab{tab:signnvalues} lists the derived $\sigmaNN$ values relevant for Glauber calculations at RHIC, LHC, and FCC energies. The value predicted for the top RHIC energy of $\snn=0.2$\,TeV, $\sigmaNN = 41.3 \pm 1.2$~mb, is consistent with (although, more precise than) the value directly measured by STAR ($\sigmaNN=43.82^{+1.39}_{-1.46}$ mb)~\cite{Adam:2020ozo}, as well as with the typical $42\pm3$~mb used so far in the RHIC literature~\cite{Miller:2007ri}. 
At the top FCC energy of $\snn=100$\,TeV, the expected cross section of $\sigmaNN=107.5\pm6.5$~mb is also in agreement with the value $105.1\pm2.0$~mb derived from the average of various model predictions~\cite{dEnterria:2016oxo}.

\ifcomment %values fits starting at 0.2 TeV
\begin{table}[t!]
\centering
\begin{tabular}{l|c|c|c|c}\hline
 Type   & $A$ & $B$ & $n$    & $\chi^2/N_{\rm dof}$ \\\hline
 $\ln s$        & $-3.38\pm1.70$   & $4.210\pm0.112$    & $1$~(fixed)      & $1.4$ \\
 $\ln^2 s$      & $25.3\pm0.9$     & $0.146\pm0.004$    & $2$~(fixed)      & $0.82$  \\
 $\ln^n s$      & $31.6\pm4.2$     & $0.023\pm0.036$    & $2.60\pm0.50$    & $0.76$ \\
 \hline
 \par
\end{tabular} 
\caption{\label{tab:snnfitvals}Fit values and $\chi^2/N_{\rm dof}$ for the collision-energy dependence of $\sigmaNN$ parametrized by $\sigmaNN(s) = A + B \,\ln^{n}(s)$, \Eq{eq:signnfunc} and displayed in \Fig{fig:sigmappvs}.}
\end{table}
\fi

\begin{table}[ht!]
\centering
\caption{\label{tab:signnvalues} Top row: Values of the nucleon-nucleon inelastic cross section $\sigmaNN$ extracted from \Eq{eq:signnfunc} and the data plotted in \Fig{fig:sigmappvs}. Central and bottom rows: Computed values of the \pA\ and \AA\ hadronic cross sections at \cm\ energies relevant for collisions at RHIC, LHC, and FCC. The quoted values are for \pPb\ and \PbPb\ collisions, except for the results at 0.2\,TeV that correspond to \pAu\ and \AuAu\ collisions. 
} 
\small
\begin{tabular}{l|ccccccc}
\hline
 $\sqrts$ (TeV)      &  0.2                    &  5.5           & 8.8      &  14                     &  39                     &  63                    &  100 \\\hline
 $\sigmaNN$ (mb) & $41.3\pm1.2$  & $68.3\pm1.2$   & $73.6\pm1.4$  & $79.2\pm1.9$    & $93.0\pm3.8$  & $100.2\pm5.1$ & $107.5\pm6.5$\\\hline
 $\sigmapA$ (b) & $1.75\pm0.03$ &$2.10\pm0.03$  &  $2.13\pm0.03$ & $2.17\pm0.03$ & $2.26\pm0.04$ & $2.30\pm0.05$ & $2.35\pm0.05$\\
 
 $\sigmaAA$ (b) & $6.84\pm0.14$ & $7.64\pm0.15$ & $7.71\pm0.16$ & $7.78\pm0.16$ & $7.93\pm0.16$ & $8.00\pm0.17$ & $8.07\pm0.17$\\
\hline
\end{tabular} 
\end{table}
\ifarxiv
\enlargethispage{0.5cm}
\fi
From $\sigmaNN$, one can then derive the corresponding values for the \pA\ and \AA\ inelastic cross sections making use of \Eq{eq:sigmaAB2}. 
The $\sigmapA$ and $\sigmaAB$ results for the \pPb\ and \PbPb\ (\pAu\ and \AuAu\ for RHIC) systems are listed in the middle and bottom rows of \Tab{tab:signnvalues}.
The quoted uncertainties account for the propagated $\sigmaNN$ uncertainties plus, in quadrature, the resulting uncertainty of independently varying the density parameters by 1~standard deviation. The latter amount to about 2\% and 1.5\% for \PbPb\ and \pPb\ collisions, respectively, and are dominated by the uncertainty of the neutron skin width.
The Glauber calculation gives $\sigmaPbPb^{\rm MC}=7.55\pm0.15$~b at $\snn=2.76$\,TeV and $\sigmapPb^{\rm MC}=2.09\pm0.03$~b at $\snn=5.02$\,TeV, in good agreement with the measured values of $\sigmaPbPb=7.7\pm0.6$~b~\cite{ALICE:2012aa}, and $\sigmapPb=2.06\pm0.08$~b~\cite{Khachatryan:2015zaa} as well as $\sigmapPb = 2.10 \pm 0.07$~b~\cite{Abelev:2014epa}, respectively.
\ifarxiv
At 5.02~TeV, $\sigmaPbPb=7.62\pm0.15$~b for $\sigmaNN=67.3\pm1.2$~mb. %For 41mb, it is 7.24
\fi
The measured ALICE and CMS inelastic \pPb\ and \PbPb\ cross sections provide therefore an inclusive validation of the Glauber model at TeV energies. %, as previously at RHIC~\cite{Chiu:2001ij}. 
This fact further justifies the common application of the Glauber approach to derive \pp\ cross section from cosmic-ray proton-air cross section~measurements~\cite{Abreu:2012wt}.

%%%%%%%%%%%%%%%%%%%%%%%%%%%%%%%%%%%%%%%%%%%%%%%%%%%%%%%%%%%%%%%%%%%%%%%%%%%%%%%%%%%
\section{DEVELOPMENTS IN THE INITIAL STATE}
\label{sec:IS}
%%%%%%%%%%%%%%%%%%%%%%%%%%%%%%%%%%%%%%%%%%%%%%%%%%%%%%%%%%%%%%%%%%%%%%%%%%%%%%%%%%%
In the past fifteen years, there have been numerous developments related to the description of the nucleon and nuclei radial profiles in Glauber models including \eg\ the incorporation of various sources of event-by-event fluctuations~(particularly relevant for collisions involving light-ions and/or protons), subnucleonic degrees of freedom, neutron skin effects, and deformed light- and heavy-ion distributions. 
The most important advances on these topics are summarized in this section.

%%%%%%%%%%%%%%%%%%%%%%%%%%%%%%%%%%%%%%%%%%%%%%%%%%%%%%%%%%%%%%%%%%%%%%%%%%%%%%%%%%%%%%%%%%%%%%%%
\subsection{Proton Transverse Profile}
The transverse profile of the proton (or, generically, of the nucleon) is of key importance in many aspects of the Glauber formalism.
First, in nuclear collisions, it determines the \NN\ interaction probability, and its realistic description is particularly important to properly describe proton-nucleus results~(where intrinsic fluctuations in the proton shape are more relevant than in \AA\ collisions). 
Second, it is a prime ingredient of MC event generators for \pp\ collisions such as \Pythia\,8~\cite{Sjostrand:2014zea} or \Herwig~\cite{Bahr:2008pv} in order to reproduce, through the underlying multiparton interactions~(MPIs), the properties of inclusive hadron production both in  minimum-bias (MB) collisions ---as a sum of the particle production activity from all, central to peripheral, \pp\ collisions--- and in the so-called ``underlying event'' (UE) accompanying hard scatterings at the LHC~\cite{Field:2011iq}. 
Third, the proton transverse distribution is also a basic element in calculations of double- and triple- parton scatterings cross sections in proton and nuclear collisions, where the effective N-parton scattering (NPS) cross section bears a simple geometric interpretation in terms of powers of the inverse of the integral of the \pp\ overlap function over all impact parameters, $\sigma_{\rm eff,\textsc{nps}}=\left[\int \dd^2b \,\Tpp^n({\bf b})\right]^{-1/(n-1)}$ (with $n=2,3,..,N$ for double-, triple-, $N$-scatterings)~\cite{dEnterria:2017yhd}.
Improving the treatment of \pp\ collisions within a Glauber-like approach based on their impact parameter and underlying parton-parton scatterings has attracted a growing interest in the last years to interpret high-multiplicity \pp\ results where collective~(QGP-like) phenomena have been observed in the data~\cite{Khachatryan:2010gv,Aad:2015gqa,Khachatryan:2016txc}, a possibility anticipated in~\cite{dEnterria:2010xip}.

The most simplistic approach used in Glauber models is to consider a fixed proton shape at all colliding energies, disregarding any varying distributions of its parton contents~(valence and sea quarks, gluons) and their correlations. 
The simplest form is a hard-sphere parametrization with uniform density, $\rho(r)={1 \over 4/3\;\pi R^3}\Theta\left(r-R\right)$ with radius consistent with electron-proton scattering fits giving a root-mean-square radius, $R_{\rm rms}\approx0.8$~fm~\cite{Hofstadter:1956qs}. 
A profile more consistent with the proton charge form factor is given by an exponential, $\rho(r)={1 \over 8\pi R^3}e^{-r/R}$, reproducing to a large extent the spatial distribution of its valence quarks, with $R=R_{\rm rms}/\sqrt{12}=0.234$~fm. 
A single Gaussian ansatz, although not very realistic, makes subsequent calculations especially transparent and hence was used in some analytical approaches. 
A double Gaussian ansatz, $\rho(r) \propto \frac{1 - \beta}{a_1^3} \, e^{-r^2/a_1^2} + \frac{\beta}{a_2^3} \,e^{-r^2/a_2^2}$, which corresponds to a distribution with a small core region of radius $a_2$ containing a fraction $\beta$ of the total matter embedded in a larger region of radius $a_1$, is a common choice, available \eg\ in the \Pythia\,8 MC generator~\cite{Corke:2011yy}.

In MC event generators for \pp\ collisions, parametrizations of the overlap function as a function of impact parameter, rather than the individual radial density of each colliding proton, are usually used. 
In \Pythia\,8, the \pp\ overlap is parametrized as
\begin{equation} 
\Tpp({\bf b})= \frac{m}{2\pi r^2_p \,\Gamma (2/m)} \exp{[-(b/r_p)^m]}\,,
\label{eq:overlapmonash}
\end{equation}
normalized to one, $\int \dd^2b\,\Tpp({\bf b}) = 1$, where $r_p$ is a characteristic ``radius'' of the proton, $\Gamma$ is the gamma function, and the exponent $m$ varies between a more Gaussian-like~($m\approx2$) to a more peaked exponential-like~($m\approx1$) distribution. 
The popular \Pythia\,8 ``Monash'' parameter settings (tune) uses $m=1.85$~\cite{Skands:2014pea}.
%THe \Herwig\ code uses an alternative parametrization~\cite{Seymour:2013qka} based on the dipole fit of the two-gluon form factor in the momentum representation, $F_{2g}({\bf q})=1/(q^2/m^2_g+1)^2$, where the gluon mass $m_g$ parameter characterizes the transverse momentum $q$ distribution of the proton~\cite{Blok:2010ge}, and the transverse density is obtained from its Fourier-transform: $f({\bf b})=\int e^{-i{\bf b}\cdot{\bf q}}F_{2g}({\bf q})\frac{\dd^2q}{(2\pi)^2}$.
%In the \Hijing~\cite{Wang:1991hta}) and \Herwig~\cite{Bahr:2008pv} generators, the overlap is given by $\TNN(\bnn) \propto \xi^3 \, K_{3}(\xi)$ with $\xi\propto\bnn/\!\snn$, where $K_3$ is the modified Bessel function of the third kind, obtained by approximating the Fourier transform of the proton electromagnetic form factor.
In the \Hijing~\cite{Wang:1991hta} and \Herwig~\cite{Bahr:2008pv} generators, the \pp\ overlap is approximated by the Fourier transform of the proton e.m.\ form factor as
\begin{equation}
\Tpp({\bf b})= \frac{\mu^2}{96\pi} \, (\mu\,b)^3\,K_3(\mu\,b) \,,
\label{eq:overlaphijing}
\end{equation}
where $\mu\propto 1/r_p$ is a free parameter\com{ on scales of the inverse of the proton radius}, and $K_3$ is the modified Bessel function of the third kind.
In \Hijing, $\mu=3.9/\!\sqrt{\sigmaNN/2\pi}$ is used to describe the dependence of $\mu$ on the effective size of the proton.

\begin{figure}[t!]
\centering
\includegraphics[width=0.49\columnwidth]{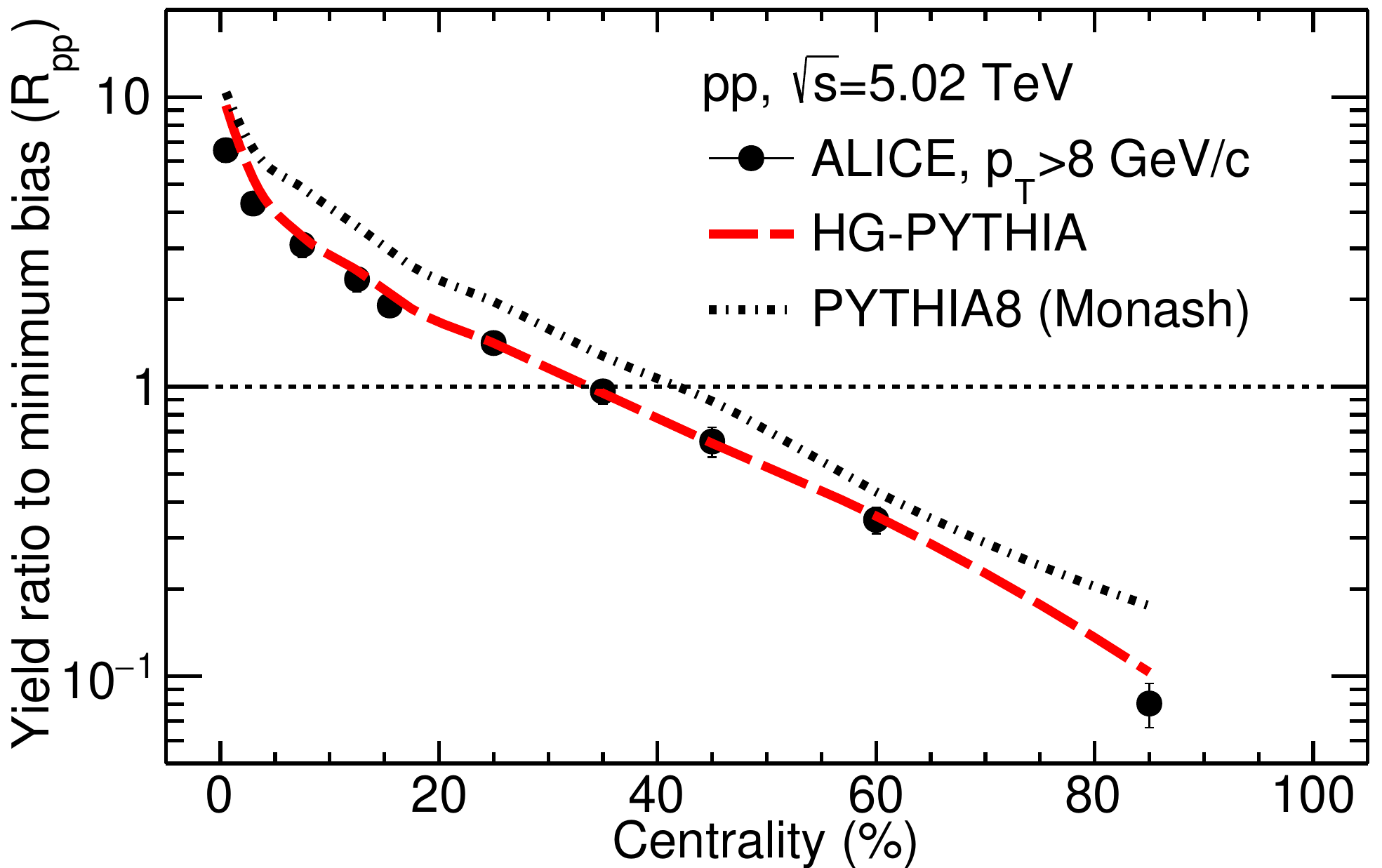}
%\hspace{0.1cm}
\includegraphics[width=0.495\columnwidth]{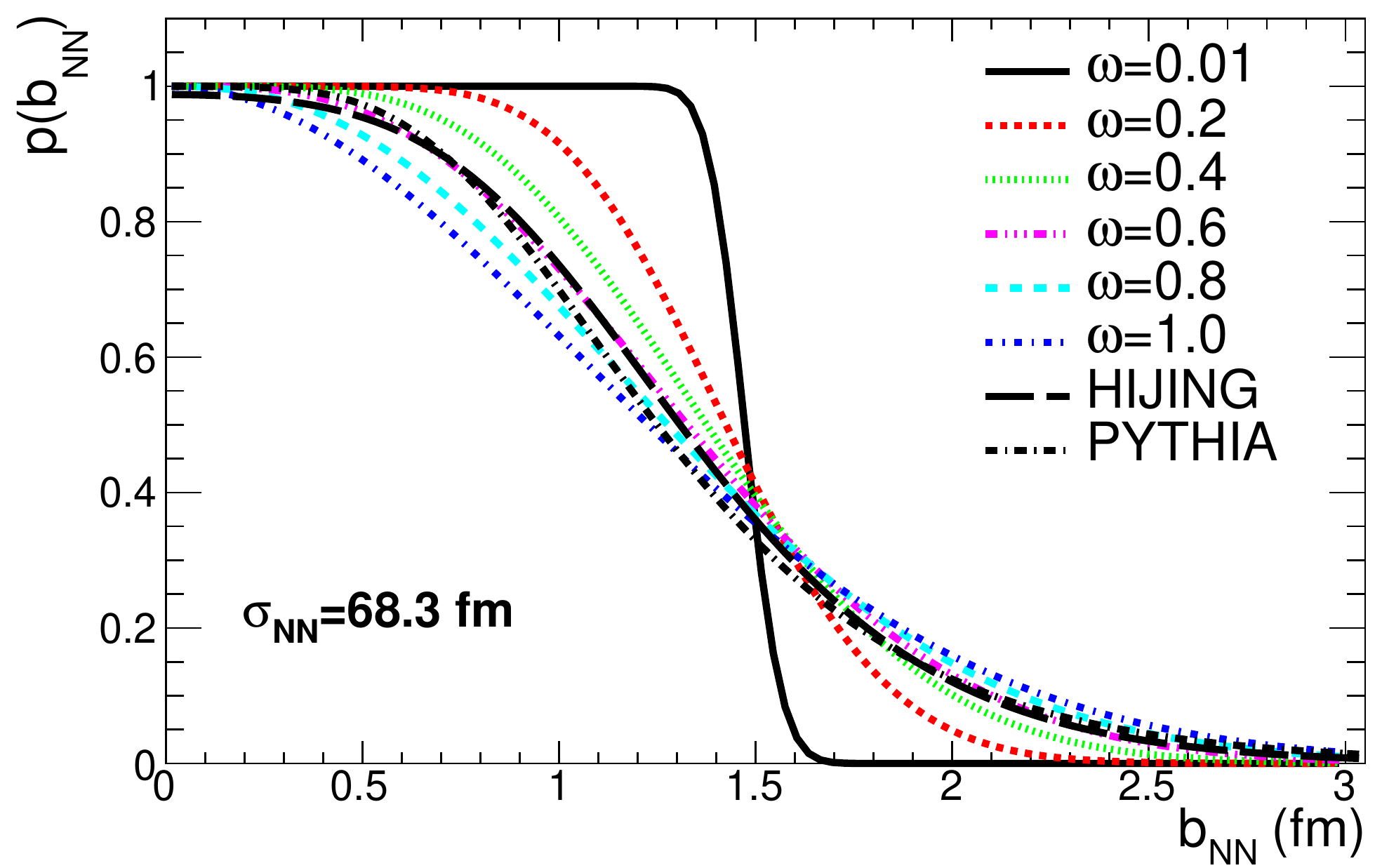}
\caption{Left:~Yield of high-\pT\ charged particles in multiplicity classes measured in \pp\ collisions at $\sqrt{s}=5.02$~TeV normalized to their MB yield~\cite{Acharya:2019mzb}, compared to \Hpythia\ and \Pythia\,8 predictions. 
Right:~Nucleon-nucleon collision probability distribution using \Eq{eq:NN_coll_profile} for various $\omega$ values in \NN\ collisions at $\sqrts=5.02$\,TeV\com{ computed with TGlauberMC}, compared to the result from the \Hijing\ and \Pythia\,8 (Monash) overlap profiles. 
\label{fig:pp_NN_plot}}
\end{figure}

The importance of properly taking into account the proton transverse profile in \pp\ collisions at LHC energies is illustrated in \Fig{fig:pp_NN_plot}~(left) where the high-$\pT$ hadron yield~(normalized to its impact-parameter-integrated value) as a function of centrality measured by the ALICE experiment at $\sqrts=5.02$~TeV, is shown. 
The multiplicity classes are obtained by ordering events according to the charged-particle response in the ALICE VZERO scintillators~($2.8<\eta<5.1$ and $-1.7<\eta<-3.7$) for events with at least one charged particle produced at midrapidity~($|\eta|<1$).
Central~(peripheral) \pp\ collisions feature about ten times larger (smaller) yields than MB collisions, consistent with the expectation from the MPI picture~\cite{Sjostrand:2017cdm}.
The centrality dependence of the yields is well reproduced by \Hpythia~\cite{Morsch:2017brb} and~(slightly less well) by \Pythia\,8~(Monash tune), where the former basically corresponds to \Pythia\ with the proton overlap profile given by \Eq{eq:overlaphijing}.

The above discussion highlights the importance in \MCG\ simulations of the choice of the \NN\ collision profile, whereby two nucleons collide if their impact parameter is less than a given distance parameter $D\approx\!\sqrt{\sigmaNN/\pi}$. 
The simplest collision profiles used are the hard sphere~(HS), $p(\bnn) = \Theta(\sqrt{\snn}-\bnn)$~(also called ``black disk''), and Gaussian, $p(b) = \exp{(-a\,b^2/D^2)}$, with $a$ fitted to reproduce the $\sigmaNN$ inelastic cross section~(\eg\ $a=0.92$ at RHIC energies)~\cite{Rybczynski:2013yba}, although more involved probabilistic ways to model the \NN\ interaction have been known for a long time~\cite{Alvioli:2011sk}. 
At LHC energies, a modification of the collision profile proposed by Refs.~\cite{Rybczynski:2013mla,Rybczynski:2013yba} uses a parametrization based on the Euler $\Gamma(z)$ and incomplete $\Gamma(\alpha,z)$ gamma functions
\begin{equation}
p(\bnn) = G\;\Gamma\left(1/\omega,\frac{G\,\bnn^2}{D^2\omega}\right)\bigg/\Gamma(1/\omega)\,,
\label{eq:NN_coll_profile}
\end{equation}
where $\omega$ interpolates between the HS ($\omega\to 0$) and Gaussian ($\omega\to 1$) cases. 
The choice $(G,\omega)=(1,0.4)$ reproduces the measured $\sigmaNN \approx 73$~mb and $\sigma_{\rm pp, el} \approx 25$~mb LHC results. 
\Figure{fig:pp_NN_plot}~(right) compares \NN\ collision profiles for varying values of the $\omega$ parameter and with the \Pythia\,8 Monash and \Hijing\ choices for \pp\ collisions at $\snn = 5.02$\,TeV. The two latter are obtained via $p(\bnn)=1-\rm{e}^{-k\Tpp(\bnn)}$ so that $2\pi\int p(\bnn)\; {\rm d}\bnn$ gives $\sigmaNN$, as commonly done for profile functions. One can see that the \Pythia\,8 Monash and \Hijing\ profiles correspond approximately to the $\omega = 0.6$ choice in Eq.~(\ref{eq:NN_coll_profile}).

An extension of the nuclear Glauber approach to \pp\ collisions at the parton-level was described in Ref.~\cite{dEnterria:2010xip} where, by analogy to the nuclear case, the thickness function of a proton with $N_g$ partons is written as $T_{p}(x,\vary)=N_g \int {\dd}z  \; \rho(x,\vary;z)$, normalized to $\int \dd^2{\rm b} \,T_{p}(b) = N_g$. 
The overlap function of a \pp\ collision at $b$ can be obtained as a convolution over the corresponding thickness functions of each proton
%$T_{pp}(b)  =  \int dx \, d\vary \; T_{p,1}\left(x+{b/2},\vary\right)\;T_{p,2}\left(x-{b/2},\vary\right),\;$
normalised to $\int \dd^2{\rm b} \,\Tpp(b) = N_g^2$. 
From the partonic cross section $\sigma_{gg}$, one can then define the number of binary parton-parton collisions in a \pp\ collision at impact parameter $b$:
%N_{coll,gg}(x,\vary;b) & =& \sigma_{gg} \; T_{p,1}\left(x+{b/2},\vary\right)T_{p,2}\left(x-{b/2},\vary\right)\;,\nonumber \\
%N_{coll,gg}(b) =  \int dx\,d\vary \; N_{coll,gg}(x,\vary;b)\; =\; \sigma_{gg} \,T_{pp}(b)$.
$N_{\rm coll,gg}(b) = \sigma_{gg} \,\Tpp(b)$.
From this, the probability of an inelastic parton-parton interaction at impact parameter $b$ can be defined as
%\begin{eqnarray}
%d^2P_{gg}^{inel}\over d^2\vec{b}}(b)  = {1-e^{-\sigma_{gg} ~T_{pp}(b)} \over \int d^2 \vec{b}\; \left(1-e^{-\sigma_{gg} ~T_{pp}(b)}\right)}, \;\mbox{ or }\;\; \nonumber \\
%{dP_{gg}^{inel}\over d b} (b) =2\pi b \; {1-e^{-\sigma_{gg} ~T_{pp}(b)} \over \int d^2 \vec{b} \;  \left(1-e^{-\sigma_{gg} ~T_{pp}(b)}\right)}, %\;\mbox{ or }\;\;\\
$P_{gg}^{\rm inel}(b) =1-e^{-\sigma_{gg}\,\Tpp(b)}\,$,
%\label{eq:dPinel_db} 
%\end{eqnarray}
where the value of $\sigma_{gg}$ is obtained by requiring that the proton-proton inelastic cross section obtained from the \pp\ overlap function,
%according to Poisson statistics. The denominator -- which sums over all events with at least one parton-parton interaction -- is just the total inelastic \pp\ cross section $\sigma_{pp}^{inel}$. The parton-number normalisation $N_g$ of the proton thickness function, Eq.~(\ref{eq:Tp}), can be obtained by requiring that the proton-proton inelastic cross section obtained from the \pp\ overlap function, Eq.~(\ref{eq:Tpp}), %in the eikonal approximation
%\begin{equation}
$\int \dd^2 \,{\rm b} \; \left(1-e^{-\sigma_{gg}\,\Tpp(b)}\right) = \sigmapp$,
%\label{eq:sigma_inel2} 
%\end{equation}
matches the $\sigmapp^{\rm inel} \approx 40$--80~mb values measured at RHIC and LHC energies~(\Tab{tab:signnvalues}). 
Values of $\sigma_{gg}\approx 6$~mb are used that are consistent with a simplistic perturbative gluon-gluon cross-section of $\sigma_{gg} = K\cdot (9/2)\, \pi\, \alpha^2_s/\pT^2$ for $\alpha_s\approx$~0.5 at a $\pT$-cutoff of the order 1\,GeV, where $K\approx2$ is a factor accounting for higher-order pQCD corrections. The final particle multiplicity density in a \pp\ collision follows the same impact-parameter-dependence as that of the number of binary parton-parton collisions, $N_{coll,gg}(b)$, and
%\begin{equation}
%{dN\over dy}(b)= {dN_0 \over dy} \cdot N_{coll,gg}(b) \;, 
%\label{eq:dNdy_vs_b} 
%\end{equation}
%with the absolute normalisation $dN_{\mbox{\tiny{\it 0}}}/dy$ chosen so as to reproduce the MB \pp\ multiplicity, namely 
the average multiplicity in a \pp\ collision, integrated over all impact-parameters, is
%\begin{equation}
$\frac{dN_{\rm MB}}{d\eta}= {dN_0\over d\eta}
% {dN_0\over dy} = \frac{dN_{\mbox{\tiny{\it MB}}} }{dy}/
\int \dd^2 {\rm b} \; N_{\rm coll,gg}(b)\,P_{gg}^{\rm inel}(b)\;$, 
%{\int d^2 \vec{b} \; N_{coll,gg}(b) \; {d^2P_{gg}^{inel}\over d^2\vec{b}}(b)}\;, 
%\label{eq:dNMB.dyb} 
%\end{equation}
with the absolute normalisation $dN_{0}/\dd\eta$ chosen so as to reproduce the MB \pp\ multiplicity of $dN_{\rm MB}/\dd\eta\approx$~10 measured at midrapidity at LHC energies~\cite{GrosseOetringhaus:2009kz}. %\cite{Khachatryan:2015jna,Adam:2015pza,Aad:2016mok}.

%\subsection{Glauber Model at the Parton Level}%~\cite{Loizides:2016djv}.
Other extensions of the nuclear \MCG\ simulations to account for subnucleonic degrees of freedom, such as \eg\ three constituent valence quarks, exist~\cite{Welsh:2016siu,Loizides:2016djv,Mitchell:2016jio,Moreland:2018gsh,Bozek:2019wyr}. The number of partonic constituents and the way to distribute them in \pA\ and \AA\ collisions ---between the two extremes of bound to individual nucleons (according \eg\ to a radially exponential nucleon form factor) or freely-distributed over the nucleus (following a global 2pF profile)--- must be chosen so as to reproduce basic experimental quantities (inelastic cross sections $\sigmaNN$, $\sigmapA$, $\sigmaAA$, overall particle multiplicities), and have a different impact on the event ellipticity and triangularity. 
Accounting for subnucleonic degrees of freedom is particularly important to generate realistic initial conditions for small QGP systems at the LHC, as discussed in~\Sec{sec:IC_hydro}.

%%%%%%%%%%%%%%%%%%%%%%%%%%%%%%%%%%%%%%%%%%%%%%%%%%%%%%%%%%%%%%%%%%%%%%%%%%%%%%%%%%%%%%%%%%%%%%%%%%%%%
\subsection{Fluctuations and Correlations}
%In QCD, event-by-event fluctuations in the interaction strength of two nucleons originate from fluctuations in their transverse size that, themselves, appear due to the quantum nature of their underlying wavefunction, that reflects in particular configurations of its partonic constituents. 
Traditionally, Glauber studies have just dealt with average nucleon and nuclear transverse densities, but in the last decade more and more studies have incorporated event-by-event shape fluctuations generated by various underlying mechanisms. Such developments have been largely motivated by an increasing number of experimental observations that pointed to the need of enhanced eccentricities fluctuations in order to explain \eg\ the large azimuthal harmonics ---direct~\cite{Teaney:2010vd}, elliptic~\cite{Alver:2006wh}, and triangular~\cite{Alver:2010gr,ALICE:2011ab,Schenke:2010rr,Adare:2011tg,ATLAS:2012at,Chatrchyan:2013nka} flows--- observed in the data.
%, such as the elliptic flow of central Cu-Cu collisions~\cite{Alver:2006wh}, directed flow~\cite{Teaney:2010vd,}, %as well as fluctuation-driven observables such as and triangular flow~\cite{Alver:2010gr,Schenke:2010rr,..}.

\begin{figure}[t!]
  \centering
  \includegraphics[width=0.50\textwidth]{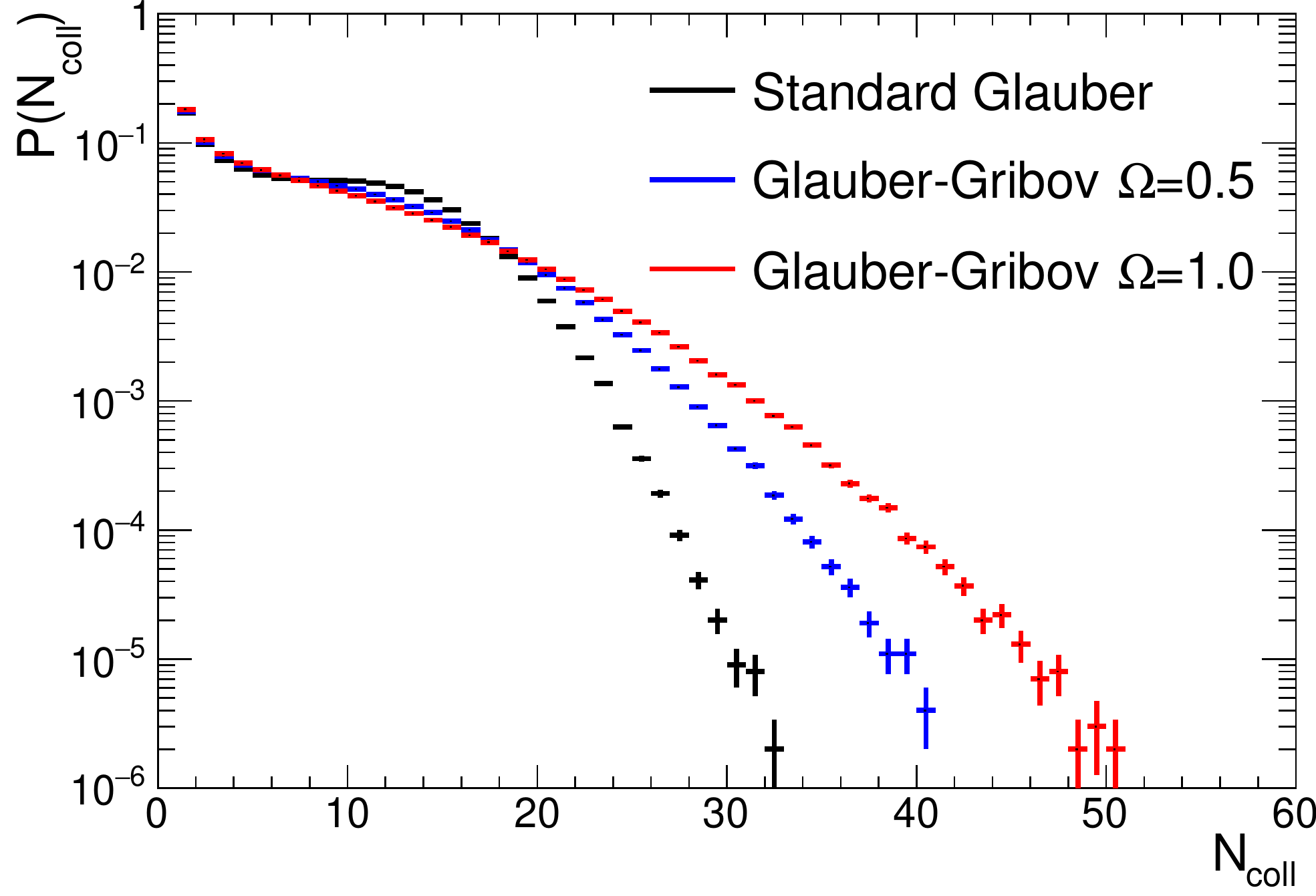}
  \hspace{0.2cm}
  \includegraphics[width=0.47\textwidth]{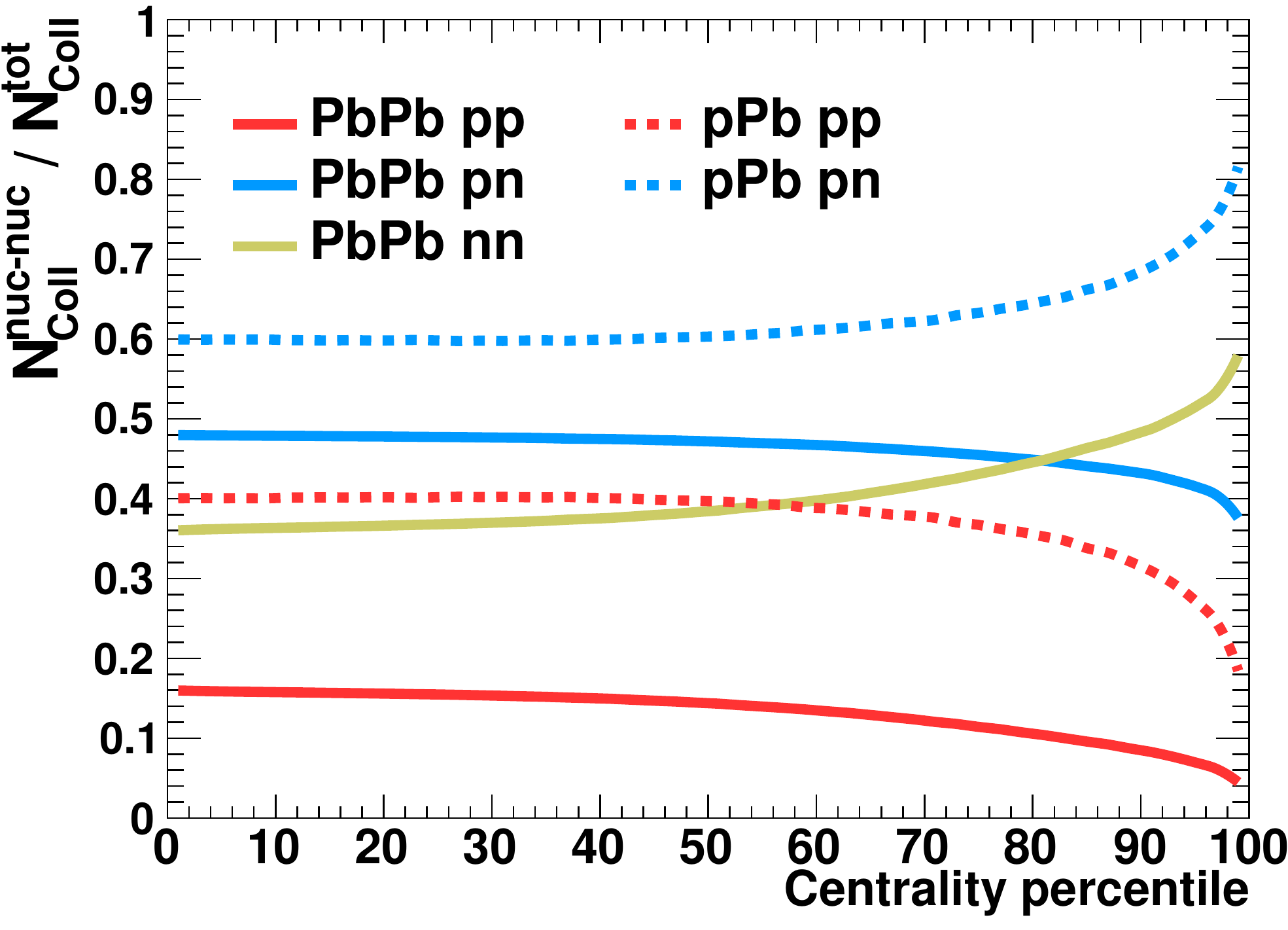} % 
  \caption{Left: Comparison of the \Ncoll\ distributions for \pPb\ collisions at the LHC with different values of the Glauber--Gribov corrections given by $\Omega=0,0.5,1$ in Eq.~(\ref{eq:fluct}).
  %Comparison of the Pb nucleus (standard) 2pF and (neutron-skin) D2pF nucleon distributions.
  %Right: Relative change in \Ncoll\ in \PbPb\ (red circles) and \pPb\ (blue squares) collisions at $\snn=5.5$ and 8.8\,TeV, respectively, after including neutron skin effects in the density profile. %The baseline uses the standard 2pF charge density, while the new results are obtained with the D2pF point charge density.
  Right:~Average fraction of pp, pn, and nn collisions for \PbPb\ (solid curves) and \pPb\ (dashed curves) obtained by incorporating a neutron skin in the Pb density profile as described in the text. Both distributions are obtained with TGlauberMC. %~\cite{Loizides:2017ack}.
\label{fig:IS}}
\end{figure}

There are two obvious sources of quantum fluctuations in \MCG\ simulations: fluctuations in the positions of nucleons within the nucleus, and fluctuations at the subnucleonic level. 
The former are in principle properly accounted for by conventional \MCG\ simulations~\cite{Blaizot:2014wba},
%are relatively small once all relevant Glauber results are projected onto the transverse plane~\cite{Blaizot:2014wba},
and we thus focus mostly on the latter. Nucleons are composite quantum-mechanical systems with varying spatial and momentum configurations of their internal quark and gluon constituents, and the overall transverse area occupied by their color fields changes event by event, a phenomenon often referred to as color fluctuations (CFs)~\cite{Heiselberg:1991is,Alvioli:2013vk,Alvioli:2017wou}.
%with  a wide range of observable consequences in the nucleon-nucleon scattering distributions.
Such fluctuations can lead to potentially large changes of the effective collision-by-collision nucleon transverse size that are often not accounted for in \MCG\ codes, notably when the \NN\ interaction is approximated by a simple ``black disk'' approach~(see above).  
Whereas such CFs tend to average out in central \AA\ collisions, they are of importance for \pp\ and \pA\ collisions. 
The CFs have been evaluated theoretically in terms of the cross section for inelastic diffractive processes in pN scattering, often referred to as ``Glauber--Gribov'' approach~\cite{Gribov:1968jf},  generalized to the nuclear case~\cite{Heiselberg:1991is}, and have been phenomenologically encoded into an event-by-event variation of $\sigmaNN$ given by a probability distribution of the form
%according to a probability $1/\lambda \, P(\sigma/\lambda)$ with
\begin{equation}
P_\sigma(\sigmaNN) = C \frac{\sigmaNN}{\sigmaNN+\sigma_0} \exp^{-\left(\frac{\sigmaNN-\sigma_0}{\sigma_0\,\Omega}\right)^2}\,,
\label{eq:fluct}
\end{equation}
where $\sigma_0$ denotes the mean $\sigmaNN$ value, and $\Omega$ its width.
The normalization $C$ %and the rescaling parameter $\lambda=\sigma_0/\left<\sigma\right>$ are 
is computed from the provided input~(mean $\sigmaNN$ and $\Omega$) requiring $\int \sigma P{\rm d}\sigma / \int P{\rm d}\sigma=\sigma_0$, with the dispersion given by the ratio of inelastic diffraction over elastic cross section at $t=0$ (zero momentum exchange).
Variations of the nucleon interaction strength naturally lead to increased fluctuations in the number of participant nucleons and binary nucleon collisions, \eg\ longer tails in the $\Npart$ and $\Ncoll$ distributions, compared to the pure eikonal picture. 
Example distributions for \pPb\ collisions at the LHC with varying width values $\Omega = 0, 0.5, 1$, are shown in \Fig{fig:IS}~(left).
Microscopic refinements in the \MCG\ treatment of fluctuations from diffractive scatterings in \pA\ collisions have been discussed in Ref.~\cite{Bierlich:2016smv}, which proposed a log-normal parametrization instead of \Eq{eq:fluct}.
In addition, CFs explain the existence of partonic configurations of the nucleon with smaller-than-average interaction strength that lead to observable differences in the centrality dependence of single jet production in \dAu\ and \pA\ collisions observed in the data at RHIC and the LHC~\cite{Adare:2015gla,Aad:2015zza}.
%a decrease in the average number of NN interactions between the projectile and target nucleus, nu, relative to that for collisions with a more typical proton configuration.
Such events are characterized by a configuration in which a large fraction of the proton momentum is carried by a single parton, \eg\ $x\gtrsim 0.1$, that is more spatially compact than the average, and their interaction strength is thereby reduced with respect to the eikonal limit~\cite{McGlinchey:2016ssj}. The standard \MCG\ models underestimate the associated number of ``peripheral'' events with low hadronic activity, while they overestimate the ``central'' ones with a large hadronic activity, although inclusive jet production rates remain unmodified.

Most Glauber-like calculations are based upon the pure eikonal approximation and disregard higher-order terms of the expansion \Eq{eq:sigmaAB} that account for short-range two-, and three-nucleon correlations (many-nucleon correlations are further suppressed). The impact of realistically correlated \NN\ configurations (centrally correlated nucleon configurations, two-body full correlations, three-body chains) on the medium eccentricity has been studied in Ref.~\cite{Alvioli:2011sk}, where it was found that their combined effects cancel out and bring the results close to the no-correlations case.
The impact of \NN\ correlations and their interplay with diffractive effects on \Ncoll\  have been estimated in Ref.~\cite{CiofidegliAtti:2011fh}. This latter work found that such correlations slightly decrease (respectively, increase) by a few percent the \Ncoll\ values (resp., $\sigmapA$ and $\sigmaAA$) compared to the pure Glauber results, but that such effects are cancelled when taking into account diffractive (aka.\ ``Gribov inelastic shadowing'') corrections that act in the opposite direction. 
The number of nucleons that are diffractively excited in the multiple collisions but revert back to their ground state before the scattering process is completed, both increases the nuclear transparency (\ie\ reduce the nuclear cross section) and reduces the \Ncoll\ results back to the values obtained with the conventional Glauber codes. 
Another possible source of correlations in \MCG\ simulations is due to the {\it recentering} procedure~\cite{Loizides:2017ack} by which the MC setup of the initial nuclear profile {\it without nucleons overlaps} is done in such a way so that the \cm\ of each nucleus is fixed at a given location in each event. Those correlations are found to be small, in particular for large nuclei~\cite{Blaizot:2014wba}.
Correlations at the parton level due to the interference of same-color gluons from different nucleons, evaluated with the \dipsy\ generator~\cite{Flensburg:2011kk} based on a BFKL resummation of small-$x$ dipoles~\cite{Salam:1998tj}, were found to result only in few percent effects for \pA\ collisions with heavy nuclei.
%The Gribov inelastic shadowing corrections generally increase the value of Ncoll, the inclusion of nucleon correlations, acting in the opposite direction, decreases it by a comparable amount. The interplay of the two effects varies with the value of the impact parameter\cite{CiofidegliAtti:2011fh}

%%%%%%%%%%%%%%%%%%%%%%%%%%%%%%%%%%%%%%%%%%%%%%%%%%%%%%%%%%%%%%%%%%%%%%%%%%%%%%
\subsection{Neutron Skin and Isospin Effects}
The transverse profile of nucleons inside a nucleus is commonly described by a single 2pF distribution.
An ansatz, based on electromagnetically probing the charge distribution (protons) of nuclei, that is however not supported by measurements with strongly-interacting probes that prefer instead two nonidentical distributions for protons and neutrons~\cite{barrett1977nuclear}, in particular at the surface of heavy stable neutron-rich nuclei, such as $^{208}$Pb with a neutron excess of $N/Z\approx 1.5$~\cite{Klos:2007is,Tarbert:2013jze}. 
These differences appear because protons around the center of the nucleus feel their common electromagnetic repulsion from all directions resulting in an electrostatic equilibrium at a constant charge density, but the outermost protons at $r\!\gtrsim\!6$~fm, where the nucleon density begins to drop, need additional ``skin'' or ``halo'' neutrons in the periphery to counteract the outward Coulomb repulsion and maintain a sufficient nuclear surface tension. %thereby increasing the overall binding energy. 
%Since the distribution of neutrons determines the underlying transverse density of $d$ quarks, such effects have a phenomenological impact on the $b$-dependence of isospin-sensitive pQCD cross sections, such as prompt photons and electroweak bosons in nuclear collisions~\cite{Paukkunen:2015bwa,De:2016ggl,Helenius:2016dsk,Alvioli:2018jls}.

The nominal heavy-ion species at the LHC is $^{208}$Pb, the heaviest stable doubly magic nucleus and one of the most intensively studied isotopes. While the average charge radius of $^{208}$Pb is known to within $\pm0.001$~fm~\cite{DeJager:1987qc,Fricke:1995zz}, past estimates placed the uncertainty in the neutron radius at about $\pm0.2$~fm~\cite{Horowitz:1999fk}. 
Neutron point density parameters of $R_{\rm n}=6.67\pm0.03$~(stat.)~fm and $a_{\rm n}=0.55\pm0.01$~(stat.)$^{+0.02}_{-0.03}$~(syst.)~fm have been measured by the Crystal Ball collaboration via coherent pion photoproduction~\cite{Tarbert:2013jze}, while the CERN LEAR experiment reports comparable values of $R_{\rm n}=6.684\pm0.02$~(stat.)~fm and $a_{\rm n}=0.571$~fm derived from antiproton-nucleus interactions coupled with radiochemistry techniques~\cite{Klos:2007is}. These data favor a peripheral neutron distribution in the form of a neutron ``skin'' rather than a neutron ``halo'', \ie\ the neutron distribution is slightly broader than the proton one because of its larger diffusivity~($a_{\rm n} - a_{\rm p} \approx 0.1$~fm), but has the same half-radius as the proton distribution~($R_{\rm p} \approx R_{\rm n} \approx 6.7$~fm).
The combined point density distribution for proton and neutrons has been implemented in recent \MCG\ simulations~\cite{Loizides:2017ack} via the weighted sum of the individual 2pF distributions.
For peripheral \PbPb\ collisions, this results in a maximum $\sim$4\% increase in \Ncoll\ and approximately half this percentage for \pPb\ collisions, largely driven by the increase of the central radius in the D2pF compared to the 2pF parametrization. 
If one focuses instead on the transverse distribution of the underlying $d$ quarks, the dominant flavor in neutrons, larger effects are expected for precise phenomenological studies of isospin-dependent gauge boson ($\gamma$, W$^\pm$, and Z) cross sections in nuclear compared to proton collisions~\cite{Paukkunen:2015bwa,De:2016ggl,Helenius:2016dsk,Alvioli:2018jls}.
\Figure{fig:IS}~(right) shows the average fraction of pp, pn, and nn scatterings for \PbPb\ and \pPb\ collisions versus centrality illustrating the increasing relevance of the neutron density for peripheral collisions.

%%%%%%%%%%%%%%%%%%%%%%%%%%%%%%%%%%%%%%%%%%%%%%%%%%%%%%%%%%%%%%%%%%%%%%%%%%%%%%%%%%%%%%%%%%%%%%%%%%%%%
\subsection{Light and Heavy Deformed Nuclei}
\label{sec:lightandheavy}
For a number of years, the RHIC machine has been providing collisions with a variety ions beyond the nominal gold nucleus, ranging from the lightest species such as deuteron and helium-3~\cite{Durham:2018btt} to the heaviest ones such as uranium~\cite{Adamczyk:2015obl}, as a means to study the system size dependence of various QGP-related signals. 
As aforementioned, for spherical nuclei, the probability density distribution in \MCG\ models is sampled from the underlying 2pF or D2pF radial probability functions, and taken to be uniform in azimuthal and polar angles, but light species such as the deuteron~(${}^{2}$H), tritium~(${}^{3}$H), \mbox{helium-3}~(${}^{3}$He), helium-4~(${}^{4}$He), carbon~(${}^{12}$C), oxygen~(${}^{16}$O), and sulfur (${}^{32}$S), have dedicated parametrizations of their transverse profiles. 
For deuteron, the Hulth\'en form $\rho(r')=\rho_0 \left(\frac{e^{-ar'}-e^{-br'}}{r'}\right)^2$,
with $a=0.228$~fm$^{-1}$ and $b=1.177$~fm$^{-1}$, and $r'=2r$ denoting the distance between the proton and neutron, is often employed~\cite{Hulthen1957,Adler:2003ii,Adler:2006xd}. 
For ${}^{3}$H and ${}^{3}$He nuclei, configurations are computed from Green's function MC calculations using the AV18/UIX Hamiltonian, which correctly sample the position of the three nucleons, including their correlations~\cite{Nagle:2013lja}. 
Similarly, results of wavefunction-based calculations are available for helium-4, carbon, and oxygen~\cite{Lim:2018huo}. 
For slightly deformed nuclei, such as \eg\ sulfur, the Fermi distribution is modified with an extra parameter $w$ and a Gaussian term, $\rho(r)=\rho_0 \frac{1+w(r/R)^2}{1+\exp(\frac{r^2-R^2}{a^2})}$.
Details on all relevant parametrizations can be found \ifarxiv in \App{sec:app} and \fi in \Refs{Loizides:2014vua,Lim:2018huo,Shou:2014eya}.

The description of the transverse profile of heavy deformed nuclei starts off with the 2pF expressions modified with an expansion of $R$ in spherical harmonics, $\rho(r)=\rho_0\left(1+\exp\frac{\left[r-R(1+\beta_2 Y_{20} +\beta_4 Y_{40})\right]}{a}\right)^{-1}$, with $Y_{20}=\sqrt{\frac{5}{16\pi}}(3\cos^2(\theta)-1)$, $Y_{40}=\frac{3}{16\sqrt{\pi}}(35\cos^4(\theta)-30 \cos^2(\theta)+3)$, and deformation parameters $\beta_2$ (quadrupole) and $\beta_4$ (hexadecapole)~\cite{DeJager:1987qc}. 
The higher harmonic eccentricities ($\epsilon_n$) of the initial QGP produced in collisions of deformed nuclei, such as U, are particularly sensitive to the parametrization of their profiles. 
%constrained from the experimental ratio of triangular over elliptic flows ($v_3/v_2$) in UU collisions, have been studied \eg\ in~\cite{Shou:2014eya,Noronha-Hostler:2019ytn}. The work~\cite{Shou:2014eya} shows that an overestimation of the $\varepsilon_3/\varepsilon_2$ ratio will lead to an enhanced amount of the viscous damping needed in the subsequent theoretical hydrodynamic evolution to match the experimental UU data. 
A proper description of collisions of heavy deformed nuclei requires also accounting for their relative, tip-on-tip and side-on-side, orientations. 
Tip-on-tip collisions produce a smaller elliptic flow but larger particle multiplicities (entropy densities), whereas on the contrary, side-on-side collisions generate a larger elliptic flow but a smaller multiplicity. 
The scaling of $v_{2,3}$ flows with multiplicity in ultracentral collisions (0--1\% centrality percentile) in small and deformed systems produced in UU, dAu, ${}^{9}$BeAu, ${}^{9}$Be${}^{9}$Be, ${}^{3}$He${}^{3}$He, and ${}^{3}$HeAu collisions at RHIC energies, including or not fluctuations from subnucleonic degrees of freedom, has been theoretically studied in~\cite{Noronha-Hostler:2019ytn,Giacalone:2019pca}. 
This work indicates that such collisions can help discriminate between different initial entropy densities of the QGP medium formed at RHIC and LHC~\cite{Huang:2019tgz,Sievert:2019zjr}. 
Implications for the extraction of QGP transport properties such as its viscosity, are further developed in \Sec{sec:IC_hydro}.

%%%%%%%%%%%%%%%%%%%%%%%%%%%%%%%%%%%%%%%%%%%%%%%%%%%%%%%%%%%%%%%%%%%%%%%%%%%%%%%%%%%%%%%%%%%%%%%%%%%%%
%\section{PHENOMENOLOGICAL USAGE OF THE GLAUBER MODEL}
\section{PHENOMENOLOGICAL APPLICATIONS}
\label{sec:pheno}

The Glauber model has many important phenomenological uses in nuclear collisions at colliders. We consider three typical cases here. First, in the definition of the baseline scalings for comparing hard-scattering cross sections in \pp, \pA, and \AA\ collisions. %, of special relevance for extractions of nuclear PDFs using colorless probes unaffected by final-state interactions in the QGP. 
Second, as an underlying framework for MC event generators used in high-energy heavy-ion and cosmic-ray physics. Third, to provide realistic initial-state conditions of the created QGP for subsequent spacetime evolution in hydrodynamics codes. We succinctly review below the basic ideas and latest progress in these three areas.

%%%%%%%%%%%%%%%%%%%%%%%%%%%%%%%%%%%%%%%%%%%%%%%%%%%%%%%%%%%%%%%%%%%%%%%%%%%%%%%%%%%
%\section{BINARY SCALING FOR HARD SCATTERINGS}
\subsection{Binary Scaling for Hard Scatterings}
\label{sec:binaryscaling}

One of the most extended uses of the Glauber model is to properly normalize the fractional cross sections or yields for the production of a given particle in hard-scattering processes (\ie\ partonic processes characterized by mass and/or $\pT$ scales above a few GeV) in \AA\ and \pA\ collisions, to be able to compare them to those expected in the simpler \pp\ collisions where no QGP formation is, in principle, expected. For pQCD observables that do not suffer any final-state effects, the assumption of binary scaling allows the extraction of %initial-state effects due to
modifications of the nuclear parton distribution functions (PDFs) compared to the free proton ones.

It is informative to recall the basic scaling rules for perturbative scatterings in nuclear collisions~\cite{Vogt:1999jp,d'Enterria:2003qs}. For a given hard process A$+$B$\,\to h+X$, from the generic~\Eq{eq:sigmaAB2} for the inclusive \AA\ cross section, one obtains the following relationship between \pp\ and nuclear collisions
\begin{equation}
\sigmaAB^{\rm hard} = \int d^2b\; \sigmaNN^{\rm hard}\;\TAB(b)\,,\; \mbox{and, therefore, }\;
\sigma_{\rm AB, MB}^{\rm hard} = \;A\cdot B\cdot\sigmaNN^{\rm hard}\;,
\label{eq:glauber_AB_2}
\end{equation}
where the second expression for the inclusive hard cross section %of a given hard process in a nucleus relative to that in a \pp\ collision, 
is obtained integrating the former over the impact parameter.
%\begin{equation}
%(\sigmaAB^{\rm hard})_{\mbox{\scriptsize{MB}}} = \;A\cdot B\cdot\sigmaNN^{\rm hard}\;.
%,\;\mbox{where $A, B$ are the corresponding numbers of nucleons.}
%\label{eq:sigma_glauber_AB_minbias}
%\end{equation}
The associated minimum-bias invariant yield per nuclear collision, $N^{\rm hard}_{\mbox{\scriptsize{\rm AB}}}=\sigma^{\rm hard}_{\mbox{\scriptsize{\rm AB}}}/\sigmaAB$, for a given hard process in an \AB\ collision compared to that of a \pp\ collision is $\langle N_{\scriptsize{\rm AB}}^{\rm hard}\rangle_{\scriptsize{\rm MB}} = \frac{A\cdot B}{\sigmaAB}\cdot\sigmaNN^{\rm hard}\;$,
where $\sigmaAB$ is the inclusive inelastic \AB\ cross section given by \Eq{eq:sigmaAB2}. 
The average nuclear overlap function at impact parameter $b$ for minimum-bias collisions is 
\begin{equation}
\langle \TAB(b)\rangle_{\scriptsize{\rm MB}} \equiv \frac{\int \dd^2b \,\TAB(b)}{\int \dd^2b} \;= \frac{A\cdot B}{\pi (R_A+R_B)^2}=\frac{A\cdot B}{\sigmaAB}\;.
\label{eq:T_AB_minbias}
\end{equation}
The corresponding expressions for a given impact parameter $b$ can be obtained by multiplying each nucleon in nucleus A with the density along the $z$ direction in nucleus B, integrated over nucleons in nucleus A, \ie\
\begin{equation}
N_{\scriptsize{\rm AB}}^{\rm hard}(b)  =  
\sigmaNN^{\rm hard} \int \dd^2\mathbf{s}  \int \rho_A({\mathbf{s},z'}) \int {\dd}z''{\dd}z'\, \rho_B(|\mathbf{b}-\mathbf{s}|,z'') \; \equiv \; \sigmaNN^{\rm hard}\cdot \TAB(b)\;.
\label{eq:N_AB}
\end{equation}
%where we have made use of expressions (\ref{eq:nuc_profile}) and (\ref{eq:nuc_overlap}).
Similarly, one obtains a useful expression for the probability of an inelastic \NN\ collision or, equivalently, for the number of binary inelastic collisions, \Ncoll, in a nucleus-nucleus collision at impact parameter $b$:
\begin{equation}
\Ncoll(b) \; = \; \sigmaNN\cdot \TAB(b)\,.
\label{eq:N_coll}
\end{equation}
From this last expression, one can see that the nuclear overlap function,  $\TAB(b)=\Ncoll(b)/\sigmaNN$ [mb$^{-1}$], can be thought of as the hard-scattering integrated luminosity~(\ie\ the number of hard collisions per unit of cross section) per \AB\ collision at a given impact parameter. 

The expressions above allow writing the standard binary~(or point-like) collision scaling formula that relates the hard-scattering yields in nuclear and proton collisions as $N_{AB}(b) = \Ncoll(b)\cdot N_{pp}$. The {\it nuclear modification factor} for hard-scattering processes is thereby defined as the ratio of \AA\ over scaled \pp\ cross sections and/or yields~(here, differential in $\pT$ and $\eta$) as
\begin{equation}
\RAB(\pT,\eta) = \frac{d\sigmaAB/d\pT d\eta}{(A\cdot B)\;d\sigmapp/d\pTd\eta}\,,
\label{eq:RAAmb}
\end{equation}
for minimum-bias collisions, and dependent on $b$ as
\begin{equation}
\RAB(\pT,\eta;b) = \frac{dN_{AB}(b)/d\pT d\eta}{\TAB(b)\;d\sigmapp/d\pT d\eta} =
\frac{dN_{AB}(b)/d\pT d\eta}{\Ncoll(b)\;dN_{pp}/d\pTd\eta}\;.
\label{eq:RAA}
\end{equation}

In the absence of any final- and/or initial-state effects, one expects $\RAB = 1$ for any hard-scattering process. 
The equality of \Eq{eq:RAAmb} or (\ref{eq:RAA}) to unity, modulo few percent nuclear PDF effects~(see below), for colorless hard probes that do not suffer final-state interactions in the produced QGP was confirmed previously in heavy-ion collisions at SPS and RHIC as well as, in the last years, at LHC energies, and constitutes a validation of the basic assumptions of the Glauber model itself. 
Prominent examples include the production yields of photons~\cite{Adler:2005ig,Chatrchyan:2012vq,Aad:2015lcb,Sirunyan:2020ycu}, and W and Z bosons~\cite{Aad:2012ew,Aad:2014bha,Aad:2015gta,Khachatryan:2015pzs,Acharya:2017wpf,Aad:2019lan,Aad:2019sfe,Sirunyan:2019dox} in \pPb\ and \PbPb\ collisions.

\begin{figure}[th!]
\includegraphics[width=0.49\textwidth]{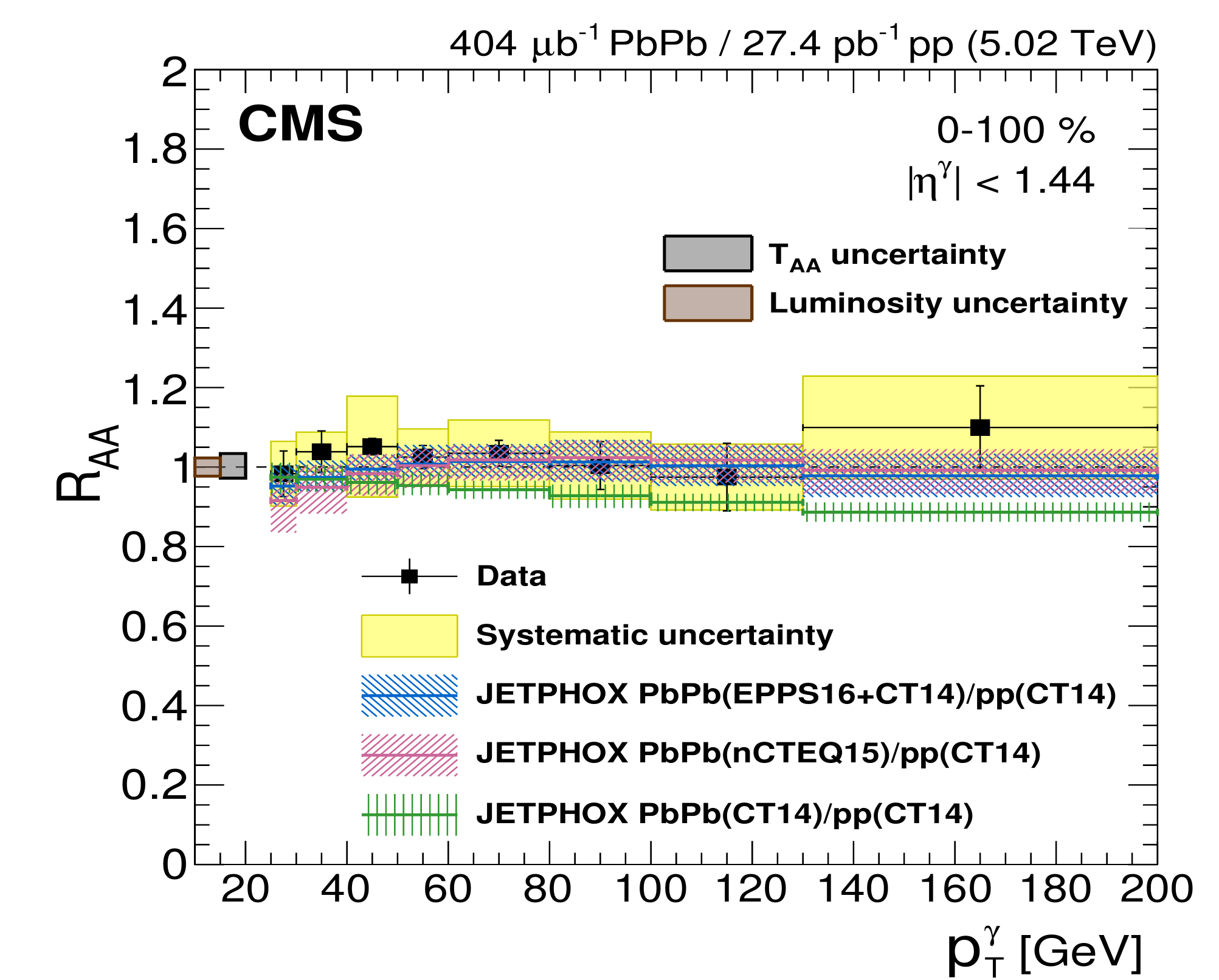}
\includegraphics[width=0.50\textwidth]{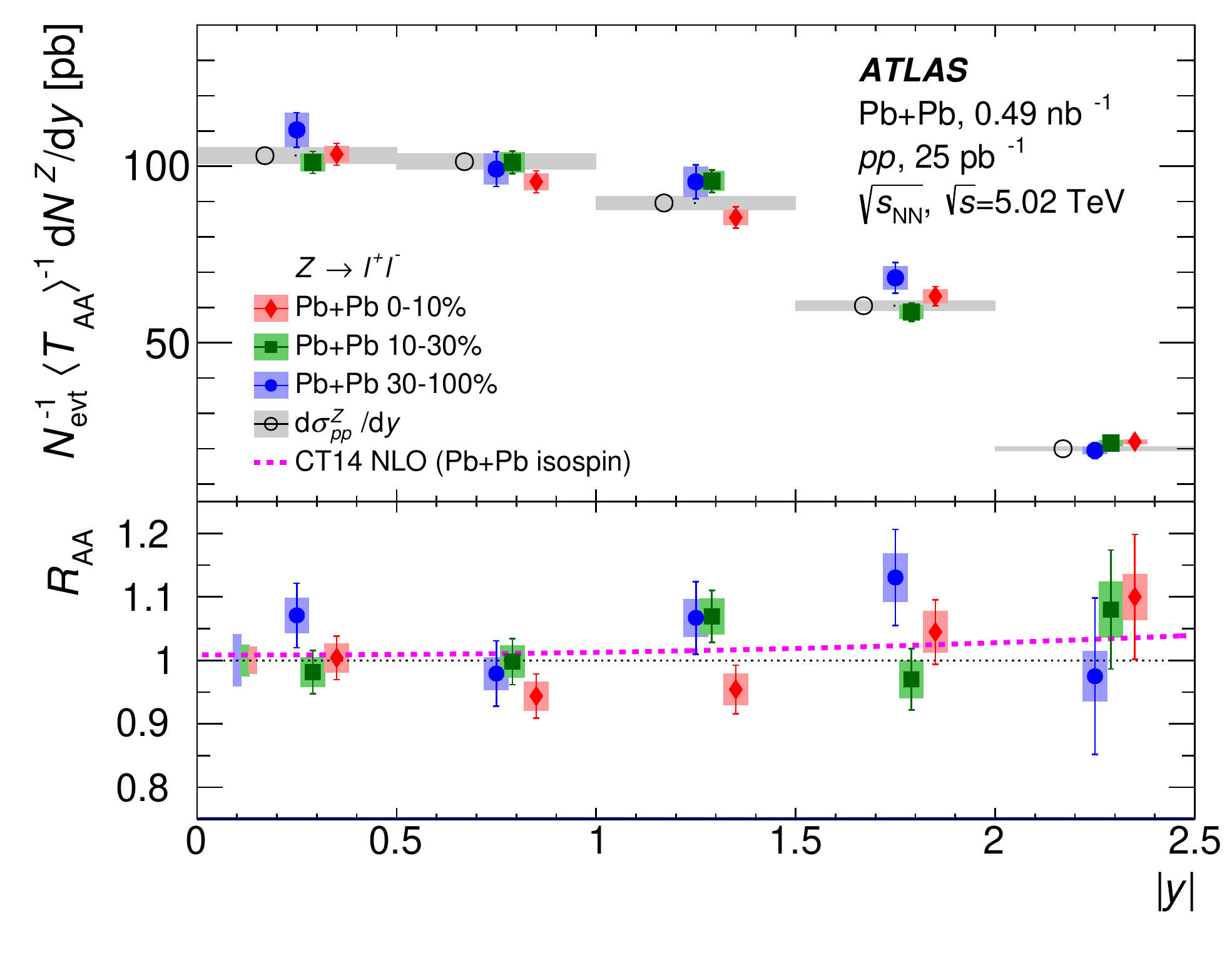}
\caption{Nuclear modification factors $\RAA$ measured for isolated photons as a function of $\pT^\gamma$ (left)~\cite{Sirunyan:2020ycu} and Z bosons as a function of rapidity (right)~\cite{Aad:2019lan} in \PbPb\ collisions at the LHC, compared to pQCD NLO calculations with nuclear PDFs.}
%measured in the 0-100\% centrality range in PbPb collisions compared to various theoretical predictions. The 3.4\% TAA uncertainty, common to all points, is indicated by the gray box centered at unity on the left side of the panel~\cite{Sirunyan:2020ycu}.}
%The three different NLO JETPHOX calculations of EPPS16+CT14 nPDFs, nCTEQ15 nPDFs, and CT14 PDFs for PbPb collisions are divided by the NLO JETPHOX calculations with CT14 PDFs for pp collisions, and compared to the data. The hatched boxes correspond to JETPHOX (n)PDF uncertainties.}
\label{fig:RAA}
\end{figure}

\Figure{fig:RAA} shows the $\RAA$ values measured for isolated photons~\cite{Sirunyan:2020ycu} and Z bosons~\cite{Aad:2019lan} as functions of  $\pT^\gamma$ and rapidity $|y^Z|$, respectively, in \PbPb\ collisions at the LHC. Both ratios are around unity with small variations due to nuclear PDF modifications related either to the increased number of $d$-quarks in the Pb nucleus compared to protons (isospin effects) and/or to few percent (anti)shadowing effects at the large virtualities ($Q^2\approx \pT^\gamma, m_\mathrm{Z}$) probed in the underlying partonic scatterings. Under the key assumption of binary scaling, the precision of the \PbPb\ data (a few percent experimental uncertainties in the case of electroweak gauge bosons) allows the derivation of the EPPS16~\cite{Eskola:2016oht}, nCTEQ15~\cite{Kusina:2016fxy}, and nNNPDF2.0~\cite{AbdulKhalek:2020yuc} nuclear PDFs at next-to-leading order (NLO), or more recently also next-to-next-to-leading order (NNLO)~\cite{Khanpour:2016pph} accuracy, through global fits combining nuclear deep-inelastic scattering (DIS) and LHC electroweak boson data.
Precision electroweak boson measurements in \PbPb\ collisions can also be used to derive a data-driven normalization, in principle independent of the Glauber model, for cross section measurements in \PbPb\ collisions, as explored in~\cite{Eskola:2020lee}. %assuming the AP can be determined with a precision of 1% or better in very peripheral collisions.

%%%%%%%%%%%%%%%%%%%%%%%%%%%%%%%%%%%%%%%%%%%%%%%%%%%%%%%%%%%%%%%%%%%%%%%%%%%%%%%%%%%%%%%%%%%%%%%%%%%%%
\subsection{Heavy-Ion Monte Carlo Event Generators}
\label{sec:HI_MCs}

All existing generic event generators of ultrarelativistic \pA\ and \AA\ collisions ---such as \Hijing~1.0~\cite{Wang:1991hta} and 2.0~\cite{Deng:2010mv}, \epos-LHC~\cite{Pierog:2013ria},  \ampt~\cite{Lin:2004en}, \qgsjet-II~\cite{Ostapchenko:2010vb}, \dpmjet-III~\cite{Roesler:2000he}, %URQMD~\cite{Bleicher:1999xi}, 
as well as the more recent \Pythia\,8-based~\cite{Sjostrand:2007gs} \angantayr~\cite{Bierlich:2018xfw} code--- internally rely on a Glauber picture to model the early stage of the collision through the proper computation of the number of inelastic subcollisions for any reaction centrality. 
The main differences among models arise from their treatment of the underlying (semi)hard scatterings: minijets in the case of the \Hijing, \angantayr, and \ampt\ codes mostly used in collider physics; and Regge--Gribov ``cut pomerons'' (aka.\ parton ladders, giving rise to one or two strings spanned between two colliding nucleons, or between a nucleon and another pomeron) in the case of the \epos-LHC, \qgsjet-II, and  \dpmjet-III codes that are mostly used in cosmic-ray physics~\cite{dEnterria:2011twh}. 
%Irrespective of the internal degrees of freedom used, 
The final hadronization of partons or strings is carried out via (variations of) the Lund fragmentation model~\cite{Andersson:1998tv} in all generators. 
%Differences of their Glauber-based modeling of multiparticle production are briefly discussed next.

The \Hijing\ generator relies on the eikonal approach to determine the number of inelastic subcollisions of two types: soft \NN\ collisions treated as in the {\sc Fritiof} approach~\cite{Andersson:1986gw}, and (multiple) hard parton-parton collisions treated perturbatively as in \Pythia.
The transverse momentum cutoff $p_{\rm T,0}\approx 2$\,GeV that separates hard from soft scatterings increases slowly  with collision energy (logarithmically, similar to the inelastic cross section \Eq{eq:signnfunc} evolution), so that the total number of minijets  per unit transverse area satisfies $p_{\rm T,0}^2/\pi>\TAA(b)\,\sigma_{\rm hard}/(\pi R_A^2)$, where $\sigma_{\rm hard}$ is the pQCD cross section for $2 \rightarrow 2$ parton scatterings, and $\TAA(b)$ is the overlap function of the \AA\ collision. 
The probability for an inelastic \NN\ collision is given by ${\rm d} \sigmaNN = 2\pi\, \bnn\, {\rm d}\bnn \, \left[1 - e^{-\left(\sigma_{\rm soft} + \sigma_{\rm hard}\right) \, \TNN(\bnn)} \right]$ with $\TNN(\bnn)$ given by \Eq{eq:overlaphijing}.
At $\snn=0.2$ and 5.02\,TeV, $\sigma_{\rm hard}=11.7$ and $124.3$~mb, respectively, and the associated number of \MPI\ per \NN\ interaction is distributed as $P(\Nhardnn) \propto e^{- \langle \Nhardnn \rangle}$, around the average number of hard scatterings determined by $\bnn$ and given by $\langle \Nhardnn \rangle = \sigma_{\rm hard} \, \TNN(\bnn)$. 
The average number of hard collisions per \NN\ collision increases from $\langle\Nhardnn\rangle=0.28$ to 1.77 between $\snn=0.2$ and 5.02\,TeV. %and becomes negligible~($<0.05$) below $0.04$\,TeV.
The total number of hard scatterings for an \AA\ collision is then obtained by summing over all \NN\ collisions in the MC Glauber, \ie\ $\Nhard=\sum_{i=1}^{\Ncoll} \left(\Nhardnn\right)_i$.

The \ampt\ code uses directly the Glauber initial conditions generated by \Hijing\ as input for its parton cascade evolution. 
The most recent heavy-ion event generator is \angantayr, which follows a Glauber approach similar to that of \Hijing\ but further takes into account Glauber--Gribov corrections, by dividing up each inelastic sub-collision as either single-, double-diffractive, or absorptive (\ie\ nondiffractive), and with CF effects implemented through a model with fluctuating nucleon radii resulting in a fluctuating \NN\ cross section inspired by the approach of Ref.~\cite{Alvioli:2013vk}. 

At variance with other generators, \epos-LHC keeps track of how many times a given nucleon interacts with nucleons from the other nucleus, and separates them event-by-event into the ``core'' (nucleons that collided more than once) and ``corona'' (nucleons that interacted exactly once). 
Event-by-event, a fraction of the string segments that do not overlap (corona) fragment into hadrons normally, following the Lund string model, whereas the other clusters with large density of strings, are used to create a (QGP-like) core that can flow and hadronize collectively. 
Such a two-component core-corona medium leads to a consistent reproduction of the particle-multiplicity dependence of a number of observables ($\mean{\pT}$, ratio of different hadron yields, etc.) in \pp, \pA, and \AA\ collisions~\cite{Pierog:2013ria}. 
The subsequent collective expansion of the medium, defined by the initial core particle density, is taken care of by relativistic hydrodynamics equations.

Apart from the hadronic MC event generators mentioned above, generators of ultraperipheral (photon-induced) nuclear collisions, such as \starlight~\cite{Klein:2016yzr} and \superchic~3~\cite{Harland-Lang:2018iur} also employ a Glauber approach to determine the non-hadronic overlap probability needed to compute the cross sections of purely exclusive final states. In \starlight, the probability of having no hadronic interactions is given by $P_{\rm no.had}(b) = \mathrm{e}^{-\sigmaNN\TA(b)}$, where $\TA(b)$ is the nuclear thickness function. For large nuclei, applying this probability function is roughly equivalent to imposing a $b > R_{\rm A}$ condition on their photon flux. This probability is also used when the photon is emitted by the proton, leading to an effective $\gamma$ flux in \pA\ collisions considerably smaller than if it was calculated directly from the e.m.\ proton form factor.

%%%%%%%%%%%%%%%%%%%%%%%%%%%%%%%%%%%%%%%%%%%%%%%%%%%%%%%%%%%%%%%%%%%%%%%%%%%%%%%%%%%%%%%%%%%%%%%%%%%%%
\subsection{Initial Conditions for Hydrodynamic QGP Evolution Calculations}
\label{sec:IC_hydro}
%The quantification of fundamental QGP properties is typically accomplished by matching experimental measurements to theoretical models that compute the full spacetime evolution of heavy-ion collisions [8, 9]. While viscous relativistic fluid dynamics provides a stable, well-tested description of the thermalized QGP medium [10–15], the initial state of the collision remains poorly constrained and constitutes the largest source of uncertainty in modern computational models [16, 17].

The strongly-interacting medium created in \AA\ collisions at the LHC is a dynamical system that expands, %longitudinally (transversely) at about (0.6 times) the speed of light, 
cools down, and transforms into a hadron gas at times around $\tau = 10$--15~fm/c. Under such conditions, the extraction of QGP properties can only be accomplished by comparing the experimental measurements of the hadronic final-state to theoretical predictions that include a model of the full spacetime evolution of the heavy-ion collision process~\cite{Heinz:2009xj}. The current state-of-the-art for the QGP evolution is given by 3-D viscous relativistic fluid dynamics calculations~\cite{Romatschke:2009im,Schenke:2010rr,Hirano:2012kj,Gale:2013da,Weller:2017tsr} ---where the plasma thermodynamic properties evolve according to the lattice QCD EoS~\cite{Borsanyi:2013bia,Bazavov:2014pvz} with nonideal corrections encoded in the medium shear viscosity $\upeta$--- matched to a hadron transport cascade (often, the UrQMD code~\cite{Petersen:2008dd}) once the energy density drops below $\upepsilon_{\rm c}$. The largest source of uncertainty in the extraction of medium properties from data-theory comparisons lies in the description of the initial state of the QGP~\cite{Luzum:2009sb}, a topic that is discussed next.
%, as shown \eg\ in~\cite{Retinskaya:2013gca} where calculations with different initial conditions are compared to a variety of azimuthal flow harmonics $v_n$ results.

Hydrodynamic models start their evolution at $\tau_0$ from a given input entropy (or energy) density in the transverse plane $s_0(\vec{r}_\perp)$. In principle, such initial conditions (ICs) should be consistently derived by solving the non-equilibrium evolution of the matter created in the first parton-parton interactions, but achieving thermalization of the interacting fields in ultrashort $\mathcal{O}(1\,{\rm fm})$ time-scales remains a difficult theoretical problem in heavy-ion physics~\cite{Strickland:2013uga}. Therefore, one assumes that the produced matter has (pre)equilibrated, and the ICs are usually given either by
\begin{description}
\item (i) the number density of produced gluons after the primary collisions, 
%linearly proportional to the initial entropy density 
via $s_0(\vec{r}_\perp) \propto dN_g/(\tau_0\dd^2r_\perp\dd\eta)$, derived analytically \eg\ in models such as MC-KLN~\cite{Kharzeev:2001yq,Drescher:2006ca} 
%\cite{Kharzeev:2000ph, Kharzeev:2002ei, Kharzeev:2004if}
and IP-Glasma~\cite{Schenke:2012hg,Schenke:2012wb} based on the Color-Glass Condensate (CGC) effective theory for parton saturation in heavy-ion collisions~\cite{Gelis:2010nm}, or by pQCD NLO calculations with {\it ad hoc} parton saturation such as EKRT~\cite{Eskola:1999fc,Niemi:2015qia}; %or 
\item (ii) Glauber MC profiles such as those shown in Fig.~\ref{fig:evt_displays} 
obtained with the ansatz given by Eq.~(\ref{eq:s0_MCG}) with $\alpha$
adjusted to match the observed multiplicity distributions, or similarly generated with one of the  MC event simulations (\Hijing, \epos, etc.) discussed in~\Sec{sec:HI_MCs} .
\end{description}
%In the latter, %the initial deposit entropy or energy density is driven by the individual collisions between the nucleons with a given distribution in the transverse plane and rapidity.
%the simplest approach is to assume that each participant nucleon adds a contribution to the initial density with Gaussian shape (in $x$ and $y$) and width 0.4 fm, a value commonly used in event-by-event hydrodynamic calculations. 
%At SPS energies,  \Npart\ scaling~\cite{Bialas:1976ed,Bialas:1977pd} is preferred, whereas at RHIC mixed \Npart+\Ncoll-scalings fitted to the data~\cite{Kharzeev:2000ph,Back:2001xy}.
%In all cases, centrality bins are assigned according to the total entropy of each Monte Carlo event, which corresponds closely to the total multiplicity that would be obtained after hydrodynamic evolution.
%NeXus, UrQMD and \Hijing\ all use binary collision between the nucleons, and thus do not take saturation in the cascade into account properly.

The corresponding local deposition of the entropy density from the underlying (parton or nucleon) collisions, and thereby the transverse area and azimuthal anisotropies of the produced medium, are then model dependent. The \trento\ approach~\cite{Moreland:2014oya} has parametrized all different approaches with a generic function of the %position-dependent 
participant target and projectile thickness functions $T_{\rm A,B}^{\rm part}$, as
\begin{equation}
  s \propto \left( \frac{\TA^{\rm part} + \TB^{\rm part}}{2} \right)^{1/p}\,,
  \label{eq:genmean}
\end{equation}
with a continuous parameter $p$ that effectively interpolates among different entropy deposition schemes. For ${p=(1, 0, -1)}$, this generalized mean reduces to arithmetic $(\TA^{\rm part} + \TB^{\rm part})/2$, geometric $\sqrt{\smash[b]{\TA^{\rm part} \TB^{\rm part}}}$, and harmonic $(2\, \TA^{\rm part} \TB^{\rm part}/(\TA^{\rm part} + \TB^{\rm part}))$ means, while for ${p \rightarrow \pm\infty}$ it asymptotes to maximum $s \propto \max(\TA^{\rm part}, \TB^{\rm part})$ and minimum $s \propto \min(\TA^{\rm part}, \TB^{\rm part})$ functions.
%The scaling of energy deposition is proportional to $\sqrt{TA TB}$ in the Trento model~\cite{Moreland:2014oya} and to $TA TB$ in CGC-based models (such as the IP-Glasma), where $T_{\rm A,B}$ are the position dependent target and projectile thickness functions.
The generalized \Eq{eq:genmean} maps different model calculations for suitable values of the parameter $p$. The Glauber wounded nucleon model (WNM), with $s \propto \TA^{\rm part} + \TB^{\rm part}$, is equivalent to the generalized mean ansatz with $p=1$. 
%A common implementation of a CGC based saturation picture is the 
The IP-Glasma approach ---which combines CGC effects of the incoming gluon distributions, %including Bjorken $x$ and impact parameter dependencies constrained by HERA DIS data~\cite{Kowalski:2003hm},
where the gluon thickness function $T_g(b)$ is a Gaussian function of the impact parameter (from the center of the probed nucleon) with a width constrained by HERA DIS data~\cite{Kowalski:2003hm}, with an event-by-event classical Yang--Mills evolution of the produced glasma gluon fields--- deposits density following $\TA^{\rm part}\cdot\TB^{\rm part}$. The default entropy deposition parameter $p = 0$ of \trento, derived from a global fit to multiple experimental data, produces similar initial eccentricities to IP-Glasma~\cite{Moreland:2014oya}.
%, are proportional to the local participant nucleon density in each nucleus.
In the KLN model, the gluon multiplicity $N_g$ can be determined perturbatively in the $k_{\rm T}$-factorization CGC approach from the parton saturation momenta of each nucleus $Q^2_{s,A} \propto \TA$, %and the entropy deposition at the QGP thermalization time is $s \propto N_g$, 
leading to $s \propto T_\text{min} \bigl[ 2 + \log(T_\text{max}/T_\text{min}) \bigr]$. This would correspond to a parameter $p\approx -0.67$ in the generalized ansatz~(\ref{eq:genmean}).
The EKRT approach combines collinear factorized pQCD minijet production with a simple model of gluon saturation, and predicts an energy density given by $\upepsilon_0 \propto \frac{K_\text{sat}}{\pi} Q_\text{sat}^3$ with saturation momentum $Q_\text{sat}(K_\text{sat}; T_A, T_B)$, corresponding to an exponent $p \approx 0$. 
%depends on nuclear thickness functions $T_A$ and $T_B$ (as well as phenomenological parameters $K_\text{sat}$ and $\beta$).

Smaller, more negative, values of $p$ pull the generalized mean towards a minimum function, and hence correspond to models with more extreme gluon saturation effects, leading to the following schematic hierarchy of more saturated ICs: WNM\,$<$\,IP-Glasma,\,\trento,\,EKRT\,$<$\,MC-KLN. Of course, such a hierarchy only accounts for average density effects, and the different approaches feature also key physics differences that lead to more (or less) QGP shape fluctuations, and thereby %result in
larger (smaller) eccentricities $\epsilon_n$ that impact significantly the extractions of \eg\ the medium viscosity-over-entropy $\upeta/s$ ratio from comparisons of azimuthal flows $v_{n}$ to the hydrodynamic predictions. In particular, the IP-Glasma model also generates the full energy-momentum tensor of the medium, with momentum anisotropies with a length scale (of the order of $Q_s^{-1}(x)= 0.1$--0.2~fm) smaller than those present in other calculations (0.4--1~fm), resulting in a finer structure of the initial entropy density compared to the MC-KLN and MC-Glauber models~\cite{Schenke:2012hg}.
%Recent works tend to incorporate additional sources of fluctuations, at the subnucleonic level [80–83], resulting in general in larger $\ep3$. The MC-rcBK model incorporates negative binomial fluctuations in nucleon-nucleon collisions~\cite{Albacete:2010ad}, the DIPSY model which incorporates a BFKL gluon cascade~\cite{Flensburg:2011kk,Flensburg:2011wx}, and the IP-Glasma model~\cite{Schenke:2012hg} which involves a classical Yang-Mills description of early-time gluon fields.

\begin{figure}[t!]
\centering
\includegraphics[width=0.99\textwidth]{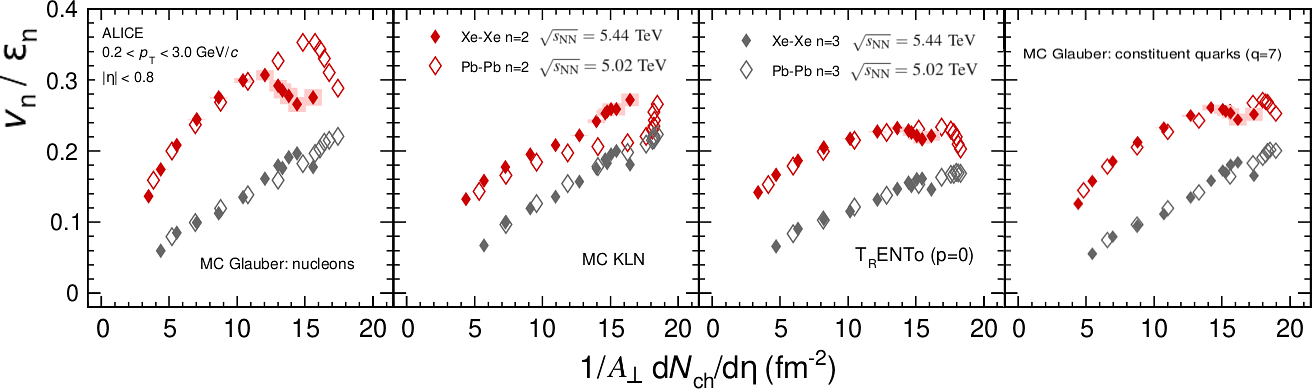}
\caption{Comparisons of the ratio $v_n/\epsilon_n$ (for $n=2,3$) as a function of particle transverse density, $(1/A_\perp)\dd N_{\rm ch}/\dd\eta$, in XeXe and \PbPb\ collisions at the LHC, where $v_n$ and $\dd N_{\rm ch}/\dd\eta$ have been measured by the ALICE experiment, and the $\epsilon_n$ and $A_\perp$ parameters are derived using the ICs of various models described in the text. Figured adapted from Ref.~\cite{Acharya:2018ihu}.
\label{fig:v2_v3}}
\end{figure}

At the LHC, elliptic $v_2$ and triangular $v_3$ flows have been studied \eg\ by the ALICE experiment in XeXe and PbPb collisions at $\snn = 5.44$ and 5.02\,TeV, respectively. The ratios of $v_{2,3}/\epsilon_{2,3}$ as a function of particle transverse density (given by $(1/A_\perp)\dd N_{\rm ch}/\dd\eta$), where $\epsilon_n$ and $A_\perp$ are derived from the ICs of various models described above, are shown in~\Fig{fig:v2_v3}. The hydrodynamic expectation is that $v_n/\epsilon_n$ increases monotonically with the transverse density across different collision energies and systems, and a violation of such a scaling may indicate an incorrect modeling of the initial transverse area $A_\perp$ and/or the azimuthal anisotropies $\epsilon_n$. The results of \Fig{fig:v2_v3} indicate that the standard \MCG\ using nucleons and the MC-KLN model (first and second panels) fail to reproduce the expected scalings for $v_2$ (red symbols), whereas the \trento\ model with $p=0$, equivalent to IP-Glasma, as well as the \MCG\ with constituent partons (third and fourth panels) feature better scaling behaviors accross flow coefficients and systems (although a drop at the largest densities is observed). These results illustrate the type of constraints imposed  by the data on IC medium models, which suggest, in this case, the need of a higher number of subnucleonic sources in order to achieve a steady increase of $v_{2,3}/\epsilon_{2,3}$ for more central collisions. %and highlight the importance of a good ICs control to precisely constrain the $\upeta/s$ of the medium.

Implications for the extraction of QGP transport properties, such as its viscosity constrained from the experimental ratio of triangular over elliptic flows ($v_3/v_2$), have also been studied \eg\ in~\cite{Shou:2014eya,Noronha-Hostler:2019ytn} for UU collisions at RHIC energies. 
The work~\cite{Shou:2014eya} shows that a model overestimation of the $\epsilon_3/\epsilon_2$ ratio will imply a larger amount of the viscous damping needed in the subsequent theoretical hydrodynamic evolution to match the experimental UU data. 
A critical comparison of ICs derived from IP-Glasma and \MCG\ models for light systems produced in \pAu\ and \dAu\ collisions at RHIC can be found in Ref.~\cite{Nagle:2018ybc}.

%%%%%%%%%%%%%%%%%%%%%%%%%%%%%%%%%%%%%%%%%%%%%%%%%%%%%%%%%%%%%%%%%%%%%%%%%%%%%%%%%%%
\section{EXPERIMENTAL DEVELOPMENTS}
\label{sec:experim}
On the experimental front, the LHC data has provided a wealth of new results that have helped to improve the extraction of relevant quantities from the Glauber approach. We review here two experimental aspects of importance for the determination of the reaction centrality in \pA\ and \AA\ collisions. The centrality determination ---a proxy for the (arguably) most important parameter of Glauber models: the collision impact parameter--- is found to be subject to stronger biases at the LHC than at lower \cm\ energies.

%%%%%%%%%%%%%%%%%%%%%%%%%%%%%%%%%%%%%%%%%%%%%%%%%%%%%%%%%%%%%%%%%%%%%%%%%%%%%%%%%%%%%%%%%%%%%%%%%%%%%
\subsection{Collision Centrality Estimates}
%\section{Relating the Glauber Model to Experimental Data}
\label{sec:centralityestimates}

\begin{figure}[t!]
\centering
\includegraphics[width=0.8\textwidth]{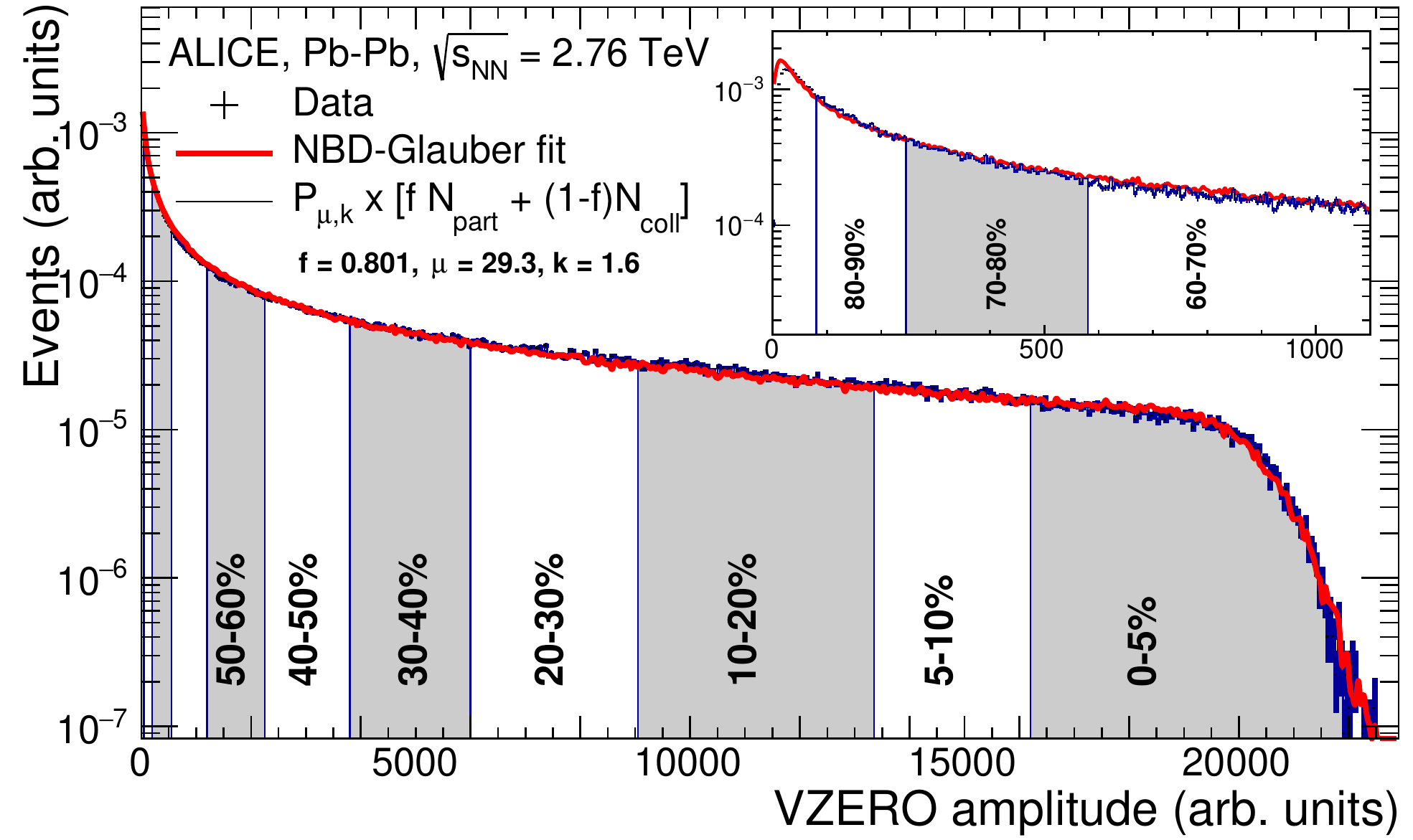} 
\caption{Distribution of the sum of amplitudes~(V0M) in the ALICE VZERO detectors measured in \PbPb\ collisions at $\snn = 2.76$\,TeV fitted with the NBD\,$+$\,Glauber model. Centrality classes are indicated by vertical lines. %, and labelled in the conventional way with smaller percentages referring to more central collisions. 
The inset focuses on the most peripheral region. Figure from \cite{Abelev:2013qoq}.
\label{fig:glauberfit}}
\end{figure}

As mentioned earlier, neither the impact parameter nor any derived Glauber quantity can be directly measured experimentally. Instead, average quantities are obtained within Glauber approaches for classes of events whose inclusive particle multiplicities and/or energy distributions can be reproduced by the corresponding calculation over a given $b$ range. Since on average the impact parameter is monotonically related to the overall particle multiplicity\com{, at mid and forward rapidity}, one typically measures multiplicity~(or energy) distributions over a suitably large phase space. The mapping to calculated quantities then proceeds in intervals of centrality or {\it centrality classes}, which are obtained by binning the distribution in fractions of its total integral.
Centrality is then typically defined as the percentile 
\begin{equation}
c_i = \frac{c_{\rm ap}}{M_{\rm tot}} \,\int_{M_{\rm i}}^{\infty} \frac{\mathrm{d}N}{\mathrm{d}M}\, \mathrm{d}M\,,
\end{equation}
of the per-event multiplicity distribution $\mathrm{d}N/\mathrm{d}M$ above $M_{\rm i}$ relative to 
\begin{equation}
M_{\rm tot}  = \int_{M_{\rm ap}}^{\infty} \frac{\mathrm{d}N}{\mathrm{d}M} \, \mathrm{d}M \,,
\end{equation}
where $M_{\rm ap}$~($<M_i$) is the multiplicity value for which the fraction of total cross section was determined at the $c_{\rm ap}$ point. 
The \emph{anchor point}~(AP) sets the absolute scale of the centrality. Clearly, one would like to achieve $M_{\rm ap}$ close to zero, resulting in $c_{\rm ap}$ close to 100\%. However, because of trigger inefficiency and the increasing background contamination from ultraperipheral photonuclear collisions, experiments typically can only set the anchor point between 80 to 90\% of the total hadronic cross section.
An example is given in \Fig{fig:glauberfit}, which shows the sum of the amplitudes in the ALICE VZERO scintillators~(at $2.8<\eta<5.1$ and $-1.7<\eta<-3.7$), called V0M centrality estimator, representing the uncorrected charged particle multiplicity distribution. 
The vertical lines indicate the typical centrality binning obtained from slicing the distribution in fractions of the total integral starting from 90\%~\footnote{The range 90--100\% is prone to potentially large contamination from photonuclear contributions, and thereby often avoided.}, where smaller fractions refer to more central collisions.

To determine the anchor point, two approaches are typically used. 
The first one involves simulation of hadronic and e.m.\ processes, including a detailed description of the detector response\com{, to study the efficiency and background contamination of the event selection}, and hence gives direct access to the fraction of hadronic events below $M_{\rm AP}$.
%Essentially this approach aims to obtain the true multiplicity distribution, corrected for efficiency and purity, which then allows direct access to the anchor point.
The second one involves \MCG\ modeling together with a simple description for particle production at detector level, to simulate the uncorrected multiplicity distribution. %, which is then compared to the data.
The calculated distribution will describe the data down to the most peripheral events up to some point where background contamination and trigger inefficiency start to matter.
The point where data and simulation start to separate can be used to set the AP.

To model multiplicity in heavy-ion collisions, one exploits that the majority of the initial-state \NN\ collisions can essentially be treated like MB \pp\ collisions, with a small perturbation from rarer hard interactions. 
The charged particle multiplicity in minimum-bias \pp\ collisions at high energy can be described by a negative binomial distribution~(NBD)~\cite{GrosseOetringhaus:2009kz}, given by
\begin{equation}
  P_{\mu,k}(N_{\rm ch}) =  \frac{\Gamma(N_{\rm ch}+k)}{\Gamma(N_{\rm ch}+1)\Gamma(k)} \cdot
  \frac{(\mu/k)^{N_{\rm ch}}}{(\mu/k+1)^{N_{\rm ch}+k}}\,,
\end{equation}
where $\mu$ is the mean and $k$ is related to the width of the multiplicity distribution.
Hence, the multiplicity for heavy-ion events can be approximated as a superposition of many NBDs, quickly approaching the Gaussian limit.
A typical approach is to assume that the number of particle emitting sources can be described by the two-component approach, $f\cdot\Npart+(1-f)\cdot\Ncoll$ as in \Eq{eq:s0_MCG}. 
A minimization procedure is then applied to the distribution of hadronic activity to determine the $\mu$, $k$, and $f$ parameters, with the values listed in \Fig{fig:glauberfit} obtained with a $\chi^2/N_{\rm dof}$ close to unity.
The AP can be determined with $\sim$1\% uncertainty~(absolute).
The centrality resolution is at the level of as few \% in peripheral collisions and better than 1\% in most central collisions.

Due to the finite kinematic acceptance, trigger inefficiency, and detector resolution, as well as the possible biases of the event selection, the details of the centrality determination differ between experiments, and even among collision systems within a given experiment.
%Hence, the details of the centrality determination differ between experiments as well as between collision systems within a given experiment.
A short introduction with references to the approaches at RHIC is given in \cite{Miller:2007ri}, while for more details on similar approaches at the LHC, see \cite{ALICE:2018tvk,Abelev:2013qoq} for ALICE, \cite{Chatrchyan:2011pb} for CMS, and \cite{ATLAS:2011ag} for ATLAS. 
Alternative centrality estimators based on the transverse energy measured at forward rapidities~(approx.~$3<|\eta|<5$ for CMS and ATLAS), as well as far-forward neutral energy in Zero Degree Calorimeters~(ZDCs) along the beam line~\cite{Wood:2012jem, Abelev:2013qoq}, have been also employed.
The same methods can be applied to determine the centrality in \pA\ or \dAu\ collisions~\cite{Adare:2013nff,Adam:2014qja,Perepelitsa:2014yta}.
However, unlike for collisions of larger nuclei, the centrality determination is often subject to larger biases due to fluctuations in the categorization of events~(see next section).

%%%%%%%%%%%%%%%%%%%%%%%%%%%%%%%%%%%%%%%%%%%%%%%%%%%%%%%%%%%%%%%%%%%%%%%%%%%%%%%%%%%%%%%%%%%%%%%%%%%%%
\subsection{Collision Centrality Biases}
\label{sec:centralitybiases}
As discussed in the previous section, medium effects on the production of perturbative QGP probes are in general quantified by the nuclear modification factor \Eq{eq:RAA}, defined as the ratio of the per-event yield measured in \AA\ collisions over the same yield expected from an incoherent superposition of \Ncoll\ binary \pp\ collisions. 
%In the absence of nuclear effects, \RAA\ is expected to be unity. 
However, event centrality classification involves selection of event samples for which the properties of the underlying binary \NN\ collisions may deviate from those of unbiased \pp\ collisions~\cite{Morsch:2017brb}. 
In this case, \RAA~(and $\RpA$) can deviate from unity even in the absence of nuclear effects. 
There are two main sources of selection biases.
First, the number of hard processes is suppressed for increasingly peripheral \AA\ collisions because of a simple geometrical bias: the probability for collisions increases proportional to $b$ while the nuclear density decreases, leading to an increased probability for more-peripheral-than-average \NN\ collisions.
Second, the centrality selection, which relies mostly on measurements dominated by soft particle production ---the relative weight of soft-to-hard contributions to the total energy distribution is given by the $f\approx0.8$ parameter in \Fig{fig:glauberfit}---
biases the average multiplicity of individual \NN\ collisions, and hence can affect the normalization of yields of collisions dominated by hard processes. 
As shown in \Fig{fig:pp_NN_plot} (left), hard scatterings are more probable in central \NN\ collisions with large partonic overlap thereby leading to many simultaneous MPIs with lots of UE activity, and since hard processes are dominated by the production of jets that fragment (or heavy resonances that decay) into a large number of final hadrons, a peripheral \AA\ event with a hard scattering often has a hadronic activity much larger than that typical of its centrality class.
%Such event-by-event correlations between hard and soft particle production 
Since the \AA~(and \pA) centrality determination is based on ordering the measured multiplicity or summed energy in the event, peripheral nuclear events with a hard scattering can thereby be wrongly assigned to a more central class.

The geometrical effect can be included into optical Glauber calculations by extending \Eq{eq:tab} with a convolution of the nuclear thickness functions (depending on the overall $b$) with the \NN\ overlap function~($\tnn$, depending on the $\bnn$ impact parameter) as
\begin{equation}
 \TAB'\left(\bf{b}\right) = \int  {\rm d}^2{\rm s}\;{\rm d}^2\bnn\, \TA\left(\bf{s}\right)\TB\left(\bf{s}-\bf{b}+\bf{\bnn}\right) \, \tnn\left(\bf{\bnn}\right)\,,
\label{eq:tab2}
\end{equation}
which effectively leads to a reduction of $\Ncoll$ in peripheral collisions compared to $\tnn=\delta(\bnn)$ because of their increased probability for less-central-than-average \NN\ collisions. Standard \MCG\ calculations, which by construction include the geometrical bias\com{ and which in addition provide an impact parameter dependent \NN\ profile in recent models~\cite{Rybczynski:2013yba,Bozek:2019wyr,Loizides:2017ack}}, however, 
do not use information about individual \NN\ collisions. 
\com{In contrast, the \Hijing\ model takes into account the possibility of multiple hard scatterings in the same \NN\ collision depending on $\bnn$.}
%As in\com{ MC models for \pp\ collisions like} \Pythia~\cite{Sjostrand:2006za}, which uses the multiparton interaction~(\MPI) picture, the mean number of hard scatterings per collision depends on \com{the \NN\ impact parameter}$\bnn$ as $\sigmahard\,\tnn(\bnn)$, where \sigmahard\ is the $\sqrts$-dependent pQCD cross-section for $2 \rightarrow 2$ parton scatterings. % rising from about 1~mb at 0.04\,TeV to about 124~mb at 5.02\,TeV. 
Even though the \NN\ collisions are still modeled as occurring incoherently, the number of hard processes for a given centrality selections is not taken proportional to $\Ncoll$, but to 
\begin{equation}
\Nhard=\Ncoll \cdot \Nhardnn / \left<\Nhardnn\right>\,, 
\label{eq:nhard}
\end{equation}
where $\Nhardnn$ is the average number of hard scatterings in a \NN\ collision for a given centrality selection and $\left<\Nhardnn\right>$ is its unbiased average value.
The mean number of hard scatterings per collision depends on $\bnn$, and can be written as $\Nhard(\bnn)=\sigmahard\,\tnn(\bnn)$, where \sigmahard\ is the pQCD cross-section for $2 \rightarrow 2$ parton scatterings. %, which strongly rises from about $1$~mb at $\sqrts=0.04$\,TeV to about $11.7$~mb at $0.2$\,TeV to about $124$~mb at $5.02$\,TeV.
Since the yield of hard and soft processes are correlated via their common $b_{\rm NN}$, the \NN\ collisions can be biased towards lower or higher than average impact parameters, when ordering the measured multiplicity or transverse energy necessary for the centrality determination.
This leads to a selection bias on $\Nhardnn$ in addition to the inherent geometrical bias.
Due to the strong dependence of $\sigmahard$ on $\sqrts$\com{, which rises from about $1$~mb at $\sqrts=0.04$\,TeV to about $11.7$~mb at $0.2$\,TeV to about $124$~mb at $5.02$\,TeV}, the selection bias is more relevant at LHC than at RHIC (and negligible at SPS) collision energies.
%Consequently, the average number of hard collisions per \NN\ collisions decreases from $\left<\Nhardnn\right>=1.77$ at $\sqrts=5.02$\,TeV to $0.28$ at $0.2$\,TeV and is negligible below $0.04$\,TeV.

\begin{figure}[t!]
\centering
\includegraphics[width=0.6\textwidth]{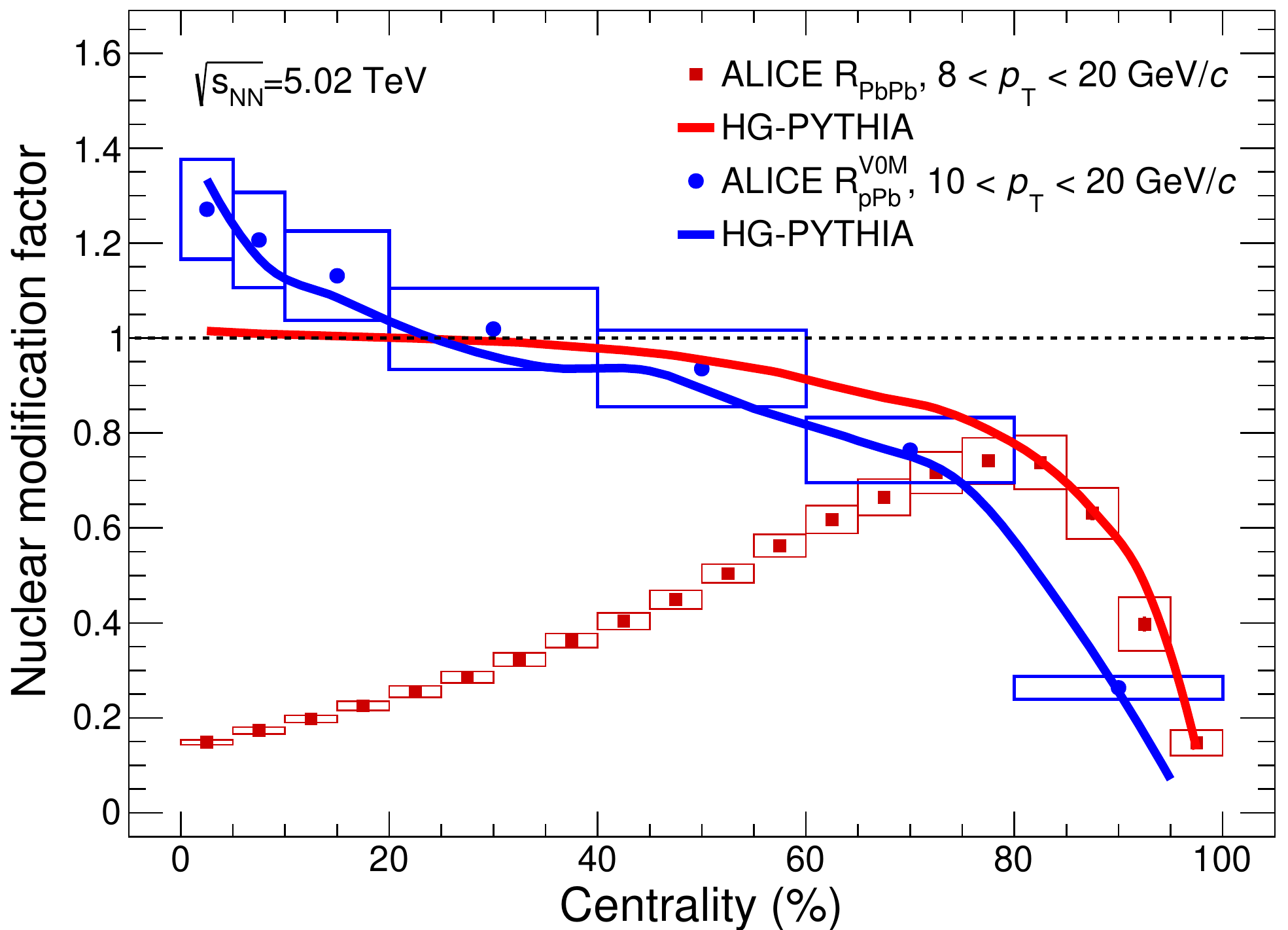} 
\caption{Nuclear modification factors for high-$\pT$ charged particles (above 8 or 10\,GeV) as a function of centrality in \pPb\ (blue circles) and \PbPb\ (red squares) collisions at $\snn=5.02$\,TeV~\cite{Adam:2014qja,Acharya:2018njl} compared to calculations with \Hpythia~\cite{Morsch:2017brb} (blue and red curves, respectively). 
%The centrality for data and calculations is estimated in the acceptance of the ALICE VZERO detector %~($2.8<\eta<5.1$ and $-1.7<\eta<-3.7$).}
\label{fig:biasraa}}
\end{figure}

The relevance of the selection bias induced by the correlation between soft and hard particles is demonstrated in \Fig{fig:biasraa} that shows the nuclear modification factors for charged particle production, integrated above a large enough $\pT$ value~(8 and 10\,GeV), in \pPb\ and \PbPb\ collisions at $\snn=5.02$\,TeV compared to calculations using \Hpythia~\cite{Morsch:2017brb}.
%For a given MC Glauber event calculated in \Hijing, we run \Pythia~(v6.28 with Perugia 2011 tune) to generate pp events with exactly $\left(\Nhardnn\right)_i$  multiple parton interactions for each \NN\ collision $i$, where $i$ runs from $0$ to \Ncoll . 
%The generated particles from all \Pythia\ events are then combined and treated like a single \AA\ event in the further analysis.
The calculation, which uses the \Hijing\ model to determine the distribution of $\Nhard(\bnn)$ in a nuclear collision and \Pythia\,6.28 (Perugia 2011 tune) to generate the corresponding \NN\ events, purposely does not include nuclear modification effects, unlike most of the models discussed in \Sec{sec:HI_MCs}.
%\todo{Models like \Hijing, Angangtyr, and others - \epos? differ in the way they deal with additional effects like shadowing, and overall-energy momentum conservation and the underlying MPI model.}
As for the data, where the V0M estimator was used, the calculation determines the centrality using charged particles in the acceptance of the ALICE VZERO detectors. %~($2.8<\eta<5.1$ and $-1.7<\eta<-3.7$).
The calculation describes well the \pPb\ and very peripheral \PbPb\ data indicating that the strong apparent suppression from unity in this region results, in both cases, from the event selection. 
For more central \PbPb\ collisions, where parton energy loss leads to the known, large suppression of particle production in \PbPb\ compared to \pp\ collisions, the $\RAA$ is not affected by such biases. %as demonstrated by the agreement between the \pPb\ data and the \Hpythia\ prediction.

The hard-soft event selection bias is particularly important when fluctuations of the centrality estimator caused by $\bnn$ are of similar size as the dynamic range of \Ncoll, as is the case in \pA\ collisions, and strongly enhanced by trivial autocorrelations if the phase space for the measurement and event categorization are close-by or overlap~\cite{Adare:2013nff,Adam:2014qja,Perepelitsa:2014yta}. 
This can be already deduced by computing the ratio between the average multiplicity of the centrality estimator and the average multiplicity per average ancestor of the Glauber fit as demonstrated in Fig.~8 of~\cite{Adam:2014qja}. %todo the arxiv version could have figure 8
In contrast, centrality measurements based on zero-degree energy should not introduce any selection bias, while the geometric bias could still play a role. 
In the so-called hybrid method, described in~\cite{Adam:2014qja}, the \pPb\ centrality selection is based on ZDC neutral energy in the Pb-going directions~(slow neutrons), and \Ncoll\ is determined from the measured charged particle multiplicity $M$ according to $\Ncoll = \langle \Ncoll \rangle \cdot M /  \langle M \rangle$, where $\langle \Ncoll \rangle$ and  $\langle M \rangle$ are, respectively, the centrality-averaged number of collisions and multiplicity. 
In case soft and hard particle yields are affected in the same way, the selection bias would cancel out in the nuclear modification factor.

%%%%%%%%%%%%%%%%%%%%%%%%%%%%%%%%%%%%%%%%%%%%%%%%%%%%%%%%%%%%%%%%%%%%%%%%%%%%%%%%%%%
\section{SUMMARY}
\label{sec:summ}
An outstanding topic in the physics of the strong interaction is the understanding of the %emergence and behavior of the 
thermodynamic and transport properties of hot and dense quark-gluon matter accessible to experimental study via high-energy collisions of nuclei. 
To correctly identify and interpret signals of collective partonic behavior in \AA\ collisions, it is necessary to have a realistic extrapolation of the baseline hadron production properties of \pp\ and \pA\ collisions where, in principle, no quark-gluon plasma (QGP) is expected to be formed. 
In this work, we have reviewed the main developments and the state-of-the-art of the Glauber approach to describe multiple scatterings in proton and nuclear collisions, after 10 years of operation with \pp, \pPb, and \PbPb\ collisions at the CERN LHC, as well as with deformed light- and heavy-ions at BNL RHIC. 
The Glauber model allows for arguably the simplest, most economical, and yet successful, understanding of collisions of extended hadronic objects based on an impact-parameter~($b$) superposition of independent elementary scatterings, each of which produces particles and thereby defines the local and global density of the precursor QGP. 
Key derived quantities in Glauber models include the nuclear overlap function $\TAA(b)$, number of participant nucleons $\Npart(b)$, number of binary collisions $\Ncoll(b)$, transverse area $A_\perp(b)$, eccentricities $\epsilon_n(b)$, average path-length $L(b)$, of the strongly-interaction medium produced at different collision centralities, which are fundamental for the extraction of the QGP properties from the data.

The new LHC measurements, performed at 50 times larger \cm\ energies than at previous nuclear collisions, and the latest precision RHIC data for a variety of colliding systems, have required to revisit and improve various ingredients of the  Monte Carlo Glauber~(MCG) simulations. 
Our review has first provided a new fit of the world measurements of the inclusive inelastic nucleon-nucleon cross sections ($\sigmaNN$), a key ingredient of \MCG\ models. 
The inelastic hadronic \pA\ and \AA\ cross sections measured at the LHC are well reproduced by the corresponding \MCG\ results derived using $\sigmaNN$, a fact that confirms the overall validity of the Glauber model at the highest \cm\ energies ever studied. 
Second, improved descriptions of the proton and nuclear density profiles, including subnucleonic degrees of freedom and neutron skin effects, and any associated sources of fluctuations and correlations, have been examined. 
Third, we have reviewed the main applications of the Glauber model for collider studies. 
The binary scaling prescription to quantitatively compare hard scattering cross sections in \pp, \pA, and \AA\ collisions, has been validated by measurements of electroweak probes at the LHC whose yields are unaffected by final-state interactions in the QGP. 
The use of the Glauber formalism in MC event generators for heavy-ion physics, as well as to provide initial entropy-density profiles as input for relativistic hydrodynamics calculations, have also been discussed. 
The importance of a realistic description of the medium eccentricities to extract key transport properties, such as the QGP shear viscosity, from comparisons of elliptic and triangular flows measurements at RHIC and LHC to viscous hydrodynamics predictions has been highlighted.
%Two particularly important ``consumers'' of such Glauber results are nuclear parton distribution functions (PDFs) and hydrodynamical studies.  aiming at precise extractions of nuclear parton distribution functions (in particular, when using isospin-sensitive observables such as $\gamma$, W, and Z gauge bosons cross sections), and  validation of the binary scaling prescription for hard scattering cross sections, and determination of initial-state conditions for relativistic hydrodynamics evolution of quark-gluon matter of relevance for viscosity/entropy v2, v3...
Last, the experimental procedures used and the inherent biases introduced by them, in the determination of the collision centrality from the data, which rely on the application of the \MCG\ model, have been briefly discussed. 

\begin{figure}[t!]
\centering
\includegraphics[width=0.49\textwidth]{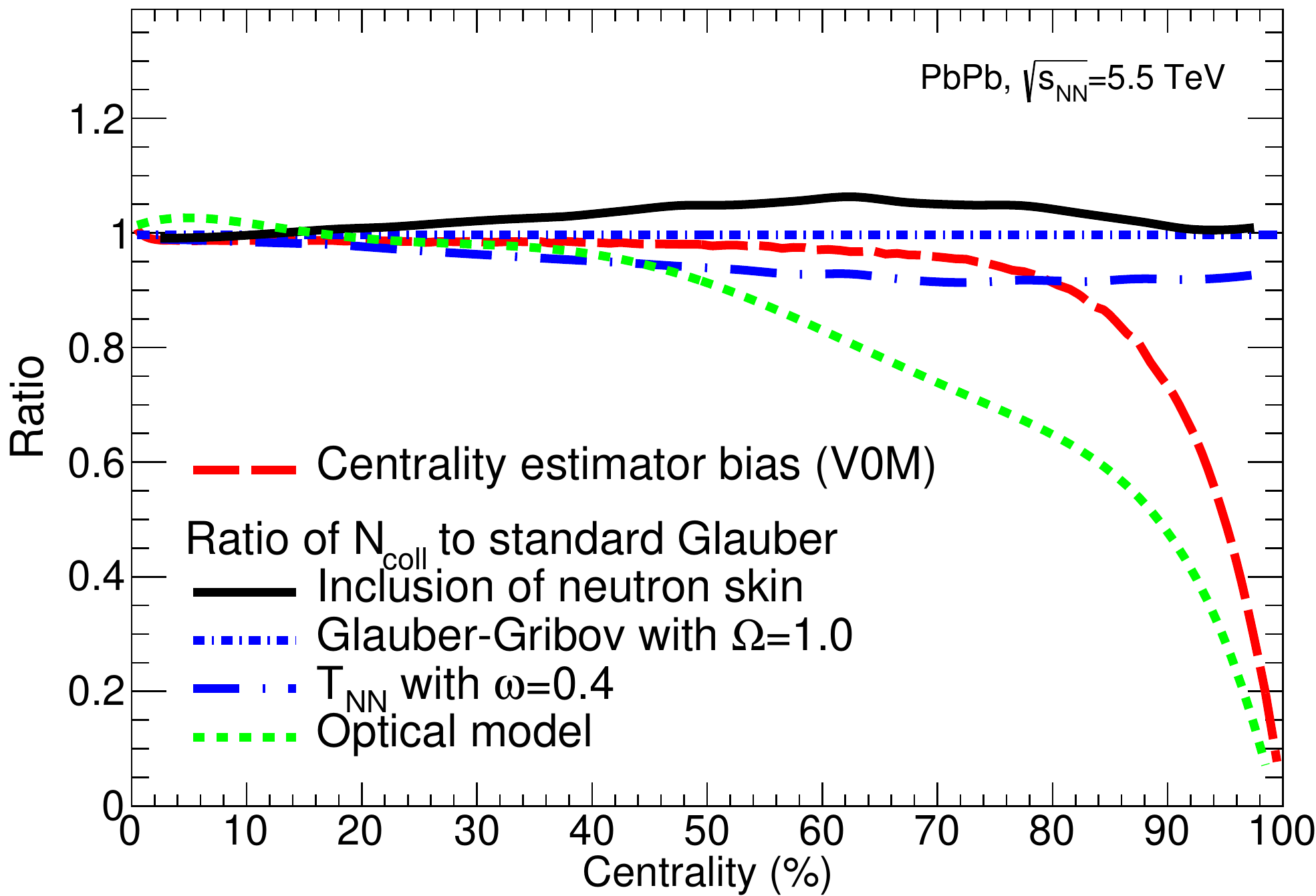} 
\includegraphics[width=0.49\textwidth]{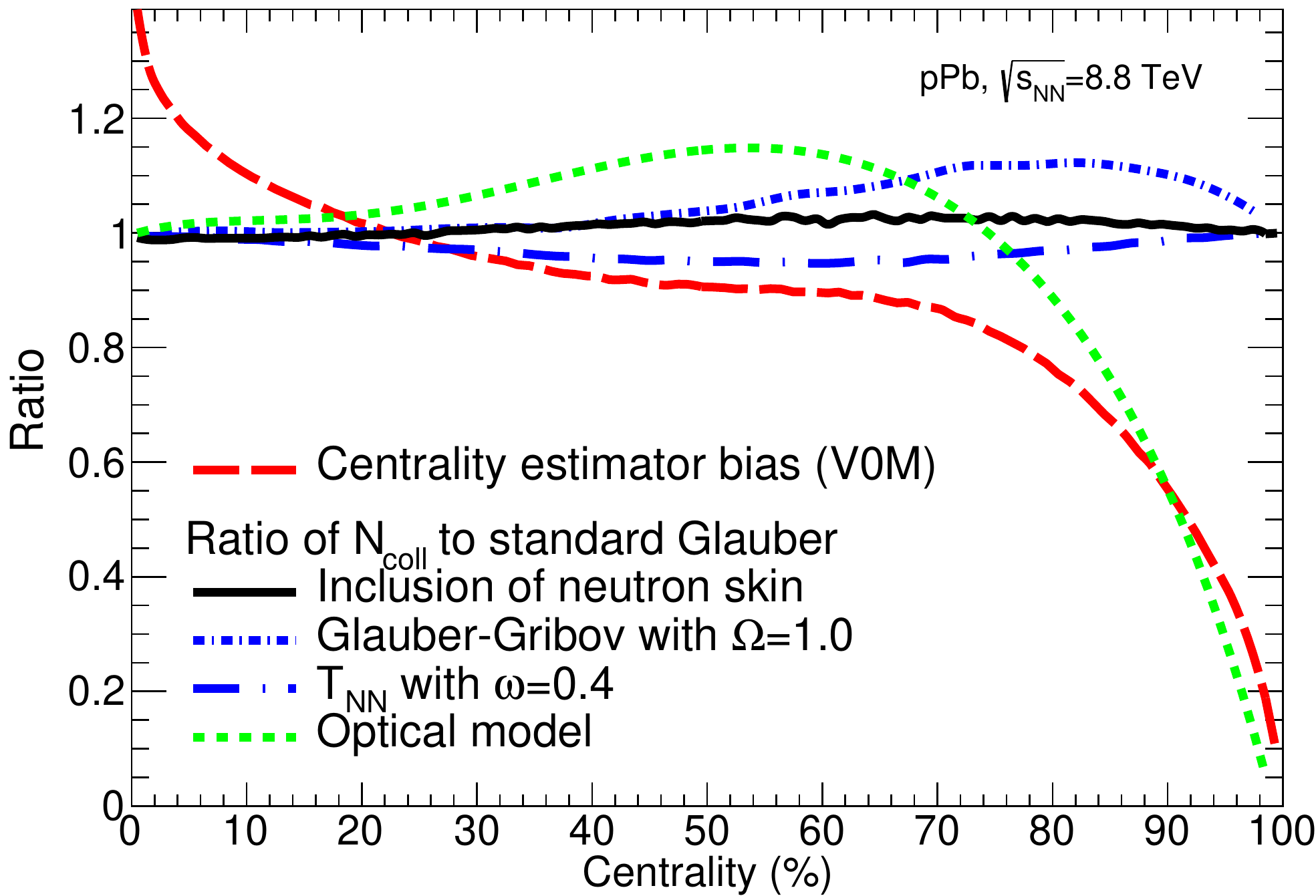} 
\caption{Ratio of $\Ncoll$~as a function of centrality in \PbPb\ (left) and \pPb\ (right) collisions at the LHC obtained with the various model developments discussed in this review, and for the optical limit~\Eq{eq:tab}, normalized to the result obtained with the conventional \MCG\com{ without such improvements}. In addition, the bias induced by the centrality determination is shown for a typical experimental estimator (ALICE~V0M, long-dashed red curves).}
\label{fig:summary}
\end{figure}

As an illustrative summary of our review, \Fig{fig:summary} shows the impact of different Glauber model ingredients and experimental biases\com{ discussed in this report}, presented as a ratio as a function of centrality of the different elements over the default standard \MCG\ calculation, for \PbPb\ (left) and \pPb\ (right) collisions at $\snn = 5.5$ and 8.8\,TeV, respectively. 
First, the red long-dashed curves indicate the magnitude of the experimental shift\com{ distortions} introduced in measured nuclear modification factors when not properly accounting for event selection~(multiplicity- and process-dependent) biases introduced by the centrality determination. 
%First, the red long-dashed curves indicate the experimental bias from ordering events according to multiplicity or transverse energy introduced by the centrality determination.
These biases are significant in the most peripheral centrality classes and need to be carefully modeled and/or corrected for, specially when aiming at precision measurements at large impact parameters. 
The other curves indicate the ratio of the values of $\Ncoll$ obtained with modified ingredients with respect to standard \MCG\ simulations. 
Inclusion of Glauber--Gribov fluctuations (via \Eq{eq:fluct} with $\Omega = 1$) , modified NN collision profiles (via \Eq{eq:NN_coll_profile} with $\omega = 0.4$), or neutron skin effects, lead to few percent modifications of the $\Ncoll$ values in different centrality ranges, in principle within the assigned Glauber model systematic uncertainties~\cite{Loizides:2017ack}. 
An analytical calculation of $\Ncoll$ in the optical Glauber limit leads, however, to significant underestimations of the number of collisions for peripheral \PbPb\ and \pPb\ collisions.
%Whereas central \PbPb\ collisions feature no significant $\TAA$ change, peripheral collisions as well as \pPb\ collisions can be largely modified compared to the standard Glauber predictions. 
%Taking into account all those effects, is mandatory in order to interpret the experimental data and improve our understanding of the initial state of quark-gluon matter produced in heavy-ion collisions. 

The results of \Fig{fig:summary} emphasize large quantitative corrections needed in the Glauber model for the most peripheral \AA\ and \pA\ collisions. 
The availability of very large data samples of electroweak bosons at the LHC, with cross section measurements with few percent experimental uncertainties similar to or smaller than those of the Glauber model, opens up the possibility of using them to define an alternative experimental proxy for the nuclear overlap function. The ratio $N_{\rm V}/(\sigmaNN^{\rm V}N_{\rm evt})$, where $\sigmaNN^{\rm V}$ is the vector boson~(V = $\sum W^\pm$, Z) production cross section in NN collisions~(that can be estimated from \pp\ measurements) and $N_{\rm V}/N_{\rm evt}$ the per-event \AA\ yields, has been suggested~\cite{CMS:2020rno} as a data-driven $\TAA(b)$ proxy that would eliminate the need for Glauber modeling, and reduce corrections for centrality and event-selection effects, while cancelling uncertainties in the determination of quantities such as $\RAA$. 
Such a proposal would require a higher level of theoretical accuracy (next-to-next-to-leading order) in the nuclear parton distribution functions, and in their centrality dependence, in order to fully exploit the high-precision $\sigmapp^{\rm V}$ measurements.

All in all, the results summarized in this review show that, despite its simplicity~(or, arguably, thanks to it), the Glauber model has stood for over 50 years as an indispensable and useful baseline approach to be able to quantitatively compare collisions of systems of varying size, from protons to uranium nuclei. 
Its continued exploitation to extract from the data key thermodynamics and transport properties of strongly interacting matter at the highest densities and temperature accessible in the laboratory, remains unchallenged for the years to come.

%%%%%%%%%%%%%%%%%%%%%%%%%%%%%%%%%%%%%%%%%%%%%%%%%%%%%%%%%%%%%%%%%%%%%%%%%%%%%%%%%%%
\ifarxiv
\section*{LICENSE}
This work is licensed under \href{https://creativecommons.org/licenses/by/4.0/}{CC-BY-4.0}.
\else
\section*{DISCLOSURE STATEMENT}
The authors are not aware of any affiliations, memberships, funding, or financial holdings that might be perceived as affecting the objectivity of this review. 
\fi

% Acknowledgements
\section*{ACKNOWLEDGMENTS}
%Discussions with ..., and ... are gratefully~acknowledged. 
We are indebted to J.~Kamin, A.~Morsch, J.~Nagle, A.~Snigirev, and P.~Steinberg for common work and discussions in the past years leading to results presented in this review.
Comments on the text by U.~Heinz, I.~Helenius, V.~Kovalenko, H.~Paukkunen, and B.~Schenke, are gratefully acknowledged.
We thank F.~Jonas for carefully reading the manuscript.
C.~Loizides is supported by the U.S. Department of Energy, Office of Science, Office of Nuclear Physics, under contract number DE-AC05-00OR22725.

%%%%%%%%%%%%%%%%%%%%%%%%%%%%%%%%%%%%%%%%%%%%%%%%%%%%%%%%%%%%%%%%%%%%%%%%%%%%%%%%%%%
\ifarxiv
\appendix
\section{Details on \MCG\ simulations and parameters}
\label{sec:app}
This appendix provides a concise summary of technical details and parameters of \MCG\ simulations.
%, and has some redundancy with \Sec{sec:intro} and \Sec{sec:lightandheavy}.
The \MCG\ calculations for \AA\ collisions are typically performed in two steps:
(i) the nucleon positions in each colliding nucleus are determined for a given event, and
(ii) the two nuclei are ``collided'' assuming that the nucleons travel in a straight-line along the beam axis (eikonal approximation). 
%such that nucleons are tagged as wounded~(participating) or spectator.
Conventional implementations can be found in the codes of Refs.~\cite{Broniowski:2007nz,Alver:2008aq,Rybczynski:2013yba,Loizides:2014vua,Loizides:2016djv,Mitchell:2016jio,Loizides:2017ack,Bozek:2019wyr}. 
Among them, \Refs{Loizides:2016djv,Mitchell:2016jio} provide a simple approach to include subnucleonic degrees of freedom.

\subsection*{Setup of nuclei}
The position of each nucleon in the nucleus is determined according to a probability density function, where one commonly requires a minimum internucleon separation~($d_{\rm min}=0.4$~fm) between the centers of the nucleons to mimic hard-core repulsion of nucleons inside the nucleus\footnote{For many years, changing this model value by $\pm0.4$~fm was done to compute the uncertainty from this apriori unknown parameter, leading to substantial uncertainties in the Glauber parameters. It was shown in \Refe{Loizides:2017ack} that, when avoiding artefacts from edge effects, the corresponding uncertainty can be neglected.}.
The probability distribution for spherical nuclei is taken to be uniform in azimuthal and polar angles, and the radial probability function is taken from nuclear charge densities extracted in low-energy electron scattering experiments~\cite{DeJager:1974liz,DeJager:1987qc}.
The nuclear charge density is usually parametrized by a Fermi distribution with three parameters 
\begin{equation}\label{eq1}
  \rho(r)=\rho_0 \frac{1+w(r/R)^2}{1+\exp(\frac{r-R}{a})}\,,
\end{equation} 
where $R$ is the nuclear radius, $a$ is the skin depth, and $w$ is a parameter to describe deviations from a spherical shape. Values used for various nuclei are listed in \Tab{tab:awR}.
The overall normalization given by the nucleon density ($\rho_0$) plays no role in the \MCG\ simulation.

Exceptions to \Eq{eq1} include the deuteron~(${}^{2}$H), tritium~(H3, ${}^{3}$H), 
helium-3~(He3, ${}^{3}$He), helium-4~(He4, ${}^{4}$He), carbon~(C), oxygen~(O), and sulfur (${}^{32}$S) nuclei. For deuteron, three typical parametrizations are used: %~(the 3rd the preferred):
\begin{enumerate}
\item The 3-parameter Fermi form, \Eq{eq1}, with values $R = 0.01$~fm, $a=0.5882$~fm, and $w=0$.  %given in \Tab{tab:awR}.
\item The Hulth\'en form:
\begin{equation}\label{eq3}
  \rho(r')=\rho_0 \left(\frac{e^{-ar'}-e^{-br'}}{r'}\right)^2,
\end{equation} 
where $a=0.228$~fm$^{-1}$ and $b=1.177$~fm$^{-1}$, and $r'$ denotes the distance between the proton and neutron, \ie\ $r=r'/2$~~\cite{Hulthen1957,Adler:2003ii,Adler:2006xd}. 
\item The proton taken from the Hulth\'en form, \Eq{eq3}, with the neutron placed opposite to it. 
\end{enumerate}

\begin{table}[t!]
\centering
  \begin{tabular}{l|c|c|c}\hline
    Name           & $R$~[fm]        &  $a$~[fm]       & $w$ \\\hline
%    p\footnotemark & 0.234         &               & \\
%    d\footnotemark & 0.01\hide{0}  & 0.5882        & 0       \\
%    H3\footnotemark&&&\\
%    He3$^{9}$&&&\\
%    He4$^{9}$&&&\\
%    C$^{9}$&&&\\
%    ${}^{16}$O$^{9}$&&&\\
%    ${}^{16}$O (par)  & 2.608         & 0.513\hide{0} & $-0.51$\hide{00}              \\
%    ${}^{16}$O (ho)   & 1.833         & 1.544\hide{0} & 0       \\
    ${}^{28}$Si    & 3.34\hide{0}  & 0.580\hide{0} & $-0.233$ \\
    ${}^{32}$S     & 2.54\hide{0}  & 2.191\hide{0} & 0.16      \\
    ${}^{40}$Ar    & 3.53\hide{0}  & 0.542\hide{0} & 0        \\
    ${}^{40}$Ca    & 3.766         & 0.586\hide{0} & $-0.161$ \\
    ${}^{58}$Ni    & 4.309         & 0.517\hide{0} & $-0.1308$  \\
    ${}^{63}$Cu    & 4.2\hide{00}  & 0.596\hide{0} & 0      \\
    ${}^{63}$CuHN  & 4.28\hide{0}  & 0.5\hide{000} & 0       \\
    ${}^{129}$Xe~\footnote{Parameters $R=5.4\pm0.1$~fm and $a=0.61^{+0.07}_{-0.09}$~fm for $^{132}$Xe~\cite{Tsukada:2017llu} are used, with the radius scaled down by $0.99 = (129/132)^{1/3}$ and $a$ reduced by $0.02$~fm, to symmetrize the uncertainty and to approximate the smaller $^{129}$Xe neutron skin.}
                  & 5.36\hide{0}  & 0.59\hide{00} & 0       \\
    ${}^{129}$Xe\footnote{Parameters $R=5.32$~fm and $a=0.57$~fm for Sb(121.7)~\cite{DeJager:1974liz,DeJager:1987qc} are used, with the radius of Sb scaled up by $1.019 = (129/122)^{1/3}$ using the average of $A\approx122$ between the two isotopes of natural Sb ($A=121$ with $57.2$\%, and $A=123$ with $42.8$\% abundances). The resulting parameters are consistent with those obtained from $^{132}$Xe.}
                  & 5.42\hide{0}  & 0.57\hide{00} & 0       \\
    ${}^{186}$W    & 6.58\hide{0}  & 0.480\hide{0} & 0       \\
    ${}^{197}$Au   & 6.38\hide{0}  & 0.535\hide{0} & 0       \\
    ${}^{197}$AuHN & 6.42\hide{0}  & 0.44\hide{00} & 0       \\
    ${}^{207}$Pb\footnote{These values are usually also used for ${}^{208}$Pb, since the Bessel--Fourier coefficients for the two nuclei are similar~\cite{DeJager:1974liz,DeJager:1987qc}: the values $R=6.624$~fm and $a=0.549$~fm for ${}^{208}$Pb~(``Pb*'') from~\cite{DeJager:1974liz} are consistent within uncertainties with the ${}^{207}$Pb values.}
                  & 6.62\hide{0}  & 0.546\hide{0} & 0       \\
    ${}^{207}$PbHN & 6.65\hide{0}  & 0.46\hide{00} & 0       \\
%    ${}^{208}$Pb*  & 6.624         & 0.549\hide{0} & 0       \\
    ${}^{208}$Pb (prot)  &  $6.68$ \com{$\pm0.02$}    &  $0.447$ \com{$\pm0.010$} & 0  \\
    ${}^{208}$Pb\footnote{Separate (point-like) profiles for protons and neutrons, as described in~\cite{Loizides:2016djv}.}       (neut) &  $6.69$ \com{$\pm0.03$}    &  $0.560$ \com{$\pm0.030$} & 0 \\
    \hline
  \end{tabular}
  \caption{\label{tab:awR}\protect Nuclear charge density parameters from \cite{DeJager:1974liz,DeJager:1987qc} for Eqs.~(\ref{eq1}) and (\ref{eq2}). %~(\ref{eq1})--(\ref{eq3}).
  The suffix 'HN' indicates rescaled values to account for finite nucleon profiles as derived in Ref.~\cite{Hirano:2009ah}.  All configurations can be found in the {\tt TGlauNucleus::Lookup} function of the TGlauberMC code~\cite{glaucode}.}
  \end{table}

For ${}^{3}$H and ${}^{3}$He, the configurations have been computed (and stored in a database) from Green's function MC calculations using the AV18\,+\,UIX model interactions, which correctly sample the position of the three nucleons, including correlations, as in \Refe{Nagle:2013lja}.
These, and the results of wavefunction-based calculations for He-4, carbon, and oxygen derived in~\cite{Lim:2018huo} are publicly available~\cite{glaucode}. There are also rescaled values\footnote{Rescaled parameters that include the {\it recentering} effect, in addition, are given in~\cite{Loizides:2017ack}.} for the radius and skin depth of  Cu~(``CuHN''), Au~(``AuHN'') and Pb~(``PbHN'') to take into account the finite nucleon profile as derived in \cite{Hirano:2009ah} and \cite{Shou:2014eya}.

For sulfur, and a few other nuclei (Si, Ca, Ni,...), a 3-parameter Gaussian-like form is typically used
\begin{equation}\label{eq2}
  \rho(r)=\rho_0 \frac{1+w(r/R)^2}{1+\exp(\frac{r^2-R^2}{a^2})}\,.
\end{equation}
%and the values for Sulfur are also given in \Tab{tab:awR}. 
For deformed nuclei, one uses
\begin{equation}\label{eq:deformed}
%  \rho(x,y,z)=\rho_0 \frac{1}{1+\exp\frac{\left(r-R(1+\beta_2 Y_{20} +\beta_4 Y_{40})\right)}{a}}\,,
\rho(x,y,z)=\rho_0 \left\{{1+\exp\frac{\left(r-R(1+\beta_2 Y_{20} +\beta_4 Y_{40})\right)}{a}} \right\}^{-1} \,,
\end{equation} 
where $Y_{20}=\sqrt{\frac{5}{16\pi}}(3{\rm cos}^2(\theta)-1)$, 
$Y_{40}=\frac{3}{16\sqrt{\pi}}(35{\rm cos}^4(\theta)-30 {\rm cos}^2(\theta)+3)$ with the
deformation parameters $\beta_2$ and $\beta_4$ taken from~\cite{DeJager:1974liz,DeJager:1987qc}. 
The values used for different nuclei~(Al, Si, Cu, Xe, Au, and U) are listed in \Tab{tab:awR2}.

\begin{table}[t!]
\centering
  \begin{tabular}{l|c|c|c|c|c}\hline
    Name            & $R$~[fm]        &  $a$~[fm]      & $w$ & $\beta_2$ & $\beta_4$    \\
    \hline
    ${}^{27}$Al     & 3.34\hide{0}  & 0.580\hide{0} & 0 & $-0.448$        & \hide{-}0.239 \\
    ${}^{28}$Si2    & 3.34\hide{0}  & 0.580\hide{0} & $-0.233$ & $-0.478$        & \hide{-}0.250 \\
    ${}^{63}$Cu2    & 4.2\hide{00}  & 0.596\hide{0} & 0 & \hide{-}0.162 & $-0.006$        \\
    ${}^{129}$Xe2\footnote{Same parameters as for ${}^{129}$Xe in Table~\ref{tab:awR} (App.) with $\beta_2$ and $\beta_4$ from \Refe{fischer:1974}.}
                    & 5.36\hide{0}  & 0.59\hide{00}  & 0 & 0.161         & $-0.003$        \\
    ${}^{129}$Xe2\footnote{Same parameters as for ${}^{129}$Xe in Table~\ref{tab:awR} (App.) with $\beta_2=0.18\pm0.02$ from interpolation between measured deformation parameters for the even-$A$ Xe isotopes~\cite{ALICE-PUBLIC-2018-003}, and $\beta_4$ set to zero.}
                    & 5.36\hide{0}  & 0.59\hide{00}  & 0 & 0.18          &  \hide{-}0    \\
    ${}^{197}$Au2   & 6.38\hide{0}  & 0.535\hide{0} & 0 & $-0.131$        & $-0.031$        \\
    ${}^{238}$U2    & 6.67\hide{0}  & 0.44\hide{00} & 0 & \hide{-}0.280 & \hide{-}0.093 \\
    ${}^{238}$U\footnote{Implementation as in \cite{Heinz:2004ir}.}
                    & 6.67\hide{0}  & 0.44\hide{00} & 0 & \hide{-}0.280 & \hide{-}0.093 \\
                    \hline
  \end{tabular}
  \caption{\label{tab:awR2}\protect Nuclear charge density parameters for deformed nuclei from \cite{DeJager:1974liz,DeJager:1987qc}. The suffix '2' for Si2, Cu2, Au2, and U2 refers to the use of \Eq{eq:deformed}. All configurations can be found in the {\tt TGlauNucleus::Lookup} function of the TGlauberMC code~\cite{glaucode}.}
\end{table}
            
\subsection*{Collision of nuclei}
After setting up the two nuclei, the impact parameter of the collision is chosen from ${\rm d}N/{\rm d}b \propto b$ up to some large maximum $b_{\rm max}$ with \mbox{$b_{\rm max}>R_{A}+R_{B}\simeq20\,$fm}. 
The centers of the nuclei are shifted to $(-b/2,0,0)$ and $(b/2,0,0)$, respectively. 
The reaction plane, defined by the impact parameter and the beam direction, is given by the $x$- and $z$-axes, while  the transverse plane is given by the $x$- and $y$-axes.
The longitudinal coordinate is irrelevant in the calculation as
%According to the eikonal formalism, 
the nucleons move along a straight trajectory along the beam axis.

The ``ball diameter'' defined as $D = \sqrt{\sigmaNN/\pi}$ defines the interaction strength of two nucleons:
Two nucleons from different nuclei are usually assumed to collide if their relative transverse distance is less than the ball diameter, \ie\ $d < D$. 
Variants to this approach use nucleon overlap profles that differ from the hard-sphere~(or ``black-disc'') approach, as discussed in \Sec{sec:IS}, see \Eq{eq:NN_coll_profile}.
If no nucleon--nucleon collision is registered for any pair of nucleons, then no nucleus--nucleus collision occurred. 
Counters for determination of the  total~(geometric) cross section are updated accordingly.

%\footnotetext[14]{Same parameters as for Xe with $\beta_2$ and $\beta_4$ from \Refe{fischer:1974}}
%\footnotetext[15]{Same parameters as for Xe with $\beta_2=0.18\pm0.02$ from interpolation between measured deformation parameters for the even-A Xe isotopes~\cite{ALICE-PUBLIC-2018-003}, and $\beta_4$ set to zero.}
%\footnotetext[15]{Implementation as done in \Refe{Heinz:2004ir}.}

\clearpage
\fi
%%%%%%%%%%%%%%%%%%%%%%%%%%%%%%%%%%%%%%%%%%%%%%%%%%%%%%%%%%%%%%%%%%%%%%%%%%%%%%%%%%%
% References
\ifdraft
\bibliographystyle{utphys}
\else
\bibliographystyle{ar-style5}
\fi
\bibliography{GlauberBiblio}

\providecommand{\href}[2]{#2}\begingroup\raggedright\begin{thebibliography}{100}

\bibitem{Busza:2018rrf}
W.~Busza, K.~Rajagopal, and W.~van~der Schee, ``{Heavy Ion Collisions: The Big
  Picture, and the Big Questions},''
  \href{http://dx.doi.org/10.1146/annurev-nucl-101917-020852}{{\em Ann. Rev.
  Nucl. Part. Sci.} {\bfseries 68} (2018) 339--376},
  \href{http://arxiv.org/abs/1802.04801}{{\ttfamily arXiv:1802.04801
  [hep-ph]}}.

\bibitem{Borsanyi:2013bia}
S.~Borsanyi, Z.~Fodor, C.~Hoelbling, S.~D. Katz, S.~Krieg, and K.~K. Szabo,
  ``{Full result for the QCD equation of state with 2+1 flavors},''
  \href{http://dx.doi.org/10.1016/j.physletb.2014.01.007}{{\em Phys. Lett. B}
  {\bfseries 730} (2014) 99--104},
  \href{http://arxiv.org/abs/1309.5258}{{\ttfamily arXiv:1309.5258 [hep-lat]}}.

\bibitem{Bazavov:2014pvz}
{\bfseries HotQCD} Collaboration, A.~Bazavov {\em et~al.}, ``{Equation of state
  in (2+1)-flavor QCD},''
  \href{http://dx.doi.org/10.1103/PhysRevD.90.094503}{{\em Phys. Rev. D}
  {\bfseries 90} (2014) 094503},
  \href{http://arxiv.org/abs/1407.6387}{{\ttfamily arXiv:1407.6387 [hep-lat]}}.

\bibitem{dEnterria:2006mtd}
D.~d'Enterria, ``{Quark-Gluon Matter},''
  \href{http://dx.doi.org/10.1088/0954-3899/34/7/S04}{{\em J. Phys. G}
  {\bfseries 34} (2007) S53--S82},
  \href{http://arxiv.org/abs/nucl-ex/0611012}{{\ttfamily
  arXiv:nucl-ex/0611012}}.

\bibitem{Martin:1958zz}
P.~Martin and R.~Glauber, ``{Relativistic theory of radiative orbital electron
  capture},'' \href{http://dx.doi.org/10.1103/PhysRev.109.1307}{{\em Phys.
  Rev.} {\bfseries 109} (1958) 1307--1325}.

\bibitem{Glauber:1970jm}
R.~J. Glauber and G.~Matthiae, ``{High-energy scattering of protons by
  nuclei},''
\href{http://dx.doi.org/10.1016/0550-3213(70)90511-0}{{\em Nucl. Phys. B}
  {\bfseries 21} (1970) 135--157}.
%%CITATION = NUPHA,B21,135;%%.

\bibitem{Czyz:1969jg}
W.~Czyz and L.~Maximon, ``{High-energy, small angle elastic scattering of
  strongly interacting composite particles},''
  \href{http://dx.doi.org/10.1016/0003-4916(69)90321-2}{{\em Annals Phys.}
  {\bfseries 52} (1969) 59--121}.

\bibitem{Bialas:1976ed}
A.~Bialas, M.~Bleszynski, and W.~Czyz, ``{Multiplicity distributions in
  nucleus-nucleus collisions at high energies},''
\href{http://dx.doi.org/10.1016/0550-3213(76)90329-1}{{\em Nucl. Phys. B}
  {\bfseries 111} (1976) 461--476}.
%%CITATION = NUPHA,B111,461;%%.

\bibitem{Bialas:1977pd}
A.~Bialas, M.~Bleszynski, and W.~Czyz, ``{Relation between the Glauber model
  and classical probability calculus},'' {\em Acta Phys. Polon. B} {\bfseries
  8} (1977) 389--392.

\bibitem{Wang:1991hta}
X.-N. Wang and M.~Gyulassy, ``{HIJING: A Monte Carlo model for multiple jet
  production in \pp, \pA\ and \AA\ collisions},''
\href{http://dx.doi.org/10.1103/PhysRevD.44.3501}{{\em Phys. Rev. D} {\bfseries
  44} (1991) 3501--3516}.
%%CITATION = PHRVA,D44,3501;%%.

\bibitem{Broniowski:2007nz}
W.~Broniowski, M.~Rybczynski, and P.~Bozek, ``{GLISSANDO: Glauber initial-state
  simulation and more...},''
  \href{http://dx.doi.org/10.1016/j.cpc.2008.07.016}{{\em Comput. Phys.
  Commun.} {\bfseries 180} (2009) 69--83},
\href{http://arxiv.org/abs/0710.5731}{{\ttfamily arXiv:0710.5731 [nucl-th]}}.
%%CITATION = ARXIV:0710.5731;%%.

\bibitem{Alver:2008aq}
B.~Alver, M.~Baker, C.~Loizides, and P.~Steinberg, ``{The PHOBOS Glauber Monte
  Carlo},'' {\em \arxiv{0805.4411}} (2008) ,
\href{http://arxiv.org/abs/0805.4411}{{\ttfamily arXiv:0805.4411 [nucl-ex]}}.
%%CITATION = ARXIV:0805.4411;%%.

\bibitem{Alvioli:2009ab}
M.~Alvioli, H.~J. Drescher, and M.~Strikman, ``{A Monte Carlo generator of
  nucleon configurations in complex nuclei including nucleon-nucleon
  correlations},'' \href{http://dx.doi.org/10.1016/j.physletb.2009.08.067}{{\em
  Phys. Lett. B} {\bfseries 680} (2009) 225--230},
\href{http://arxiv.org/abs/0905.2670}{{\ttfamily arXiv:0905.2670 [nucl-th]}}.
%%CITATION = ARXIV:0905.2670;%%.

\bibitem{Rybczynski:2013yba}
M.~Rybczynski, G.~Stefanek, W.~Broniowski, and P.~Bozek, ``{GLISSANDO 2:
  GLauber Initial-State Simulation AND mOre..., ver. 2},''
  \href{http://dx.doi.org/10.1016/j.cpc.2014.02.016}{{\em Comput. Phys.
  Commun.} {\bfseries 185} (2014) 1759--1772},
\href{http://arxiv.org/abs/1310.5475}{{\ttfamily arXiv:1310.5475 [nucl-th]}}.
%%CITATION = ARXIV:1310.5475;%%.

\bibitem{Loizides:2014vua}
C.~Loizides, J.~Nagle, and P.~Steinberg, ``{Improved version of the PHOBOS
  Glauber Monte Carlo},''
  \href{http://dx.doi.org/10.1016/j.softx.2015.05.001}{{\em SoftwareX}
  {\bfseries 1-2} (2015) 13--18},
\href{http://arxiv.org/abs/1408.2549}{{\ttfamily arXiv:1408.2549 [nucl-ex]}}.
%%CITATION = ARXIV:1408.2549;%%.

\bibitem{Loizides:2016djv}
C.~Loizides, ``{Glauber modeling of high-energy nuclear collisions at the
  subnucleon level},'' \href{http://dx.doi.org/10.1103/PhysRevC.94.024914}{{\em
  Phys. Rev. C} {\bfseries 94} (2016) 024914},
\href{http://arxiv.org/abs/1603.07375}{{\ttfamily arXiv:1603.07375 [nucl-ex]}}.
%%CITATION = ARXIV:1603.07375;%%.

\bibitem{Mitchell:2016jio}
J.~T. Mitchell, D.~V. Perepelitsa, M.~J. Tannenbaum, and P.~W. Stankus,
  ``{Tests of constituent-quark generation methods which maintain both the
  nucleon center of mass and the desired radial distribution in Monte Carlo
  Glauber models},'' \href{http://dx.doi.org/10.1103/PhysRevC.93.054910}{{\em
  Phys. Rev. C} {\bfseries 93} (2016) 054910},
\href{http://arxiv.org/abs/1603.08836}{{\ttfamily arXiv:1603.08836 [nucl-ex]}}.
%%CITATION = ARXIV:1603.08836;%%.

\bibitem{Loizides:2017ack}
C.~Loizides, J.~Kamin, and D.~d'Enterria, ``{Improved Monte Carlo Glauber
  predictions at present and future nuclear colliders},''
  \href{http://dx.doi.org/10.1103/PhysRevC.97.054910,
  10.1103/PhysRevC.99.019901}{{\em Phys. Rev. C} {\bfseries 97} (2018) 054910},
  \href{http://arxiv.org/abs/1710.07098}{{\ttfamily arXiv:1710.07098
  [nucl-ex]}}.
[Erratum: Phys. Rev. C 99 (2019) 019901] \url{http://tglaubermc.hepforge.org/}.
%%CITATION = ARXIV:1710.07098;%%.

\bibitem{Bozek:2019wyr}
P.~Bo\.zek, W.~Broniowski, M.~Rybczynski, and G.~Stefanek, ``{GLISSANDO 3:
  GLauber Initial-State Simulation AND mOre..., ver. 3},''
  \href{http://dx.doi.org/10.1016/j.cpc.2019.07.014}{{\em Comput. Phys.
  Commun.} {\bfseries 245} (2019) 106850},
  \href{http://arxiv.org/abs/1901.04484}{{\ttfamily arXiv:1901.04484
  [nucl-th]}}.

\bibitem{DeJager:1974liz}
C.~W. De~Jager, H.~De~Vries, and C.~De~Vries, ``{Nuclear charge and
  magnetization density distribution parameters from elastic electron
  scattering},''
\href{http://dx.doi.org/10.1016/S0092-640X(74)80002-1}{{\em Atom. Data Nucl.
  Data Tabl.} {\bfseries 14} (1974) 479--508}.
%%CITATION = ADNDA,14,479;%%.

\bibitem{DeJager:1987qc}
H.~De~Vries, C.~W. De~Jager, and C.~De~Vries, ``{Nuclear charge and
  magnetization density distribution parameters from elastic electron
  scattering},''
\href{http://dx.doi.org/10.1016/0092-640X(87)90013-1}{{\em Atom. Data Nucl.
  Data Tabl.} {\bfseries 36} (1987) 495--536}.
%%CITATION = ADNDA,36,495;%%.

\bibitem{Wang:2000bf}
X.-N. Wang and M.~Gyulassy, ``{Energy and centrality dependence of rapidity
  densities at RHIC},''
  \href{http://dx.doi.org/10.1103/PhysRevLett.86.3496}{{\em Phys. Rev. Lett.}
  {\bfseries 86} (2001) 3496--3499},
  \href{http://arxiv.org/abs/nucl-th/0008014}{{\ttfamily
  arXiv:nucl-th/0008014}}.

\bibitem{Ollitrault:1992bk}
J.-Y. Ollitrault, ``{Anisotropy as a signature of transverse collective
  flow},'' \href{http://dx.doi.org/10.1103/PhysRevD.46.229}{{\em Phys. Rev. D}
  {\bfseries 46} (1992) 229--245}.

\bibitem{Romatschke:2009im}
P.~Romatschke, ``{New developments in relativistic viscous hydrodynamics},''
  \href{http://dx.doi.org/10.1142/S0218301310014613}{{\em Int. J. Mod. Phys. E}
  {\bfseries 19} (2010) 1--53},
  \href{http://arxiv.org/abs/0902.3663}{{\ttfamily arXiv:0902.3663 [hep-ph]}}.

\bibitem{Teaney:2009qa}
D.~A. Teaney, {\em {Viscous Hydrodynamics and the Quark Gluon Plasma}},
  \href{http://dx.doi.org/10.1142/9789814293297_0004}{pp.~207--266}.
\newblock 2010.
\newblock \href{http://arxiv.org/abs/0905.2433}{{\ttfamily arXiv:0905.2433
  [nucl-th]}}.

\bibitem{Luzum:2009sb}
M.~Luzum and P.~Romatschke, ``{Viscous hydrodynamic predictions for nuclear
  collisions at the LHC},''
  \href{http://dx.doi.org/10.1103/PhysRevLett.103.262302}{{\em Phys. Rev.
  Lett.} {\bfseries 103} (2009) 262302},
  \href{http://arxiv.org/abs/0901.4588}{{\ttfamily arXiv:0901.4588 [nucl-th]}}.

\bibitem{Schenke:2010rr}
B.~Schenke, S.~Jeon, and C.~Gale, ``{Elliptic and triangular flow in
  event-by-event (3+1)D viscous hydrodynamics},''
  \href{http://dx.doi.org/10.1103/PhysRevLett.106.042301}{{\em Phys. Rev.
  Lett.} {\bfseries 106} (2011) 042301},
  \href{http://arxiv.org/abs/1009.3244}{{\ttfamily arXiv:1009.3244 [hep-ph]}}.

\bibitem{Heinz:2013th}
U.~Heinz and R.~Snellings, ``{Collective flow and viscosity in relativistic
  heavy-ion collisions},''
  \href{http://dx.doi.org/10.1146/annurev-nucl-102212-170540}{{\em Ann. Rev.
  Nucl. Part. Sci.} {\bfseries 63} (2013) 123--151},
  \href{http://arxiv.org/abs/1301.2826}{{\ttfamily arXiv:1301.2826 [nucl-th]}}.

\bibitem{Weller:2017tsr}
R.~D. Weller and P.~Romatschke, ``{One fluid to rule them all: viscous
  hydrodynamic description of event-by-event central \pp, \pPb\ and \PbPb\
  collisions at $\sqrt{s}=5.02$ TeV},''
  \href{http://dx.doi.org/10.1016/j.physletb.2017.09.077}{{\em Phys. Lett. B}
  {\bfseries 774} (2017) 351--356},
  \href{http://arxiv.org/abs/1701.07145}{{\ttfamily arXiv:1701.07145
  [nucl-th]}}.

\bibitem{Alver:2010gr}
B.~Alver and G.~Roland, ``{Collision geometry fluctuations and triangular flow
  in heavy-ion collisions},''
  \href{http://dx.doi.org/10.1103/PhysRevC.82.039903,
  10.1103/PhysRevC.81.054905}{{\em Phys. Rev. C} {\bfseries 81} (2010) 054905},
  \href{http://arxiv.org/abs/1003.0194}{{\ttfamily arXiv:1003.0194 [nucl-th]}}.
[Erratum: Phys. Rev.C82,039903(2010)].
%%CITATION = ARXIV:1003.0194;%%.

\bibitem{ALICE:2011ab}
{\bfseries ALICE} Collaboration, K.~Aamodt {\em et~al.}, ``{Higher harmonic
  anisotropic flow measurements of charged particles in Pb-Pb collisions at
  $\snn$=2.76 TeV},''
  \href{http://dx.doi.org/10.1103/PhysRevLett.107.032301}{{\em Phys. Rev.
  Lett.} {\bfseries 107} (2011) 032301},
  \href{http://arxiv.org/abs/1105.3865}{{\ttfamily arXiv:1105.3865 [nucl-ex]}}.

\bibitem{Adare:2011tg}
{\bfseries PHENIX} Collaboration, A.~Adare {\em et~al.}, ``{Measurements of
  higher-order flow harmonics in \AuAu\ collisions at $\snn=200$ GeV},''
  \href{http://dx.doi.org/10.1103/PhysRevLett.107.252301}{{\em Phys. Rev.
  Lett.} {\bfseries 107} (2011) 252301},
  \href{http://arxiv.org/abs/1105.3928}{{\ttfamily arXiv:1105.3928 [nucl-ex]}}.

\bibitem{ATLAS:2012at}
{\bfseries ATLAS} Collaboration, G.~Aad {\em et~al.}, ``{Measurement of the
  azimuthal anisotropy for charged particle production in $\snn=2.76$ TeV
  lead-lead collisions with the ATLAS detector},''
  \href{http://dx.doi.org/10.1103/PhysRevC.86.014907}{{\em Phys. Rev. C}
  {\bfseries 86} (2012) 014907},
  \href{http://arxiv.org/abs/1203.3087}{{\ttfamily arXiv:1203.3087 [hep-ex]}}.

\bibitem{Chatrchyan:2013nka}
{\bfseries CMS} Collaboration, S.~Chatrchyan {\em et~al.}, ``{Multiplicity and
  transverse momentum Dependence of two- and four-particle correlations in
  \pPb\ and \PbPb\ collisions},''
  \href{http://dx.doi.org/10.1016/j.physletb.2013.06.028}{{\em Phys. Lett. B}
  {\bfseries 724} (2013) 213--240},
  \href{http://arxiv.org/abs/1305.0609}{{\ttfamily arXiv:1305.0609 [nucl-ex]}}.

\bibitem{Dainese:2004te}
A.~Dainese, C.~Loizides, and G.~Paic, ``{Leading-particle suppression in high
  energy nucleus-nucleus collisions},''
  \href{http://dx.doi.org/10.1140/epjc/s2004-02077-x}{{\em Eur. Phys. J. C}
  {\bfseries 38} (2005) 461--474},
\href{http://arxiv.org/abs/hep-ph/0406201}{{\ttfamily arXiv:hep-ph/0406201
  [hep-ph]}}.
%%CITATION = HEP-PH/0406201;%%.

\bibitem{Lokhtin:2005px}
I.~Lokhtin and A.~Snigirev, ``{A model of jet quenching in ultrarelativistic
  heavy ion collisions and high-$\pt$ hadron spectra at RHIC},''
  \href{http://dx.doi.org/10.1140/epjc/s2005-02426-3}{{\em Eur. Phys. J. C}
  {\bfseries 45} (2006) 211--217},
  \href{http://arxiv.org/abs/hep-ph/0506189}{{\ttfamily arXiv:hep-ph/0506189}}.

\bibitem{Djordjevic:2018ita}
M.~Djordjevic, D.~Zigic, M.~Djordjevic, and J.~Auvinen, ``{How to test
  path-length dependence in energy loss mechanisms: analysis leading to a new
  observable},'' \href{http://dx.doi.org/10.1103/PhysRevC.99.061902}{{\em Phys.
  Rev. C} {\bfseries 99} (2019) 061902},
  \href{http://arxiv.org/abs/1805.04030}{{\ttfamily arXiv:1805.04030
  [nucl-th]}}.

\bibitem{dEnterria:2009xfs}
D.~d'Enterria, ``{Jet quenching},''
  \href{http://dx.doi.org/10.1007/978-3-642-01539-7_16}{{\em Landolt-Bornstein}
  {\bfseries 23} (2010) 471}, \href{http://arxiv.org/abs/0902.2011}{{\ttfamily
  arXiv:0902.2011 [nucl-ex]}}.

\bibitem{Baltz:2007kq}
A.~Baltz, ``{The Physics of ultraperipheral collisions at the LHC},''
  \href{http://dx.doi.org/10.1016/j.physrep.2007.12.001}{{\em Phys. Rept.}
  {\bfseries 458} (2008) 1--171},
  \href{http://arxiv.org/abs/0706.3356}{{\ttfamily arXiv:0706.3356 [nucl-ex]}}.

\bibitem{Miller:2007ri}
M.~L. Miller, K.~Reygers, S.~J. Sanders, and P.~Steinberg, ``{Glauber modeling
  in high energy nuclear collisions},''
  \href{http://dx.doi.org/10.1146/annurev.nucl.57.090506.123020}{{\em Ann. Rev.
  Nucl. Part. Sci.} {\bfseries 57} (2007) 205--243},
\href{http://arxiv.org/abs/nucl-ex/0701025}{{\ttfamily arXiv:nucl-ex/0701025
  [nucl-ex]}}.
%%CITATION = NUCL-EX/0701025;%%.

\bibitem{Alver:2010ck}
{\bfseries PHOBOS} Collaboration, B.~Alver {\em et~al.}, ``{Phobos results on
  charged particle multiplicity and pseudorapidity distributions in Au+Au,
  Cu+Cu, d+Au, and p+p collisions at ultra-relativistic energies},''
  \href{http://dx.doi.org/10.1103/PhysRevC.83.024913}{{\em Phys. Rev. C}
  {\bfseries 83} (2011) 024913},
  \href{http://arxiv.org/abs/1011.1940}{{\ttfamily arXiv:1011.1940 [nucl-ex]}}.

\bibitem{Benedikt:2018csr}
{\bfseries FCC} Collaboration, A.~Abada {\em et~al.}, ``{FCC-hh: The Hadron
  Collider}: {Future Circular Collider Conceptual Design Report Volume 3},''
  \href{http://dx.doi.org/10.1140/epjst/e2019-900087-0}{{\em Eur. Phys. J. ST}
  {\bfseries 228} (2019) 755--1107}.

\bibitem{Tanabashi:2018oca}
{\bfseries Particle Data Group} Collaboration, M.~Tanabashi {\em et~al.},
  ``{Review of Particle Physics},''
  \href{http://dx.doi.org/10.1103/PhysRevD.98.030001}{{\em Phys. Rev. D}
  {\bfseries 98} (2018) 030001}.

\bibitem{Alner:1986iy}
{\bfseries UA5} Collaboration, G.~J. Alner {\em et~al.}, ``{Antiproton-proton
  cross sections at 200 and 900 GeV c.m. energy},''
\href{http://dx.doi.org/10.1007/BF01552491}{{\em Z. Phys. C} {\bfseries 32}
  (1986) 153--161}.
%%CITATION = ZEPYA,C32,153;%%.

\bibitem{Amos:1990jh}
{\bfseries E710} Collaboration, N.~A. Amos {\em et~al.}, ``{A luminosity
  independent measurement of the $\bar{\rm p}$p total cross section at
  $\sqrt{s}=1.8$ TeV},''
\href{http://dx.doi.org/10.1016/0370-2693(90)90973-A}{{\em Phys. Lett. B}
  {\bfseries 243} (1990) 158--164}.
%%CITATION = PHLTA,B243,158;%%.

\bibitem{Amos:1991bp}
{\bfseries E710} Collaboration, N.~A. Amos {\em et~al.}, ``{Measurement of
  $\rho$, the ratio of the real to imaginary part of the $\bar{\rm p}$p forward
  elastic scattering amplitude, at $\sqrt{s}=1.8$ TeV},''
\href{http://dx.doi.org/10.1103/PhysRevLett.68.2433}{{\em Phys. Rev. Lett.}
  {\bfseries 68} (1992) 2433--2436}.
%%CITATION = PRLTA,68,2433;%%.

\bibitem{Abe:1993xy}
{\bfseries CDF} Collaboration, F.~Abe {\em et~al.}, ``{Measurement of the
  $\bar{p}p$ total cross-section at $\sqrt{s}=546$ GeV and 1800 GeV},''
\href{http://dx.doi.org/10.1103/PhysRevD.50.5550}{{\em Phys. Rev. D} {\bfseries
  50} (1994) 5550--5561}.
%%CITATION = PHRVA,D50,5550;%%.

\bibitem{Abe:1993xx}
{\bfseries CDF} Collaboration, F.~Abe {\em et~al.}, ``{Measurement of small
  angle $\bar{p}p$ elastic scattering at $\sqrt{s}=546$ GeV and 1800 GeV},''
\href{http://dx.doi.org/10.1103/PhysRevD.50.5518}{{\em Phys. Rev. D} {\bfseries
  50} (1994) 5518--5534}.
%%CITATION = PHRVA,D50,5518;%%.

\bibitem{Adam:2020ozo}
{\bfseries STAR} Collaboration, J.~Adam {\em et~al.}, ``{Results on total and
  elastic cross sections in proton-proton collisions at $\sqrt{s}=200$ GeV},''
  \href{http://dx.doi.org/10.1016/j.physletb.2020.135663}{{\em Phys. Lett. B}
  {\bfseries 808} (2020) 135663},
  \href{http://arxiv.org/abs/2003.12136}{{\ttfamily arXiv:2003.12136
  [hep-ex]}}.

\bibitem{Abelev:2012sea}
{\bfseries ALICE} Collaboration, B.~Abelev {\em et~al.}, ``{Measurement of
  inelastic, single- and double-diffraction cross sections in proton-proton
  collisions at the LHC with ALICE},''
  \href{http://dx.doi.org/10.1140/epjc/s10052-013-2456-0}{{\em Eur. Phys. J. C}
  {\bfseries 73} (2013) 2456},
\href{http://arxiv.org/abs/1208.4968}{{\ttfamily arXiv:1208.4968 [hep-ex]}}.
%%CITATION = ARXIV:1208.4968;%%.

\bibitem{Aad:2011eu}
{\bfseries ATLAS} Collaboration, G.~Aad {\em et~al.}, ``{Measurement of the
  inelastic proton-proton cross section at $\sqrt{s}=7$ TeV with the ATLAS
  Detector},'' \href{http://dx.doi.org/10.1038/ncomms1472}{{\em Nature Commun.}
  {\bfseries 2} (2011) 463},
\href{http://arxiv.org/abs/1104.0326}{{\ttfamily arXiv:1104.0326 [hep-ex]}}.
%%CITATION = ARXIV:1104.0326;%%.

\bibitem{Aad:2014dca}
{\bfseries ATLAS} Collaboration, G.~Aad {\em et~al.}, ``{Measurement of the
  total cross section from elastic scattering in \pp\ collisions at
  $\sqrt{s}=7$ TeV with the ATLAS detector},''
  \href{http://dx.doi.org/10.1016/j.nuclphysb.2014.10.019}{{\em Nucl. Phys. B}
  {\bfseries 889} (2014) 486--548},
\href{http://arxiv.org/abs/1408.5778}{{\ttfamily arXiv:1408.5778 [hep-ex]}}.
%%CITATION = ARXIV:1408.5778;%%.

\bibitem{Aaboud:2016ijx}
{\bfseries ATLAS} Collaboration, M.~Aaboud {\em et~al.}, ``{Measurement of the
  total cross section from elastic scattering in \pp\ collisions at
  $\sqrt{s}=8$ TeV with the ATLAS detector},''
  \href{http://dx.doi.org/10.1016/j.physletb.2016.08.020}{{\em Phys. Lett. B}
  {\bfseries 761} (2016) 158--178},
\href{http://arxiv.org/abs/1607.06605}{{\ttfamily arXiv:1607.06605 [hep-ex]}}.
%%CITATION = ARXIV:1607.06605;%%.

\bibitem{Aaboud:2016mmw}
{\bfseries ATLAS} Collaboration, M.~Aaboud {\em et~al.}, ``{Measurement of the
  inelastic proton-proton cross section at $\sqrt{s}=13$ TeV with the ATLAS
  detector at the LHC},''
  \href{http://dx.doi.org/10.1103/PhysRevLett.117.182002}{{\em Phys. Rev.
  Lett.} {\bfseries 117} (2016) 182002},
\href{http://arxiv.org/abs/1606.02625}{{\ttfamily arXiv:1606.02625 [hep-ex]}}.
%%CITATION = ARXIV:1606.02625;%%.

\bibitem{Chatrchyan:2012nj}
{\bfseries CMS} Collaboration, S.~Chatrchyan {\em et~al.}, ``{Measurement of
  the inelastic proton-proton cross section at $\sqrt{s}=7$ TeV},''
  \href{http://dx.doi.org/10.1016/j.physletb.2013.03.024}{{\em Phys. Lett. B}
  {\bfseries 722} (2013) 5--27},
\href{http://arxiv.org/abs/1210.6718}{{\ttfamily arXiv:1210.6718 [hep-ex]}}.
%%CITATION = ARXIV:1210.6718;%%.

\bibitem{Sirunyan:2018nqx}
{\bfseries CMS} Collaboration, A.~M. Sirunyan {\em et~al.}, ``{Measurement of
  the inelastic proton-proton cross section at $ \sqrt{s}=13$ TeV},''
  \href{http://dx.doi.org/10.1007/JHEP07(2018)161}{{\em JHEP} {\bfseries 07}
  (2018) 161}, \href{http://arxiv.org/abs/1802.02613}{{\ttfamily
  arXiv:1802.02613 [hep-ex]}}.

\bibitem{Aaij:2014vfa}
{\bfseries LHCb} Collaboration, R.~Aaij {\em et~al.}, ``{Measurement of the
  inelastic pp cross-section at a centre-of-mass energy of $\sqrt{s}=7$ TeV},''
  \href{http://dx.doi.org/10.1007/JHEP02(2015)129}{{\em JHEP} {\bfseries 02}
  (2015) 129},
\href{http://arxiv.org/abs/1412.2500}{{\ttfamily arXiv:1412.2500 [hep-ex]}}.
%%CITATION = ARXIV:1412.2500;%%.

\bibitem{Aaij:2018okq}
{\bfseries LHCb} Collaboration, R.~Aaij {\em et~al.}, ``{Measurement of the
  inelastic \pp\ cross-section at a centre-of-mass energy of 13 TeV},''
  \href{http://dx.doi.org/10.1007/JHEP06(2018)100}{{\em JHEP} {\bfseries 06}
  (2018) 100}, \href{http://arxiv.org/abs/1803.10974}{{\ttfamily
  arXiv:1803.10974 [hep-ex]}}.

\bibitem{Antchev:2011vs}
{\bfseries TOTEM} Collaboration, G.~Antchev {\em et~al.}, ``{First measurement
  of the total proton-proton cross section at the LHC energy of $\sqrt{s}=7$
  TeV},'' \href{http://dx.doi.org/10.1209/0295-5075/96/21002}{{\em Europhys.
  Lett.} {\bfseries 96} (2011) 21002},
\href{http://arxiv.org/abs/1110.1395}{{\ttfamily arXiv:1110.1395 [hep-ex]}}.
%%CITATION = ARXIV:1110.1395;%%.

\bibitem{Antchev:2013iaa}
{\bfseries TOTEM} Collaboration, G.~Antchev {\em et~al.},
  ``{Luminosity-independent measurements of total, elastic and inelastic
  cross-sections at $\sqrt{s}=7$ TeV},''
\href{http://dx.doi.org/10.1209/0295-5075/101/21004}{{\em Europhys. Lett.}
  {\bfseries 101} (2013) 21004}.
%%CITATION = EULEE,101,21004;%%.

\bibitem{Antchev:2013paa}
{\bfseries TOTEM} Collaboration, G.~Antchev {\em et~al.},
  ``{Luminosity-independent measurement of the proton-proton total cross
  section at $\sqrt{s}=8$ TeV},''
\href{http://dx.doi.org/10.1103/PhysRevLett.111.012001}{{\em Phys. Rev. Lett.}
  {\bfseries 111} (2013) 012001}.
%%CITATION = PRLTA,111,012001;%%.

\bibitem{Antchev:2017dia}
{\bfseries TOTEM} Collaboration, G.~Antchev {\em et~al.}, ``{First measurement
  of elastic, inelastic and total cross-section at $\sqrt{s}=13$ TeV by TOTEM
  and overview of cross-section data at LHC energies},''
  \href{http://dx.doi.org/10.1140/epjc/s10052-019-6567-0}{{\em Eur. Phys. J. C}
  {\bfseries 79} (2019) 103}, \href{http://arxiv.org/abs/1712.06153}{{\ttfamily
  arXiv:1712.06153 [hep-ex]}}.

\bibitem{Cafagna:2020izf}
{\bfseries TOTEM} Collaboration, F.~Cafagna, ``{Latest results for
  proton-proton cross section measurements with the TOTEM experiment at LHC},''
  \href{http://dx.doi.org/10.22323/1.358.0207}{{\em PoS} {\bfseries ICRC2019}
  (2020) 207}.

\bibitem{Abreu:2012wt}
{\bfseries Pierre Auger} Collaboration, P.~Abreu {\em et~al.}, ``{Measurement
  of the proton-air cross-section at $\sqrt{s}=57$ TeV with the Pierre Auger
  Observatory},'' \href{http://dx.doi.org/10.1103/PhysRevLett.109.062002}{{\em
  Phys. Rev. Lett.} {\bfseries 109} (2012) 062002},
\href{http://arxiv.org/abs/1208.1520}{{\ttfamily arXiv:1208.1520 [hep-ex]}}.
%%CITATION = ARXIV:1208.1520;%%.

\bibitem{Cudell:2002xe}
{\bfseries COMPETE} Collaboration, J.~R. Cudell {\em et~al.}, ``{Benchmarks for
  the forward observables at RHIC, the Tevatron Run II and the LHC},''
  \href{http://dx.doi.org/10.1103/PhysRevLett.89.201801}{{\em Phys. Rev. Lett.}
  {\bfseries 89} (2002) 201801},
\href{http://arxiv.org/abs/hep-ph/0206172}{{\ttfamily arXiv:hep-ph/0206172
  [hep-ph]}}.
%%CITATION = HEP-PH/0206172;%%.

\bibitem{dEnterria:2016oxo}
D.~d'Enterria and T.~Pierog, ``{Global properties of proton-proton collisions
  at $\sqrt{s}$ = 100 TeV},''
  \href{http://dx.doi.org/10.1007/JHEP08(2016)170}{{\em JHEP} {\bfseries 08}
  (2016) 170},
\href{http://arxiv.org/abs/1604.08536}{{\ttfamily arXiv:1604.08536 [hep-ph]}}.
%%CITATION = ARXIV:1604.08536;%%.

\bibitem{ALICE:2012aa}
{\bfseries ALICE} Collaboration, B.~Abelev {\em et~al.}, ``{Measurement of the
  cross section for electromagnetic dissociation with neutron emission in
  \PbPb\ collisions at $\snn=2.76$ TeV},''
  \href{http://dx.doi.org/10.1103/PhysRevLett.109.252302}{{\em Phys. Rev.
  Lett.} {\bfseries 109} (2012) 252302},
\href{http://arxiv.org/abs/1203.2436}{{\ttfamily arXiv:1203.2436 [nucl-ex]}}.
%%CITATION = ARXIV:1203.2436;%%.

\bibitem{Khachatryan:2015zaa}
{\bfseries CMS} Collaboration, V.~Khachatryan {\em et~al.}, ``{Measurement of
  the inelastic cross section in proton–lead collisions at $\snn=$5.02
  TeV},'' \href{http://dx.doi.org/10.1016/j.physletb.2016.06.027}{{\em Phys.
  Lett. B} {\bfseries 759} (2016) 641--662},
\href{http://arxiv.org/abs/1509.03893}{{\ttfamily arXiv:1509.03893 [hep-ex]}}.
%%CITATION = ARXIV:1509.03893;%%.

\bibitem{Abelev:2014epa}
{\bfseries ALICE} Collaboration, B.~B. Abelev {\em et~al.}, ``{Measurement of
  visible cross sections in proton-lead collisions at $\snn$ = 5.02 TeV in van
  der Meer scans with the ALICE detector},''
  \href{http://dx.doi.org/10.1088/1748-0221/9/11/P11003}{{\em JINST} {\bfseries
  9} (2014) P11003},
\href{http://arxiv.org/abs/1405.1849}{{\ttfamily arXiv:1405.1849 [nucl-ex]}}.
%%CITATION = ARXIV:1405.1849;%%.

\bibitem{Sjostrand:2014zea}
T.~Sj\"ostrand, S.~Ask, J.~R. Christiansen, R.~Corke, N.~Desai, P.~Ilten,
  S.~Mrenna, S.~Prestel, C.~O. Rasmussen, and P.~Z. Skands, ``{An introduction
  to PYTHIA 8.2},'' \href{http://dx.doi.org/10.1016/j.cpc.2015.01.024}{{\em
  Comput. Phys. Commun.} {\bfseries 191} (2015) 159--177},
  \href{http://arxiv.org/abs/1410.3012}{{\ttfamily arXiv:1410.3012 [hep-ph]}}.

\bibitem{Bahr:2008pv}
M.~Bahr {\em et~al.}, ``{Herwig++ Physics and Manual},''
  \href{http://dx.doi.org/10.1140/epjc/s10052-008-0798-9}{{\em Eur. Phys. J. C}
  {\bfseries 58} (2008) 639--707},
  \href{http://arxiv.org/abs/0803.0883}{{\ttfamily arXiv:0803.0883 [hep-ph]}}.

\bibitem{Field:2011iq}
R.~Field, ``{Min-Bias and the underlying event at the LHC},''
  \href{http://dx.doi.org/10.5506/APhysPolB.42.2631}{{\em Acta Phys. Polon. B}
  {\bfseries 42} (2011) 2631--2656},
  \href{http://arxiv.org/abs/1110.5530}{{\ttfamily arXiv:1110.5530 [hep-ph]}}.

\bibitem{dEnterria:2017yhd}
D.~d'Enterria and A.~Snigirev, ``{Double, triple, and $n$-parton scatterings in
  high-energy proton and nuclear collisions},''
  \href{http://dx.doi.org/10.1142/9789813227767_0009}{{\em Adv. Ser. Direct.
  High Energy Phys.} {\bfseries 29} (2018) 159--187},
  \href{http://arxiv.org/abs/1708.07519}{{\ttfamily arXiv:1708.07519
  [hep-ph]}}.

\bibitem{Khachatryan:2010gv}
{\bfseries CMS} Collaboration, V.~Khachatryan {\em et~al.}, ``{Observation of
  long-range near-side angular correlations in proton-proton collisions at the
  LHC},'' \href{http://dx.doi.org/10.1007/JHEP09(2010)091}{{\em JHEP}
  {\bfseries 09} (2010) 091}, \href{http://arxiv.org/abs/1009.4122}{{\ttfamily
  arXiv:1009.4122 [hep-ex]}}.

\bibitem{Aad:2015gqa}
{\bfseries ATLAS} Collaboration, G.~Aad {\em et~al.}, ``{Observation of
  long-range elliptic azimuthal anisotropies in $\sqrt{s}=$13 and 2.76 TeV \pp\
  collisions with the ATLAS Detector},''
  \href{http://dx.doi.org/10.1103/PhysRevLett.116.172301}{{\em Phys. Rev.
  Lett.} {\bfseries 116} (2016) 172301},
  \href{http://arxiv.org/abs/1509.04776}{{\ttfamily arXiv:1509.04776
  [hep-ex]}}.

\bibitem{Khachatryan:2016txc}
{\bfseries CMS} Collaboration, V.~Khachatryan {\em et~al.}, ``{Evidence for
  collectivity in \pp\ collisions at the LHC},''
  \href{http://dx.doi.org/10.1016/j.physletb.2016.12.009}{{\em Phys. Lett. B}
  {\bfseries 765} (2017) 193--220},
  \href{http://arxiv.org/abs/1606.06198}{{\ttfamily arXiv:1606.06198
  [nucl-ex]}}.

\bibitem{dEnterria:2010xip}
D.~d'Enterria {\em et~al.}, ``{Estimates of hadron azimuthal anisotropy from
  multiparton interactions in proton-proton collisions at $\sqrt{s}=14$ TeV},''
  \href{http://dx.doi.org/10.1140/epjc/s10052-009-1232-7}{{\em Eur. Phys. J. C}
  {\bfseries 66} (2010) 173--185},
\href{http://arxiv.org/abs/0910.3029}{{\ttfamily arXiv:0910.3029 [hep-ph]}}.
%%CITATION = ARXIV:0910.3029;%%.

\bibitem{Hofstadter:1956qs}
R.~Hofstadter, ``{Electron scattering and nuclear structure},''
\href{http://dx.doi.org/10.1103/RevModPhys.28.214}{{\em Rev. Mod. Phys.}
  {\bfseries 28} (1956) 214--254}.
%%CITATION = RMPHA,28,214;%%.

\bibitem{Corke:2011yy}
R.~Corke and T.~Sj{\"o}strand, ``{Multiparton Interactions with an
  $x$-dependent proton size},''
  \href{http://dx.doi.org/10.1007/JHEP05(2011)009}{{\em JHEP} {\bfseries 05}
  (2011) 009},
\href{http://arxiv.org/abs/1101.5953}{{\ttfamily arXiv:1101.5953 [hep-ph]}}.
%%CITATION = ARXIV:1101.5953;%%.

\bibitem{Skands:2014pea}
P.~Skands, S.~Carrazza, and J.~Rojo, ``{Tuning PYTHIA 8.1: the Monash 2013
  tune},'' \href{http://dx.doi.org/10.1140/epjc/s10052-014-3024-y}{{\em Eur.
  Phys. J. C} {\bfseries 74} (2014) 3024},
  \href{http://arxiv.org/abs/1404.5630}{{\ttfamily arXiv:1404.5630 [hep-ph]}}.

\bibitem{Acharya:2019mzb}
{\bfseries ALICE} Collaboration, S.~Acharya {\em et~al.}, ``{Charged-particle
  production as a function of multiplicity and transverse spherocity in \pp\
  collisions at $\sqrt{s}=5.02$ and 13 TeV},''
  \href{http://dx.doi.org/10.1140/epjc/s10052-019-7350-y}{{\em Eur. Phys. J. C}
  {\bfseries 79} (2019) 857}, \href{http://arxiv.org/abs/1905.07208}{{\ttfamily
  arXiv:1905.07208 [nucl-ex]}}.

\bibitem{Sjostrand:2017cdm}
T.~Sj{\"o}strand, ``{The development of MPI modeling in PYTHIA},''
  \href{http://dx.doi.org/10.1142/9789813227767_0010}{{\em Adv. Ser. Direct.
  High Energy Phys.} {\bfseries 29} (2018) 191--225},
  \href{http://arxiv.org/abs/1706.02166}{{\ttfamily arXiv:1706.02166
  [hep-ph]}}.

\bibitem{Morsch:2017brb}
C.~Loizides and A.~Morsch, ``{Absence of jet quenching in peripheral
  nucleus-–nucleus collisions},''
  \href{http://dx.doi.org/10.1016/j.physletb.2017.09.002}{{\em Phys. Lett. B}
  {\bfseries 773} (2017) 408--411},
\href{http://arxiv.org/abs/1705.08856}{{\ttfamily arXiv:1705.08856 [nucl-ex]}}.
%%CITATION = ARXIV:1705.08856;%%.

\bibitem{Alvioli:2011sk}
M.~Alvioli, H.~Holopainen, K.~J. Eskola, and M.~Strikman, ``{Initial state
  anisotropies and their uncertainties in ultrarelativistic heavy-ion
  collisions from the Monte Carlo Glauber model},''
  \href{http://dx.doi.org/10.1103/PhysRevC.85.034902}{{\em Phys. Rev. C}
  {\bfseries 85} (2012) 034902},
\href{http://arxiv.org/abs/1112.5306}{{\ttfamily arXiv:1112.5306 [hep-ph]}}.
%%CITATION = ARXIV:1112.5306;%%.

\bibitem{Rybczynski:2013mla}
M.~Rybczyński and Z.~Wlodarczyk, ``{The nucleon-nucleon collision profile and
  cross section fluctuations},''
  \href{http://dx.doi.org/10.1088/0954-3899/41/1/015106}{{\em J. Phys. G}
  {\bfseries 41} (2013) 015106},
\href{http://arxiv.org/abs/1307.0636}{{\ttfamily arXiv:1307.0636 [nucl-th]}}.
%%CITATION = ARXIV:1307.0636;%%.

\bibitem{GrosseOetringhaus:2009kz}
J.~F. Grosse-Oetringhaus and K.~Reygers, ``{Charged-Particle Multiplicity in
  Proton-Proton Collisions},''
  \href{http://dx.doi.org/10.1088/0954-3899/37/8/083001}{{\em J. Phys. G}
  {\bfseries 37} (2010) 083001},
  \href{http://arxiv.org/abs/0912.0023}{{\ttfamily arXiv:0912.0023 [hep-ex]}}.

\bibitem{Welsh:2016siu}
K.~Welsh, J.~Singer, and U.~W. Heinz, ``{Initial state fluctuations in
  collisions between light and heavy ions},''
  \href{http://dx.doi.org/10.1103/PhysRevC.94.024919}{{\em Phys. Rev. C}
  {\bfseries 94} (2016) 024919},
  \href{http://arxiv.org/abs/1605.09418}{{\ttfamily arXiv:1605.09418
  [nucl-th]}}.

\bibitem{Moreland:2018gsh}
J.~S. Moreland, J.~E. Bernhard, and S.~A. Bass, ``{Bayesian calibration of a
  hybrid nuclear collision model using \pPb\ and \PbPb\ data at energies
  available at the CERN Large Hadron Collider},''
  \href{http://dx.doi.org/10.1103/PhysRevC.101.024911}{{\em Phys. Rev. C}
  {\bfseries 101} (2020) 024911},
  \href{http://arxiv.org/abs/1808.02106}{{\ttfamily arXiv:1808.02106
  [nucl-th]}}.

\bibitem{Teaney:2010vd}
D.~Teaney and L.~Yan, ``{Triangularity and dipole asymmetry in heavy ion
  collisions},'' \href{http://dx.doi.org/10.1103/PhysRevC.83.064904}{{\em Phys.
  Rev. C} {\bfseries 83} (2011) 064904},
\href{http://arxiv.org/abs/1010.1876}{{\ttfamily arXiv:1010.1876 [nucl-th]}}.
%%CITATION = ARXIV:1010.1876;%%.

\bibitem{Alver:2006wh}
{\bfseries PHOBOS} Collaboration, B.~Alver {\em et~al.}, ``{System size,
  energy, pseudorapidity, and centrality dependence of elliptic flow},''
  \href{http://dx.doi.org/10.1103/PhysRevLett.98.242302}{{\em Phys. Rev. Lett.}
  {\bfseries 98} (2007) 242302},
\href{http://arxiv.org/abs/nucl-ex/0610037}{{\ttfamily arXiv:nucl-ex/0610037
  [nucl-ex]}}.
%%CITATION = NUCL-EX/0610037;%%.

\bibitem{Blaizot:2014wba}
J.-P. Blaizot, W.~Broniowski, and J.-Y. Ollitrault, ``{Correlations in the
  Monte Carlo Glauber model},''
  \href{http://dx.doi.org/10.1103/PhysRevC.90.034906}{{\em Phys. Rev. C}
  {\bfseries 90} (2014) 034906},
  \href{http://arxiv.org/abs/1405.3274}{{\ttfamily arXiv:1405.3274 [nucl-th]}}.

\bibitem{Heiselberg:1991is}
H.~Heiselberg, G.~Baym, B.~Blaettel, L.~Frankfurt, and M.~Strikman, ``{Color
  transparency, color opacity, and fluctuations in nuclear collisions},''
  \href{http://dx.doi.org/10.1103/PhysRevLett.67.2946}{{\em Phys. Rev. Lett.}
  {\bfseries 67} (1991) 2946--2949}.

\bibitem{Alvioli:2013vk}
M.~Alvioli and M.~Strikman, ``{Color fluctuation effects in proton-nucleus
  collisions},'' \href{http://dx.doi.org/10.1016/j.physletb.2013.04.042}{{\em
  Phys. Lett. B} {\bfseries 722} (2013) 347--354},
\href{http://arxiv.org/abs/1301.0728}{{\ttfamily arXiv:1301.0728 [hep-ph]}}.
%%CITATION = ARXIV:1301.0728;%%.

\bibitem{Alvioli:2017wou}
M.~Alvioli, L.~Frankfurt, D.~Perepelitsa, and M.~Strikman, ``{Global analysis
  of color fluctuation effects in proton\textendash{} and
  deuteron\textendash{}nucleus collisions at RHIC and the LHC},''
  \href{http://dx.doi.org/10.1103/PhysRevD.98.071502}{{\em Phys. Rev. D}
  {\bfseries 98} (2018) 071502},
  \href{http://arxiv.org/abs/1709.04993}{{\ttfamily arXiv:1709.04993
  [hep-ph]}}.

\bibitem{Gribov:1968jf}
V.~Gribov, ``{Glauber corrections and the interaction between high-energy
  hadrons and nuclei},'' {\em Sov. Phys. JETP} {\bfseries 29} (1969) 483--487.

\bibitem{Bierlich:2016smv}
C.~Bierlich, G.~Gustafson, and L.~L\"onnblad, ``{Diffractive and
  non-diffractive wounded nucleons and final states in \pA\ collisions},''
  \href{http://dx.doi.org/10.1007/JHEP10(2016)139}{{\em JHEP} {\bfseries 10}
  (2016) 139}, \href{http://arxiv.org/abs/1607.04434}{{\ttfamily
  arXiv:1607.04434 [hep-ph]}}.

\bibitem{Adare:2015gla}
{\bfseries PHENIX} Collaboration, A.~Adare {\em et~al.},
  ``{Centrality-dependent modification of jet-production rates in deuteron-gold
  collisions at $\snn$=200 GeV},''
  \href{http://dx.doi.org/10.1103/PhysRevLett.116.122301}{{\em Phys. Rev.
  Lett.} {\bfseries 116} (2016) 122301},
  \href{http://arxiv.org/abs/1509.04657}{{\ttfamily arXiv:1509.04657
  [nucl-ex]}}.

\bibitem{Aad:2015zza}
{\bfseries ATLAS} Collaboration, G.~Aad {\em et~al.}, ``{Measurement of the
  centrality dependence of the charged-particle pseudorapidity distribution in
  \pPb\ collisions at $\snn=5.02$ TeV with the ATLAS detector},''
  \href{http://dx.doi.org/10.1140/epjc/s10052-016-4002-3}{{\em Eur. Phys. J. C}
  {\bfseries 76} (2016) 199}, \href{http://arxiv.org/abs/1508.00848}{{\ttfamily
  arXiv:1508.00848 [hep-ex]}}.

\bibitem{McGlinchey:2016ssj}
D.~McGlinchey, J.~Nagle, and D.~Perepelitsa, ``{Consequences of high-$x$ proton
  size fluctuations in small collision systems at $\snn=$200 GeV},''
  \href{http://dx.doi.org/10.1103/PhysRevC.94.024915}{{\em Phys. Rev. C}
  {\bfseries 94} (2016) 024915},
  \href{http://arxiv.org/abs/1603.06607}{{\ttfamily arXiv:1603.06607
  [nucl-th]}}.

\bibitem{CiofidegliAtti:2011fh}
C.~Ciofi~degli Atti, B.~Kopeliovich, C.~Mezzetti, I.~Potashnikova, and
  I.~Schmidt, ``{Number of collisions in the Glauber Model and beyond},''
  \href{http://dx.doi.org/10.1103/PhysRevC.84.025205}{{\em Phys. Rev. C}
  {\bfseries 84} (2011) 025205},
  \href{http://arxiv.org/abs/1105.1080}{{\ttfamily arXiv:1105.1080 [nucl-th]}}.

\bibitem{Flensburg:2011kk}
C.~Flensburg, G.~Gustafson, and L.~L{\"o}nnblad, ``{Inclusive and exclusive
  observables from dipoles in high energy collisions},''
  \href{http://dx.doi.org/10.1007/JHEP08(2011)103}{{\em JHEP} {\bfseries 08}
  (2011) 103}, \href{http://arxiv.org/abs/1103.4321}{{\ttfamily arXiv:1103.4321
  [hep-ph]}}.

\bibitem{Salam:1998tj}
G.~Salam, ``{A resummation of large subleading corrections at small $x$},''
  \href{http://dx.doi.org/10.1088/1126-6708/1998/07/019}{{\em JHEP} {\bfseries
  07} (1998) 019}, \href{http://arxiv.org/abs/hep-ph/9806482}{{\ttfamily
  arXiv:hep-ph/9806482}}.

\bibitem{barrett1977nuclear}
R.~C. Barrett and D.~F. Jackson, {\em Nuclear sizes and structure}.
\newblock Oxford University Press, 1977.

\bibitem{Klos:2007is}
B.~Klos {\em et~al.}, ``{Neutron density distributions from antiprotonic
  $^{208}$Pb and $^{209}$Bi atoms},''
  \href{http://dx.doi.org/10.1103/PhysRevC.76.014311}{{\em Phys. Rev. C}
  {\bfseries 76} (2007) 014311},
\href{http://arxiv.org/abs/nucl-ex/0702016}{{\ttfamily arXiv:nucl-ex/0702016
  [NUCL-EX]}}.
%%CITATION = NUCL-EX/0702016;%%.

\bibitem{Tarbert:2013jze}
C.~M. Tarbert {\em et~al.}, ``{Neutron skin of $^{208}$Pb from coherent pion
  photoproduction},''
  \href{http://dx.doi.org/10.1103/PhysRevLett.112.242502}{{\em Phys. Rev.
  Lett.} {\bfseries 112} (2014) 242502},
\href{http://arxiv.org/abs/1311.0168}{{\ttfamily arXiv:1311.0168 [nucl-ex]}}.
%%CITATION = ARXIV:1311.0168;%%.

\bibitem{Fricke:1995zz}
G.~Fricke, C.~Bernhardt, K.~Heilig, L.~A. Schaller, L.~Schellenberg, E.~B.
  Shera, and C.~W. de~Jager, ``{Nuclear Ground State Charge Radii from
  Electromagnetic Interactions},''
\href{http://dx.doi.org/10.1006/adnd.1995.1007}{{\em Atom. Data Nucl. Data
  Tabl.} {\bfseries 60} (1995) 177}.
%%CITATION = ADNDA,60,177;%%.

\bibitem{Horowitz:1999fk}
C.~J. Horowitz, S.~J. Pollock, P.~A. Souder, and R.~Michaels, ``{Parity
  violating measurements of neutron densities},''
  \href{http://dx.doi.org/10.1103/PhysRevC.63.025501}{{\em Phys. Rev. C}
  {\bfseries 63} (2001) 025501},
\href{http://arxiv.org/abs/nucl-th/9912038}{{\ttfamily arXiv:nucl-th/9912038
  [nucl-th]}}.
%%CITATION = NUCL-TH/9912038;%%.

\bibitem{Paukkunen:2015bwa}
H.~Paukkunen, ``{Neutron skin and centrality classification in high-energy
  heavy-ion collisions at the LHC},''
  \href{http://dx.doi.org/10.1016/j.physletb.2015.04.037}{{\em Phys. Lett. B}
  {\bfseries 745} (2015) 73--78},
\href{http://arxiv.org/abs/1503.02448}{{\ttfamily arXiv:1503.02448 [hep-ph]}}.
%%CITATION = ARXIV:1503.02448;%%.

\bibitem{De:2016ggl}
S.~De, ``{The effect of neutron skin on inclusive prompt photon production in
  \PbPb\ collisions at Large Hadron Collider energies},''
  \href{http://dx.doi.org/10.1088/1361-6471/aa5689}{{\em J. Phys. G} {\bfseries
  44} (2017) 045104},
\href{http://arxiv.org/abs/1609.09608}{{\ttfamily arXiv:1609.09608 [nucl-th]}}.
%%CITATION = ARXIV:1609.09608;%%.

\bibitem{Helenius:2016dsk}
I.~Helenius, H.~Paukkunen, and K.~J. Eskola, ``{Neutron-skin effect in
  direct-photon and charged hadron-production in \PbPb\ collisions at the
  LHC},'' \href{http://dx.doi.org/10.1140/epjc/s10052-017-4709-9}{{\em Eur.
  Phys. J. C} {\bfseries 77} (2017) 148},
\href{http://arxiv.org/abs/1606.06910}{{\ttfamily arXiv:1606.06910 [hep-ph]}}.
%%CITATION = ARXIV:1606.06910;%%.

\bibitem{Alvioli:2018jls}
M.~Alvioli and M.~Strikman, ``{Spin-isospin correlated configurations in
  complex nuclei and neutron skin effect in W$^\pm$ production in high-energy
  proton-lead collisions},''
  \href{http://dx.doi.org/10.1103/PhysRevC.100.024912}{{\em Phys. Rev. C}
  {\bfseries 100} (2019) 024912},
  \href{http://arxiv.org/abs/1811.10078}{{\ttfamily arXiv:1811.10078
  [hep-ph]}}.

\bibitem{Durham:2018btt}
{\bfseries PHENIX} Collaboration, J.~M. Durham, ``{PHENIX results on system
  size dependence of J/$\psi$ nuclear modification in $p, d,$ $^3$He+A
  collisions at $\sqrt{s_{\rm NN}}$=200 GeV},''
  \href{http://dx.doi.org/10.22323/1.345.0165}{{\em PoS} {\bfseries
  HardProbes2018} (2018) 165},
  \href{http://arxiv.org/abs/1812.08889}{{\ttfamily arXiv:1812.08889
  [nucl-ex]}}.

\bibitem{Adamczyk:2015obl}
{\bfseries STAR} Collaboration, L.~Adamczyk {\em et~al.}, ``{Azimuthal
  anisotropy in U$+$U and Au$+$Au collisions at RHIC},''
  \href{http://dx.doi.org/10.1103/PhysRevLett.115.222301}{{\em Phys. Rev.
  Lett.} {\bfseries 115} (2015) 222301},
  \href{http://arxiv.org/abs/1505.07812}{{\ttfamily arXiv:1505.07812
  [nucl-ex]}}.

\bibitem{Hulthen1957}
L.~Hulth{\'e}n and M.~Sugawara, {\em The two-nucleon problem},
  \href{http://dx.doi.org/10.1007/978-3-642-45872-9_1}{pp.~1--143}.
\newblock Springer Berlin Heidelberg, Berlin, Heidelberg, 1957.

\bibitem{Adler:2003ii}
{\bfseries PHENIX} Collaboration, S.~S. Adler {\em et~al.}, ``{Absence of
  suppression in particle production at large transverse momentum in $\snn =
  200$ GeV \dAu\ collisions},''
  \href{http://dx.doi.org/10.1103/PhysRevLett.91.072303}{{\em Phys. Rev. Lett.}
  {\bfseries 91} (2003) 072303},
\href{http://arxiv.org/abs/nucl-ex/0306021}{{\ttfamily arXiv:nucl-ex/0306021
  [nucl-ex]}}.
%%CITATION = NUCL-EX/0306021;%%.

\bibitem{Adler:2006xd}
{\bfseries PHENIX} Collaboration, S.~S. Adler {\em et~al.}, ``{Nuclear effects
  on hadron production in \dAu\ and \pp\ collisions at $\snn=200$ GeV},''
  \href{http://dx.doi.org/10.1103/PhysRevC.74.024904}{{\em Phys. Rev. C}
  {\bfseries 74} (2006) 024904},
\href{http://arxiv.org/abs/nucl-ex/0603010}{{\ttfamily arXiv:nucl-ex/0603010
  [nucl-ex]}}.
%%CITATION = NUCL-EX/0603010;%%.

\bibitem{Nagle:2013lja}
J.~L. Nagle {\em et~al.}, ``{Exploiting intrinsic triangular geometry in
  relativistic \HeAu\ collisions to disentangle medium properties},''
  \href{http://dx.doi.org/10.1103/PhysRevLett.113.112301}{{\em Phys. Rev.
  Lett.} {\bfseries 113} (2014) 112301},
\href{http://arxiv.org/abs/1312.4565}{{\ttfamily arXiv:1312.4565 [nucl-th]}}.
%%CITATION = ARXIV:1312.4565;%%.

\bibitem{Lim:2018huo}
S.~Lim {\em et~al.}, ``{Exploring new small system geometries in heavy ion
  collisions},'' \href{http://dx.doi.org/10.1103/PhysRevC.99.044904}{{\em Phys.
  Rev. C} {\bfseries 99} (2019) 044904},
  \href{http://arxiv.org/abs/1812.08096}{{\ttfamily arXiv:1812.08096
  [nucl-th]}}.

\bibitem{Shou:2014eya}
Q.~Y. Shou, Y.~G. Ma, P.~Sorensen, A.~H. Tang, F.~Videbæk, and H.~Wang,
  ``{Parameterization of deformed nuclei for Glauber modeling in relativistic
  heavy-ion collisions},''
  \href{http://dx.doi.org/10.1016/j.physletb.2015.07.078}{{\em Phys. Lett. B}
  {\bfseries 749} (2015) 215--220},
\href{http://arxiv.org/abs/1409.8375}{{\ttfamily arXiv:1409.8375 [nucl-th]}}.
%%CITATION = ARXIV:1409.8375;%%.

\bibitem{Noronha-Hostler:2019ytn}
J.~Noronha-Hostler, N.~Paladino, S.~Rao, M.~D. Sievert, and D.~E. Wertepny,
  ``{Ultracentral collisions of small and deformed systems at RHIC: UU, dAu,
  ${}^{9}$BeAu, ${}^{9}$Be${}^{9}$Be ${}^{3}$He${}^{3}$He, and ${}^{3}$HeAu
  collisions},'' \href{http://arxiv.org/abs/1905.13323}{{\ttfamily
  arXiv:1905.13323 [hep-ph]}}.

\bibitem{Giacalone:2019pca}
G.~Giacalone, ``{Observing the deformation of nuclei with relativistic nuclear
  collisions},'' \href{http://dx.doi.org/10.1103/PhysRevLett.124.202301}{{\em
  Phys. Rev. Lett.} {\bfseries 124} (2020) 202301},
  \href{http://arxiv.org/abs/1910.04673}{{\ttfamily arXiv:1910.04673
  [nucl-th]}}.

\bibitem{Huang:2019tgz}
S.~Huang, Z.~Chen, J.~Jia, and W.~Li, ``{Disentangling contributions to
  small-system collectivity via scans of light nucleus-nucleus collisions},''
  \href{http://dx.doi.org/10.1103/PhysRevC.101.021901}{{\em Phys. Rev. C}
  {\bfseries 101} (2020) 021901},
  \href{http://arxiv.org/abs/1904.10415}{{\ttfamily arXiv:1904.10415
  [nucl-ex]}}.

\bibitem{Sievert:2019zjr}
M.~D. Sievert and J.~Noronha-Hostler, ``{CERN Large Hadron Collider system size
  scan predictions for \PbPb, \XeXe, \ArAr, and \OO\ with relativistic
  hydrodynamics},'' \href{http://dx.doi.org/10.1103/PhysRevC.100.024904}{{\em
  Phys. Rev. C} {\bfseries 100} (2019) 024904},
  \href{http://arxiv.org/abs/1901.01319}{{\ttfamily arXiv:1901.01319
  [nucl-th]}}.

\bibitem{Vogt:1999jp}
R.~Vogt, ``{Relation of hard and total cross-sections to centrality},'' {\em
  Acta Phys. Hung. A} {\bfseries 9} (1999) 339--348,
  \href{http://arxiv.org/abs/nucl-th/9903051}{{\ttfamily
  arXiv:nucl-th/9903051}}.

\bibitem{d'Enterria:2003qs}
D.~d'Enterria, ``{Hard scattering cross-sections at LHC in the Glauber
  approach: From \pp\ to \pA\ and \AA\ collisions},''
\href{http://arxiv.org/abs/nucl-ex/0302016}{{\ttfamily arXiv:nucl-ex/0302016
  [nucl-ex]}}.
%%CITATION = NUCL-EX/0302016;%%.

\bibitem{Adler:2005ig}
{\bfseries PHENIX} Collaboration, S.~Adler {\em et~al.}, ``{Centrality
  dependence of direct photon production in $\snn$=200 GeV \AuAu\
  collisions},'' \href{http://dx.doi.org/10.1103/PhysRevLett.94.232301}{{\em
  Phys. Rev. Lett.} {\bfseries 94} (2005) 232301},
  \href{http://arxiv.org/abs/nucl-ex/0503003}{{\ttfamily
  arXiv:nucl-ex/0503003}}.

\bibitem{Chatrchyan:2012vq}
{\bfseries CMS} Collaboration, S.~Chatrchyan {\em et~al.}, ``{Measurement of
  isolated photon production in \pp\ and \PbPb\ collisions at $\snn=2.76$
  TeV},'' \href{http://dx.doi.org/10.1016/j.physletb.2012.02.077}{{\em Phys.
  Lett. B} {\bfseries 710} (2012) 256--277},
  \href{http://arxiv.org/abs/1201.3093}{{\ttfamily arXiv:1201.3093 [nucl-ex]}}.

\bibitem{Aad:2015lcb}
{\bfseries ATLAS} Collaboration, G.~Aad {\em et~al.}, ``{Centrality, rapidity
  and transverse momentum dependence of isolated prompt photon production in
  lead-lead collisions at $\snn=2.76$ TeV measured with the ATLAS detector},''
  \href{http://dx.doi.org/10.1103/PhysRevC.93.034914}{{\em Phys. Rev. C}
  {\bfseries 93} (2016) 034914},
  \href{http://arxiv.org/abs/1506.08552}{{\ttfamily arXiv:1506.08552
  [hep-ex]}}.

\bibitem{Sirunyan:2020ycu}
{\bfseries CMS} Collaboration, A.~M. Sirunyan {\em et~al.}, ``{The production
  of isolated photons in \PbPb\ and \pp\ collisions at $\snn=5.02$ TeV},''
  \href{http://dx.doi.org/10.1007/JHEP07(2020)116}{{\em JHEP} {\bfseries 07}
  (2020) 116}, \href{http://arxiv.org/abs/2003.12797}{{\ttfamily
  arXiv:2003.12797 [hep-ex]}}.

\bibitem{Aad:2012ew}
{\bfseries ATLAS} Collaboration, G.~Aad {\em et~al.}, ``{Measurement of $Z$
  boson production in \PbPb\ collisions at $\snn=2.76$ TeV with the ATLAS
  detector},'' \href{http://dx.doi.org/10.1103/PhysRevLett.110.022301}{{\em
  Phys. Rev. Lett.} {\bfseries 110} (2013) 022301},
  \href{http://arxiv.org/abs/1210.6486}{{\ttfamily arXiv:1210.6486 [hep-ex]}}.

\bibitem{Aad:2014bha}
{\bfseries ATLAS} Collaboration, G.~Aad {\em et~al.}, ``{Measurement of the
  production and lepton charge asymmetry of $W$ bosons in \PbPb\ collisions at
  $\snn=2.76$ TeV with the ATLAS detector},''
  \href{http://dx.doi.org/10.1140/epjc/s10052-014-3231-6}{{\em Eur. Phys. J. C}
  {\bfseries 75} (2015) 23}, \href{http://arxiv.org/abs/1408.4674}{{\ttfamily
  arXiv:1408.4674 [hep-ex]}}.

\bibitem{Aad:2015gta}
{\bfseries ATLAS} Collaboration, G.~Aad {\em et~al.}, ``{$Z$ boson production
  in \pPb\ collisions at $\snn=5.02$ TeV measured with the ATLAS detector},''
  \href{http://dx.doi.org/10.1103/PhysRevC.92.044915}{{\em Phys. Rev. C}
  {\bfseries 92} (2015) 044915},
  \href{http://arxiv.org/abs/1507.06232}{{\ttfamily arXiv:1507.06232
  [hep-ex]}}.

\bibitem{Khachatryan:2015pzs}
{\bfseries CMS} Collaboration, V.~Khachatryan {\em et~al.}, ``{Study of Z boson
  production in \pPb\ collisions at $\snn=5.02$ TeV},''
  \href{http://dx.doi.org/10.1016/j.physletb.2016.05.044}{{\em Phys. Lett. B}
  {\bfseries 759} (2016) 36--57},
  \href{http://arxiv.org/abs/1512.06461}{{\ttfamily arXiv:1512.06461
  [hep-ex]}}.

\bibitem{Acharya:2017wpf}
{\bfseries ALICE} Collaboration, S.~Acharya {\em et~al.}, ``{Measurement of
  Z$^0$-boson production at large rapidities in Pb-Pb collisions at
  $\sqrt{s_{\rm NN}}=5.02$ TeV},''
  \href{http://dx.doi.org/10.1016/j.physletb.2018.03.010}{{\em Phys. Lett. B}
  {\bfseries 780} (2018) 372--383},
  \href{http://arxiv.org/abs/1711.10753}{{\ttfamily arXiv:1711.10753
  [nucl-ex]}}.

\bibitem{Aad:2019lan}
{\bfseries ATLAS} Collaboration, G.~Aad {\em et~al.}, ``{$Z$ boson production
  in \PbPb\ collisions at $\snn$=5.02 TeV measured by the ATLAS experiment},''
  \href{http://dx.doi.org/10.1016/j.physletb.2020.135262}{{\em Phys. Lett. B}
  {\bfseries 802} (2020) 135262},
  \href{http://arxiv.org/abs/1910.13396}{{\ttfamily arXiv:1910.13396
  [nucl-ex]}}.

\bibitem{Aad:2019sfe}
{\bfseries ATLAS} Collaboration, G.~Aad {\em et~al.}, ``{Measurement of $W^\pm
  $ boson production in \PbPb\ collisions at $\snn=5.02$ TeV with the ATLAS
  detector},'' \href{http://dx.doi.org/10.1140/epjc/s10052-019-7439-3}{{\em
  Eur. Phys. J. C} {\bfseries 79} (2019) 935},
  \href{http://arxiv.org/abs/1907.10414}{{\ttfamily arXiv:1907.10414
  [nucl-ex]}}.

\bibitem{Sirunyan:2019dox}
{\bfseries CMS} Collaboration, A.~M. Sirunyan {\em et~al.}, ``{Observation of
  nuclear modifications in W$^\pm$ boson production in \pPb\ collisions at
  $\snn=8.16$ TeV},''
  \href{http://dx.doi.org/10.1016/j.physletb.2019.135048}{{\em Phys. Lett. B}
  {\bfseries 800} (2020) 135048},
  \href{http://arxiv.org/abs/1905.01486}{{\ttfamily arXiv:1905.01486
  [hep-ex]}}.

\bibitem{Eskola:2016oht}
K.~J. Eskola, P.~Paakkinen, H.~Paukkunen, and C.~A. Salgado, ``{EPPS16: Nuclear
  parton distributions with LHC data},''
  \href{http://dx.doi.org/10.1140/epjc/s10052-017-4725-9}{{\em Eur. Phys. J. C}
  {\bfseries 77} (2017) 163}, \href{http://arxiv.org/abs/1612.05741}{{\ttfamily
  arXiv:1612.05741 [hep-ph]}}.

\bibitem{Kusina:2016fxy}
A.~Kusina {\em et~al.}, ``{Vector boson production in \pPb\ and \PbPb\
  collisions at the LHC and its impact on nCTEQ15 PDFs},''
  \href{http://dx.doi.org/10.1140/epjc/s10052-017-5036-x}{{\em Eur. Phys. J. C}
  {\bfseries 77} (2017) 488}, \href{http://arxiv.org/abs/1610.02925}{{\ttfamily
  arXiv:1610.02925 [nucl-th]}}.

\bibitem{AbdulKhalek:2020yuc}
R.~Abdul~Khalek, J.~J. Ethier, J.~Rojo, and G.~van Weelden, ``{nNNPDF2.0: quark
  flavor separation in nuclei from LHC data},''
  \href{http://dx.doi.org/10.1007/JHEP09(2020)183}{{\em JHEP} {\bfseries 09}
  (2020) 183}, \href{http://arxiv.org/abs/2006.14629}{{\ttfamily
  arXiv:2006.14629 [hep-ph]}}.

\bibitem{Khanpour:2016pph}
H.~Khanpour and S.~Atashbar~Tehrani, ``{Global analysis of nuclear parton
  distribution functions and their uncertainties at next-to-next-to-leading
  order},'' \href{http://dx.doi.org/10.1103/PhysRevD.93.014026}{{\em Phys. Rev.
  D} {\bfseries 93} (2016) 014026},
  \href{http://arxiv.org/abs/1601.00939}{{\ttfamily arXiv:1601.00939
  [hep-ph]}}.

\bibitem{Eskola:2020lee}
K.~J. Eskola, I.~Helenius, M.~Kuha, and H.~Paukkunen, ``{Shadowing in inelastic
  nucleon-nucleon cross section?},''
  \href{http://dx.doi.org/10.1103/PhysRevLett.125.212301}{{\em Phys. Rev.
  Lett.} {\bfseries 125} (2020) 212301},
  \href{http://arxiv.org/abs/2003.11856}{{\ttfamily arXiv:2003.11856
  [hep-ph]}}.

\bibitem{Deng:2010mv}
W.-T. Deng, X.-N. Wang, and R.~Xu, ``{Hadron production in \pp, \pPb, and
  \PbPb\ collisions with the HIJING 2.0 model at energies available at the CERN
  Large Hadron Collider},''
  \href{http://dx.doi.org/10.1103/PhysRevC.83.014915}{{\em Phys. Rev. C}
  {\bfseries 83} (2011) 014915},
  \href{http://arxiv.org/abs/1008.1841}{{\ttfamily arXiv:1008.1841 [hep-ph]}}.

\bibitem{Pierog:2013ria}
T.~Pierog, I.~Karpenko, J.~Katzy, E.~Yatsenko, and K.~Werner, ``{EPOS LHC: Test
  of collective hadronization with data measured at the CERN Large Hadron
  Collider},'' \href{http://dx.doi.org/10.1103/PhysRevC.92.034906}{{\em Phys.
  Rev. C} {\bfseries 92} (2015) 034906},
  \href{http://arxiv.org/abs/1306.0121}{{\ttfamily arXiv:1306.0121 [hep-ph]}}.

\bibitem{Lin:2004en}
Z.-W. Lin, C.~M. Ko, B.-A. Li, B.~Zhang, and S.~Pal, ``{A Multi-phase transport
  model for relativistic heavy ion collisions},''
  \href{http://dx.doi.org/10.1103/PhysRevC.72.064901}{{\em Phys. Rev. C}
  {\bfseries 72} (2005) 064901},
  \href{http://arxiv.org/abs/nucl-th/0411110}{{\ttfamily
  arXiv:nucl-th/0411110}}.

\bibitem{Ostapchenko:2010vb}
S.~Ostapchenko, ``{Monte Carlo treatment of hadronic interactions in enhanced
  Pomeron scheme: I. QGSJET-II model},''
  \href{http://dx.doi.org/10.1103/PhysRevD.83.014018}{{\em Phys. Rev. D}
  {\bfseries 83} (2011) 014018},
  \href{http://arxiv.org/abs/1010.1869}{{\ttfamily arXiv:1010.1869 [hep-ph]}}.

\bibitem{Roesler:2000he}
S.~Roesler, R.~Engel, and J.~Ranft,
  \href{http://dx.doi.org/10.1007/978-3-642-18211-2_166}{``{The Monte Carlo
  event generator DPMJET-III},''} in {\em {Intl. Conf. on Adv. Monte Carlo for
  Rad. Phys., Part. Transp. Simul. and Appl. (MC 2000)}}, pp.~1033--1038.
\newblock 12, 2000.
\newblock \href{http://arxiv.org/abs/hep-ph/0012252}{{\ttfamily
  arXiv:hep-ph/0012252}}.

\bibitem{Sjostrand:2007gs}
T.~Sj{\"o}strand, S.~Mrenna, and P.~Z. Skands, ``{A Brief Introduction to
  PYTHIA 8.1},'' \href{http://dx.doi.org/10.1016/j.cpc.2008.01.036}{{\em
  Comput. Phys. Commun.} {\bfseries 178} (2008) 852--867},
  \href{http://arxiv.org/abs/0710.3820}{{\ttfamily arXiv:0710.3820 [hep-ph]}}.

\bibitem{Bierlich:2018xfw}
C.~Bierlich, G.~Gustafson, L.~L{\"o}nnblad, and H.~Shah, ``{The Angantyr model
  for heavy-ion collisions in PYTHIA8},''
  \href{http://dx.doi.org/10.1007/JHEP10(2018)134}{{\em JHEP} {\bfseries 10}
  (2018) 134}, \href{http://arxiv.org/abs/1806.10820}{{\ttfamily
  arXiv:1806.10820 [hep-ph]}}.

\bibitem{dEnterria:2011twh}
D.~d'Enterria, R.~Engel, T.~Pierog, S.~Ostapchenko, and K.~Werner,
  ``{Constraints from the first LHC data on hadronic event generators for
  ultra-high energy cosmic-ray physics},''
  \href{http://dx.doi.org/10.1016/j.astropartphys.2011.05.002}{{\em Astropart.
  Phys.} {\bfseries 35} (2011) 98--113},
  \href{http://arxiv.org/abs/1101.5596}{{\ttfamily arXiv:1101.5596
  [astro-ph.HE]}}.

\bibitem{Andersson:1998tv}
B.~Andersson, {\em {The Lund model}}, vol.~7.
\newblock Cambridge University Press, 7, 2005.

\bibitem{Andersson:1986gw}
B.~Andersson, G.~Gustafson, and B.~Nilsson-Almqvist, ``{A model for low $\pT$
  hadronic reactions, with generalizations to hadron--nucleus and
  nucleus--nucleus collisions},''
  \href{http://dx.doi.org/10.1016/0550-3213(87)90257-4}{{\em Nucl. Phys. B}
  {\bfseries 281} (1987) 289--309}.

\bibitem{Klein:2016yzr}
S.~R. Klein, J.~Nystrand, J.~Seger, Y.~Gorbunov, and J.~Butterworth,
  ``{STARlight: A Monte Carlo simulation program for ultra-peripheral
  collisions of relativistic ions},''
  \href{http://dx.doi.org/10.1016/j.cpc.2016.10.016}{{\em Comput. Phys.
  Commun.} {\bfseries 212} (2017) 258--268},
  \href{http://arxiv.org/abs/1607.03838}{{\ttfamily arXiv:1607.03838
  [hep-ph]}}.

\bibitem{Harland-Lang:2018iur}
L.~Harland-Lang, V.~Khoze, and M.~Ryskin, ``{Exclusive LHC physics with heavy
  ions: SuperChic 3},''
  \href{http://dx.doi.org/10.1140/epjc/s10052-018-6530-5}{{\em Eur. Phys. J. C}
  {\bfseries 79} (2019) 39}, \href{http://arxiv.org/abs/1810.06567}{{\ttfamily
  arXiv:1810.06567 [hep-ph]}}.

\bibitem{Heinz:2009xj}
U.~W. Heinz, {\em {Early collective expansion: Relativistic hydrodynamics and
  the transport properties of QCD matter}}, vol.~23,
  \href{http://dx.doi.org/10.1007/978-3-642-01539-7_9}{p.~240}.
\newblock 2010.
\newblock \href{http://arxiv.org/abs/0901.4355}{{\ttfamily arXiv:0901.4355
  [nucl-th]}}.

\bibitem{Hirano:2012kj}
T.~Hirano, P.~Huovinen, K.~Murase, and Y.~Nara, ``{Integrated dynamical
  approach to relativistic heavy ion collisions},''
  \href{http://dx.doi.org/10.1016/j.ppnp.2013.02.002}{{\em Prog. Part. Nucl.
  Phys.} {\bfseries 70} (2013) 108--158},
  \href{http://arxiv.org/abs/1204.5814}{{\ttfamily arXiv:1204.5814 [nucl-th]}}.

\bibitem{Gale:2013da}
C.~Gale, S.~Jeon, and B.~Schenke, ``{Hydrodynamic modeling of heavy-ion
  collisions},'' \href{http://dx.doi.org/10.1142/S0217751X13400113}{{\em Int.
  J. Mod. Phys. A} {\bfseries 28} (2013) 1340011},
  \href{http://arxiv.org/abs/1301.5893}{{\ttfamily arXiv:1301.5893 [nucl-th]}}.

\bibitem{Petersen:2008dd}
H.~Petersen, J.~Steinheimer, G.~Burau, M.~Bleicher, and H.~Stocker, ``{A fully
  integrated transport approach to heavy ion reactions with an intermediate
  hydrodynamic sStage},''
  \href{http://dx.doi.org/10.1103/PhysRevC.78.044901}{{\em Phys. Rev. C}
  {\bfseries 78} (2008) 044901},
  \href{http://arxiv.org/abs/0806.1695}{{\ttfamily arXiv:0806.1695 [nucl-th]}}.

\bibitem{Strickland:2013uga}
M.~Strickland, ``{Thermalization and isotropization in heavy-ion collisions},''
  \href{http://dx.doi.org/10.1007/s12043-015-0972-1}{{\em Pramana} {\bfseries
  84} (2015) 671--684}, \href{http://arxiv.org/abs/1312.2285}{{\ttfamily
  arXiv:1312.2285 [hep-ph]}}.

\bibitem{Kharzeev:2001yq}
D.~Kharzeev, E.~Levin, and M.~Nardi, ``{The Onset of classical QCD dynamics in
  relativistic heavy ion collisions},''
  \href{http://dx.doi.org/10.1103/PhysRevC.71.054903}{{\em Phys. Rev. C}
  {\bfseries 71} (2005) 054903},
  \href{http://arxiv.org/abs/hep-ph/0111315}{{\ttfamily arXiv:hep-ph/0111315}}.

\bibitem{Drescher:2006ca}
H.-J. Drescher and Y.~Nara, ``{Effects of fluctuations on the initial
  eccentricity from the Color Glass Condensate in heavy ion collisions},''
  \href{http://dx.doi.org/10.1103/PhysRevC.75.034905}{{\em Phys. Rev. C}
  {\bfseries 75} (2007) 034905},
  \href{http://arxiv.org/abs/nucl-th/0611017}{{\ttfamily
  arXiv:nucl-th/0611017}}.

\bibitem{Schenke:2012hg}
B.~Schenke, P.~Tribedy, and R.~Venugopalan, ``{Event-by-event gluon
  multiplicity, energy density, and eccentricities in ultrarelativistic
  heavy-ion collisions},''
  \href{http://dx.doi.org/10.1103/PhysRevC.86.034908}{{\em Phys. Rev. C}
  {\bfseries 86} (2012) 034908},
  \href{http://arxiv.org/abs/1206.6805}{{\ttfamily arXiv:1206.6805 [hep-ph]}}.

\bibitem{Schenke:2012wb}
B.~Schenke, P.~Tribedy, and R.~Venugopalan, ``{Fluctuating Glasma initial
  conditions and flow in heavy ion collisions},''
  \href{http://dx.doi.org/10.1103/PhysRevLett.108.252301}{{\em Phys. Rev.
  Lett.} {\bfseries 108} (2012) 252301},
  \href{http://arxiv.org/abs/1202.6646}{{\ttfamily arXiv:1202.6646 [nucl-th]}}.

\bibitem{Gelis:2010nm}
F.~Gelis, E.~Iancu, J.~Jalilian-Marian, and R.~Venugopalan, ``{The Color Glass
  Condensate},''
  \href{http://dx.doi.org/10.1146/annurev.nucl.010909.083629}{{\em Ann. Rev.
  Nucl. Part. Sci.} {\bfseries 60} (2010) 463--489},
  \href{http://arxiv.org/abs/1002.0333}{{\ttfamily arXiv:1002.0333 [hep-ph]}}.

\bibitem{Eskola:1999fc}
K.~Eskola, K.~Kajantie, P.~Ruuskanen, and K.~Tuominen, ``{Scaling of transverse
  energies and multiplicities with atomic number and energy in
  ultrarelativistic nuclear collisions},''
  \href{http://dx.doi.org/10.1016/S0550-3213(99)00720-8}{{\em Nucl. Phys. B}
  {\bfseries 570} (2000) 379--389},
  \href{http://arxiv.org/abs/hep-ph/9909456}{{\ttfamily arXiv:hep-ph/9909456}}.

\bibitem{Niemi:2015qia}
H.~Niemi, K.~Eskola, and R.~Paatelainen, ``{Event-by-event fluctuations in a
  perturbative QCD + saturation + hydrodynamics model: Determining QCD matter
  shear viscosity in ultrarelativistic heavy-ion collisions},''
  \href{http://dx.doi.org/10.1103/PhysRevC.93.024907}{{\em Phys. Rev. C}
  {\bfseries 93} (2016) 024907},
  \href{http://arxiv.org/abs/1505.02677}{{\ttfamily arXiv:1505.02677
  [hep-ph]}}.

\bibitem{Moreland:2014oya}
J.~S. Moreland, J.~E. Bernhard, and S.~A. Bass, ``{Alternative ansatz to
  wounded nucleon and binary collision scaling in high-energy nuclear
  collisions},'' \href{http://dx.doi.org/10.1103/PhysRevC.92.011901}{{\em Phys.
  Rev. C} {\bfseries 92} (2015) 011901},
  \href{http://arxiv.org/abs/1412.4708}{{\ttfamily arXiv:1412.4708 [nucl-th]}}.

\bibitem{Kowalski:2003hm}
H.~Kowalski and D.~Teaney, ``{An Impact parameter dipole saturation model},''
  \href{http://dx.doi.org/10.1103/PhysRevD.68.114005}{{\em Phys. Rev. D}
  {\bfseries 68} (2003) 114005},
  \href{http://arxiv.org/abs/hep-ph/0304189}{{\ttfamily arXiv:hep-ph/0304189}}.

\bibitem{Acharya:2018ihu}
{\bfseries ALICE} Collaboration, S.~Acharya {\em et~al.}, ``{Anisotropic flow
  in \XeXe\ collisions at $\snn=5.44$ TeV},''
  \href{http://dx.doi.org/10.1016/j.physletb.2018.06.059}{{\em Phys. Lett. B}
  {\bfseries 784} (2018) 82--95},
  \href{http://arxiv.org/abs/1805.01832}{{\ttfamily arXiv:1805.01832
  [nucl-ex]}}.

\bibitem{Nagle:2018ybc}
J.~Nagle and W.~Zajc, ``{Assessing saturation physics explanations of
  collectivity in small collision systems with the IP-Jazma model},''
  \href{http://dx.doi.org/10.1103/PhysRevC.99.054908}{{\em Phys. Rev. C}
  {\bfseries 99} (2019) 054908},
  \href{http://arxiv.org/abs/1808.01276}{{\ttfamily arXiv:1808.01276
  [nucl-th]}}.

\bibitem{Abelev:2013qoq}
{\bfseries ALICE} Collaboration, B.~Abelev {\em et~al.}, ``{Centrality
  determination of \PbPb\ collisions at $\snn=2.76$ TeV with ALICE},''
  \href{http://dx.doi.org/10.1103/PhysRevC.88.044909}{{\em Phys. Rev. C}
  {\bfseries 88} (2013) 044909},
\href{http://arxiv.org/abs/1301.4361}{{\ttfamily arXiv:1301.4361 [nucl-ex]}}.
%%CITATION = ARXIV:1301.4361;%%.

\bibitem{ALICE:2018tvk}
{\bfseries ALICE} Collaboration, S.~Acharya, ``{Centrality determination in
  heavy ion collisions},'' {\em ALICE-PUBLIC-2018-011} (8, 2018) .
  \url{https://cds.cern.ch/record/2636623}.

\bibitem{Chatrchyan:2011pb}
{\bfseries CMS} Collaboration, S.~Chatrchyan {\em et~al.}, ``{Dependence on
  pseudorapidity and centrality of charged hadron production in \PbPb\
  collisions at a nucleon-nucleon centre-of-mass energy of 2.76 TeV},''
  \href{http://dx.doi.org/10.1007/JHEP08(2011)141}{{\em JHEP} {\bfseries 08}
  (2011) 141},
\href{http://arxiv.org/abs/1107.4800}{{\ttfamily arXiv:1107.4800 [nucl-ex]}}.
%%CITATION = ARXIV:1107.4800;%%.

\bibitem{ATLAS:2011ag}
{\bfseries ATLAS} Collaboration, G.~Aad {\em et~al.}, ``{Measurement of the
  centrality dependence of the charged particle pseudorapidity distribution in
  lead-lead collisions at $\snn=2.76$ TeV with the ATLAS detector},''
  \href{http://dx.doi.org/10.1016/j.physletb.2012.02.045}{{\em Phys. Lett. B}
  {\bfseries 710} (2012) 363--382},
\href{http://arxiv.org/abs/1108.6027}{{\ttfamily arXiv:1108.6027 [hep-ex]}}.
%%CITATION = ARXIV:1108.6027;%%.

\bibitem{Wood:2012jem}
J.~S. Wood, \href{http://dx.doi.org/10.2172/1369234}{{\em {The development of
  the CMS Zero Degree Calorimeters to derive the centrality of \AA\
  collisions}}}.
\newblock PhD thesis, Kansas U., 2012.

\bibitem{Adare:2013nff}
{\bfseries PHENIX} Collaboration, A.~Adare {\em et~al.}, ``{Centrality
  categorization for $R_{p(d)+A}$ in high-energy collisions},''
  \href{http://dx.doi.org/10.1103/PhysRevC.90.034902}{{\em Phys. Rev. C}
  {\bfseries 90} (2014) 034902},
  \href{http://arxiv.org/abs/1310.4793}{{\ttfamily arXiv:1310.4793 [nucl-ex]}}.

\bibitem{Adam:2014qja}
{\bfseries ALICE} Collaboration, J.~Adam {\em et~al.}, ``{Centrality dependence
  of particle production in \pPb\ collisions at $\snn$= 5.02 TeV},''
  \href{http://dx.doi.org/10.1103/PhysRevC.91.064905}{{\em Phys. Rev. C}
  {\bfseries 91} (2015) 064905},
  \href{http://arxiv.org/abs/1412.6828}{{\ttfamily arXiv:1412.6828 [nucl-ex]}}.

\bibitem{Perepelitsa:2014yta}
D.~V. Perepelitsa and P.~A. Steinberg, ``{Calculation of centrality bias
  factors in \pA\ collisions based on a positive correlation of hard process
  yields with underlying event activity},''
  \href{http://arxiv.org/abs/1412.0976}{{\ttfamily arXiv:1412.0976 [nucl-ex]}}.

\bibitem{Acharya:2018njl}
{\bfseries ALICE} Collaboration, S.~Acharya {\em et~al.}, ``{Analysis of the
  apparent nuclear modification in peripheral \PbPb\ collisions at 5.02 TeV},''
  \href{http://dx.doi.org/10.1016/j.physletb.2019.04.047}{{\em Phys. Lett. B}
  {\bfseries 793} (2019) 420--432},
  \href{http://arxiv.org/abs/1805.05212}{{\ttfamily arXiv:1805.05212
  [nucl-ex]}}.

\bibitem{CMS:2020rno}
{\bfseries CMS} Collaboration, ``{New constraints of initial states in \PbPb\
  collisions with Z boson yields and azimuthal anisotropy at $\snn=5.02$
  TeV},'' {\em \url{http://inspirehep.net/literature/1799323}} (6, 2020) .

\bibitem{Tsukada:2017llu}
K.~Tsukada {\em et~al.}, ``{First elastic electron scattering from $^{132}$Xe
  at the SCRIT facility},''
  \href{http://dx.doi.org/10.1103/PhysRevLett.118.262501}{{\em Phys. Rev.
  Lett.} {\bfseries 118} (2017) 262501},
\href{http://arxiv.org/abs/1703.04278}{{\ttfamily arXiv:1703.04278 [nucl-ex]}}.
%%CITATION = ARXIV:1703.04278;%%.

\bibitem{Hirano:2009ah}
T.~Hirano and Y.~Nara, ``{Eccentricity fluctuation effects on elliptic flow in
  relativistic heavy ion collisions},''
  \href{http://dx.doi.org/10.1103/PhysRevC.79.064904}{{\em Phys. Rev. C}
  {\bfseries 79} (2009) 064904},
\href{http://arxiv.org/abs/0904.4080}{{\ttfamily arXiv:0904.4080 [nucl-th]}}.
%%CITATION = ARXIV:0904.4080;%%.

\bibitem{glaucode}
``{TGlauberMC on HepForge}.'' \url{http://tglaubermc.hepforge.org/}.
\newblock {v3.2}.

\bibitem{fischer:1974}
{W. Fischer and others}, ``{Isotope shifts in the atomic spectrum of xenon and
  nuclear deformation effects},'' {\em Z. Physik} {\bfseries 270} (1974) 113.

\bibitem{ALICE-PUBLIC-2018-003}
{\bfseries ALICE} Collaboration, {S. Acharya and others}, ``{Centrality
  determination using the Glauber model in \XeXe\ collisions at $\snn=5.44$
  TeV},''. \url{https://cds.cern.ch/record/2315401}.

\bibitem{Heinz:2004ir}
U.~W. Heinz and A.~Kuhlman, ``{Anisotropic flow and jet quenching in
  ultrarelativistic \UU\ collisions},''
  \href{http://dx.doi.org/10.1103/PhysRevLett.94.132301}{{\em Phys. Rev. Lett.}
  {\bfseries 94} (2005) 132301},
\href{http://arxiv.org/abs/nucl-th/0411054}{{\ttfamily arXiv:nucl-th/0411054
  [nucl-th]}}.
%%CITATION = NUCL-TH/0411054;%%.

\end{thebibliography}\endgroup
\end{document}